\documentclass[aps,pre,twocolumn,superscriptaddress,showpacs]{revtex4-1}
\usepackage{graphicx,amsbsy,amsmath,epsfig,natbib,float}
\usepackage{amssymb,latexsym,wasysym,pifont,mathrsfs}
\usepackage{comment,verbatim,multirow,array,sidecap,color}
\usepackage{lineno,setspace,soul,cancel}
\usepackage[normalem]{ulem}

\begin{document}

\title{Yielding transition of a two dimensional glass former under
  athermal cyclic shear deformation}

\author{Himangsu Bhaumik} \affiliation{Jawaharlal
  Nehru Center for Advanced Scientific Research, Jakkur Campus, Bengaluru 560064, India.}

\author{Giuseppe Foffi} \affiliation{Universit\'e Paris-Saclay, CNRS, Laboratoire de Physique des Solides, 91405 Orsay, France}
\author{Srikanth Sastry} \email{sastry@jncasr.ac.in} \affiliation{Jawaharlal
  Nehru Center for Advanced Scientific Research, Jakkur Campus, Bengaluru 560064, India.}

%\date{\today}
 
\begin{abstract}  
We study numerically the yielding transition of a two dimensional model glass  subjected to athermal quasi-static cyclic shear deformation, with the aim of investigating the effect on the yielding behaviour of the degree of annealing, which in turn depends on the preparation protocol. We find two distinct regimes of annealing separated by a threshold energy. Poorly annealed glasses progressively evolve towards the threshold energy as the strain amplitude is increased towards the yielding value. Well annealed glasses with initial energies below the threshold energy exhibit stable behaviour, with negligible change in energy with increasing strain amplitude, till they yield. Discontinuities in energy and stress at yielding increase with the degree of annealing, consistently with recent results found in three dimensions. We observe significant structural change with strain amplitude that closely mirrors the changes in energy and stresses. We investigate groups of particles that are involved in plastic rearrangements. We analyse the distributions of avalanche sizes, of clusters of connected rearranging particles, and related quantities, employing finite size scaling analysis. We verify previously investigated relations between exponents characterising these distributions, and  a newly proposed relation between exponents describing avalanche and cluster size distributions. 
\end{abstract}

\maketitle 

\section{Introduction}

The response of amorphous solids to external deformation or applied stress is of obvious importance in characterising their behaviour,  both from a  fundamental and an applied point of view \cite{Bonn2017c,Nicolas2018}. For small deformations, such response is characterized by elastic-like behaviour, but for large enough deformation the response begins to display plasticity and  irreversible flow-like behaviour. Yielding behaviour, and the transition to elasto-plastic flow, can be broadly classified into two different categories. Upon deformation, the stress may increase gradually and monotonically, till a steady value is reached for large deformation. Foams\cite{LauridsenPRL02}, emulsions\cite{BecuPRL06} and colloidal suspensions \cite{SchallSCI07} have been shown to display such phenomenology. On the other hand, one may observe a stress overshoot and a drop, accompanied by strain localization. This behaviour has been observed, for example,  in  metallic glasses \cite{GreerMSE13,HufnagelACMT16} and window glasses \cite{GuinPRL04}. The phenomenon of yielding is central to understanding  the mechanical properties in different materials spanning several scales, ranging from   nano-structured materials \cite{ZhaoNATCOM2016} to large scale events like earthquakes, landslides, and  avalanches~\cite{ChesterTEC1998,UhlSR2015}. In recent years, an increasing number of studies have explored the fundamental aspects of the yielding transition  in amorphous solids through  experiments\cite{KeimSM13,KnowltonSM14,NagamanasaPRE14,DiMicheleSM14,KeimPRR20}, numerical simulations \cite{maloneyPRE06,KarmakarPRE10,JaiswalPRL16,fioccoPRE13,ShrivastavPRE16,regev2015reversibility,leishangthemNAT2017,ParisiPNAS17,parmarPRX2019,BarbotPRE20,LembergPRE20} and theoretical models \cite{Urbani2017b,Popovic2018a,BarlowPRL20,liu2020oscillatory,sastryPRL20}. 
Differently from crystals~\cite{sethna2017deformation}, amorphous solid plasticity is not driven by specific defects and thus the correlation of plastic rearrangements and specific structural motifs has  been a subject of several investigations, as well as the characterisation of structural changes that arise from plastic events ~\cite{Richard2020,Bonfanti2019,parmarPRX2019,DenisovSR15,VasishtPRE20,Mitra_2021}. 
The distribution of sizes of events -- avalanches -- involved in plastic rearrangements
\cite{dahmenPRL09,LinPNAS14,regev2015reversibility,leishangthemNAT2017,oyama2020unified} have also been widely investigated, which display characteristic power law forms whose origins have been explored through elasto-plastic models and mean-field theories \cite{LinPNAS14,JaglaPRE15,liu2016driving,BouchbinderPRE07,DahmenNATPHY2011,FranzPRB17}.

Recent computer simulations, performed largely  using the athermal quasi-static shear deformation (AQS) implementing both cyclic and uniform shear, support the idea that  the nature of the yielding transition is highly sensitive on the initial degree of annealing of the amorphous \cite{ozawaPNAS2018,BhaumikPNAS21,yehprl20,OzawaPRR20}. For uniform shear, a threshold degree of annealing that separates the gradual and sharp/discontinuous yielding behaviours (termed in this context as ductile and brittle behaviour) has been argued to correspond to a random critical point~\cite{ozawaPNAS2018,OzawaPRR20}. Investigations of yielding under cyclic shear reveal that, for a well annealed glass (above a threshold degree of annealing or below a threshold energy), the drop in the maximum stress over a cycle is discontinuous at the yielding transition, and the magnitude of the stress jump increases with an increase in the degree of annealing. On the other hand, for poorly annealed glasses, irrespective of the initial energy, cyclic shear progressively anneals the glasses as the strain amplitude increases. In all such cases, independently of the initial (low) degree of annealing, the glasses approach  a common energy -- the threshold energy -- at the yielding point~\cite{BhaumikPNAS21,yehprl20}. The energy and stress jumps observed for initially poorly annealed glasses is small \cite{parmarPRX2019,BhaumikPNAS21}. In previous work, we have investigated such behaviour both in a strong glass (silica) and a fragile glass (the Kob-Andersen binary Lennard-Jones mixture) in three dimensions. 

Here, we investigate the corresponding properties of yielding under cyclic shear for a model two-dimensional glass. In particular, we explore whether the yielding transition displays the same kind of dependence on annealing as the three dimensional glasses. In addition to studying the behaviour of energy and stress, we also characterise the mechanically induced annealing and yielding by analysing the structural changes involved. 
%however, an apparently different picture emerged in the recent study of two dimensional amorphous solids under cyclic deformation where rejuvenation occurs for well annealed glasses before they yield \cite{LembergPRE20} an observation that  merits further investigations.
Above yielding, strain has been observed to become localised in shear bands under cyclic shear 
\cite{RadhakrishnanPRL16,fioccoPRE13,parmarPRX2019} as well as for annealed glasses under uniform shear \cite{shiprb06,shi2007,ozawaPNAS2018}. We investigate strain localisation in the model we study, focusing attention on the nature of structural change within the shear bands, compared to regions outside the shear band. Finally, we investigate the statistics of avalanches~\cite{dahmenPRL09,LinPNAS14,regev2015reversibility,leishangthemNAT2017}, {\it i.e.} the collection of particles undergoing plastic deformation, both below and above yielding. 
In previous work, the nature of rearrangements under cyclic shear has been investigated through the statistics of clusters of connected particles that are displaced beyond a threshold value \cite{leishangthemNAT2017}, for a three dimensional glass. The statistics of sizes of such clusters is qualitatively different below and above yielding. In particular, the characteristic size of such clusters was shown \cite{leishangthemNAT2017} to remain finite below the yielding transition, whereas above, their sizes scale with the system size. These observations are in contrast with simulations of a two-dimensional model \cite{regev2015reversibility} wherein the sizes of avalanches diverges as the yielding transition is approached from below. We investigate whether such a difference arises from dimensionality. Further, we perform a careful investigation of avalanches above yielding, by distinguishing the number  of all particles participating in a plastic rearrangement, for which we use the term {\it avalanche}, {\it vs.} spatially connected clusters of rearranging particles. We further investigate the distribution of the number of clusters that participate in an avalanche, to understand the clustering properties of avalanches recently discussed in \cite{PriolPRL21}. These analyses confirm earlier conclusions for three dimensional glasses \cite{leishangthemNAT2017}, and provide additional insights into the nature of avalanches in the two dimensional model we investigate. 
In particular, we verify a relation between exponents characterising the distribution of avalanches and clusters, proposed in an accompanying paper \cite{bhaumiksilica2021} and verified for three dimensional glasses.

The rest of the paper is organized as follows: In Section II we describe the model potential (II.A), methods of initial glass preparation (II.B) and protocol of cyclic shear that we use in this study (II.C). In Sec. III we present the results starting with the yielding transition (III.A), followed by results concerning structural change across the yielding transition (III.B) and strain localization (III.C). Next, we present a detailed statistical investigation of avalanches, analyzing various quantities related to plastic rearrangements (III.D). A rigorous finite size scaling analysis will be provided in support of various exponents and scaling relation we describe (III.E). Finally, Sec. IV contains a summary and discussion of our results. Various supporting results are presented in the Appendix. 

\section{Simulation Methods}
\subsection{System potentials}
\begin{figure}
 \centering{ 
  \includegraphics[width = 0.4\textwidth]{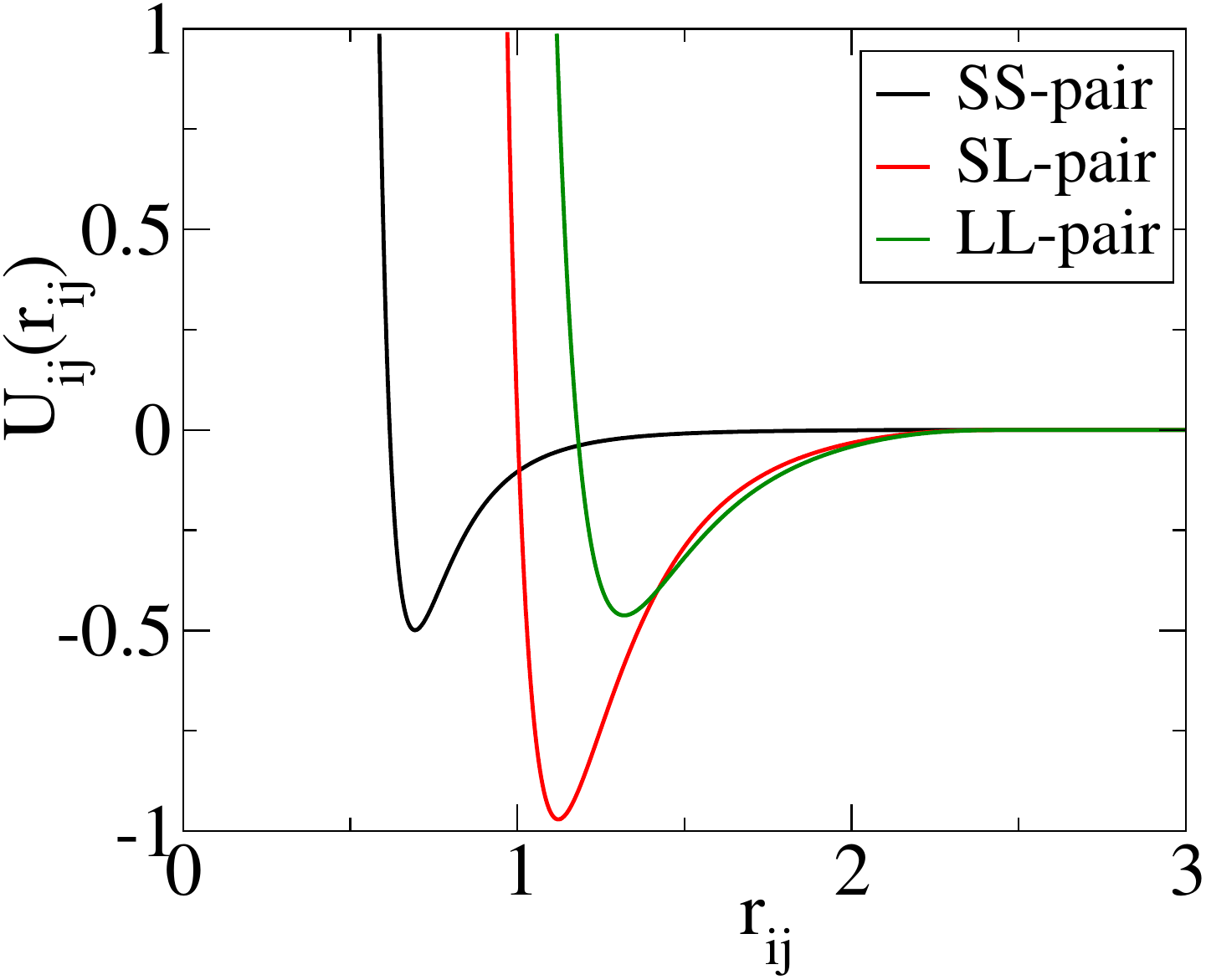}}
\caption{\label{fig_potential} The interaction potential (Eq. (\ref{eqnpot})) used in this study, for the different pairs ($SS$,$SL$ and $LL$) of particles.}
\label{Uij}
\end{figure}

We  perform  molecular dynamics simulations of a binary Lennard-Jones mixture  in two dimensions, introduced in Ref.~\cite{LanconJPF1984}, that has been extensively used to investigate the mechanical properties of amorphous solids \cite{WidomPRL87,shi2005strain,ShiPRL07,barbotPRE18,BarbotPRE20}. In this model, the two types of atoms interact {\it via.} the Lennard-Jones inter-atomic potential, with a cut-off, given by,
\begin{equation}
U_{ij}=\left\{
\begin{array}{lr}
4 \epsilon_{ij}[(\frac{\sigma_{ij}}{r_{ij}})^{12}-(\frac{\sigma_{ij}}{r_{ij}})^6]+A, &r_{ij} < R_{in}\\ \sum_{k=0}^{4}C_k(r_{ij}-R_{in})^k,&R_{in} < r_{ij}< R_{\rm{cut}}\\0,&r_{ij}>R_{cut}
\end{array}
\right.
\label{eqnpot}
\end{equation}
where the constants (A, C's) can be found in Ref. \cite{barbotPRE18}. A plot of the interaction potentials is shown in Fig. \ref{Uij}. The density of the system is kept constant and equals $1.02$. The reduced units, used throughout, are expressed in terms of the mass $m = m_S = m_L$, and the energy scale $\epsilon_{SL}$, length scale $\sigma_{SL}$ defining the inter species interaction. For this model $\sigma_{SS}/\sigma_{SL}=2\sin(\pi/10)$, $\sigma_{SS}/\sigma_{SL}=2\sin(\pi/5)$ and $\epsilon_{SS}/\epsilon_{SL}=\epsilon_{LL}/\epsilon_{SL}=0.5$. Time is expressed in  units of $\tau = \sigma_{SL} \sqrt{m/\epsilon_{SL}}$ and temperature is expressed in units of $\epsilon_{SL}/k_B$ where $k_B$ is the Boltzmann constant. The composition of the mixture is given by the number ratio of large ($L$) and small ($S$) particles, $N_L : N_S = (1 + \sqrt 5):4$. The potential is modified so that its value as well as the first two derivatives go continuously to zero at  a cutoff distance $R_{cut} = 2.5\sigma$, and be twice continuously differentiable  at  $R_{in} = 2 \sigma $. 

%\sri{Why is this relevant? Tg etc. Plese remove also in the appendix} 
%The  non-smoothed version of this model display a transition temperature $T_g$ at approximately $0.325$~\cite{shi2005strain} and a mode-coupling analysis provides the MCT $T_c=0.373$~\cite{BarbotPRE20}.

\subsection{Initial glass preparation}
We perform constant temperature and volume (NVT) molecular dynamics simulations using the Nos\'e-Hoover thermostat with an integration time step $0.001$. Two different temperatures, $T=2.98$ and $T=0.35$, are considered. The system was equilibrated in the liquid state for $10^4\tau_\alpha$ for $T=2.98$  and $100\tau_\alpha$ for  $T=0.35$, where $\tau_\alpha$ is the structural relaxation time.
%obtained from the self intermediate scattering function. 
We consider different system sizes, i.e. $N=1024,2500,4900,10000,22500$. We employ two different protocols to prepare glasses in four different states. The first two sets of  glasses are obtained from the well equilibrated  high-temperature liquid (HTL, T=2.98) and  the supercooled liquid (ESL, $T=0.35$) configurations, by performing a local energy minimization, to obtain the corresponding {\it inherent structures}. Two sets of well annealed glasses with lower energies are obtained by the  the finite temperature and shear rate annealing protocol described in \cite{das2018annealing}. Following this procedure, we subject equilibrated configurations at $T = 0.35$  to  cyclic shear  deformation following the SLLOD equations of motion \cite{evansPRE84} with a shear rate $\dot{\gamma}=10^{-3}$, strain amplitude $\gamma_{max}= 0.035$ and simulation temperature $T = 0.3$. The energies of the inherent structures obtained at the end of each cycle of shear decrease with cycles, as shown in Fig. \ref{sllod} of the Appendix, which also shows the inherent structure energy as a function of temperature for $T = 2.98$ to $T = 0.35$ obtained from MD simulations.  We obtain two different sets of samples of such glasses whose energies on average are equal to $E_{IS}=-2.41$ and $-2.45$, which we refer to as  WAL1 and WAL2, respectively. These energies, based on extrapolation of the inherent structure energies from the MD simulations, corresponds to  $T = 0.271$ and $T = 0.241$ respectively. We analyze $10-12$ configurations for HTL glass for different system sizes and $3$ configurations for each of ESL, WAL1 and WAL2. 

\subsection{Athermal cyclic shear}
The different initial configurations obtained from the protocols described above, are subjected to the athermal quasi-static (AQS) shearing protocol which involves two steps: (i) an affine transformation that increases the strain by a small amount $d\gamma$ in the $xy$-direction (with coordinate transformations $x^\prime\to x+d\gamma ~y$, $y^\prime\to y$) and (ii) an energy minimization. For a given strain amplitude $\gamma_{max}$, the strain $\gamma$ is varied cyclically as : $0\to \gamma_{\rm {max}}\to -\gamma_{\rm{max}}\to 0$ in each cycle. The accumulated strain $\gamma_{acc}=\sum_i|d\gamma|$ serves as an effective time variable in the analysis. The accumulated strain at $\gamma = 0$ at the end of each cycle is given by $\gamma_{acc}= 4 \times \gamma_{max}\times N_{cyc}$, $N_{cyc}$ being the number of performed cycles. Repeating the deformation cycle for a fixed amplitude $\gamma_{\rm{max}}$, the shear deformed glasses evolve through a transient into a  steady state, in which their properties either become invariant from cycle to cycle (termed an {\it absorbing} stated, obtained for $\gamma_{\rm{max}}$ below the yield value) or in which their properties fluctuate around a mean value, but the particles as a function of $\gamma_{acc}$ exhibit diffusive behaviour ({\it e. g.} \cite{leishangthemNAT2017}). 
The steady states are obtained and studied for different samples for strain amplitudes ranging from  $\gamma_{\rm{max}}=0.02$ to $0.12$. We employ the conjugate-gradient algorithm for energy minimization and execute all the numerical simulations in LAMMPS \cite{plimptonjcp1995}.

The variation of the energy, stress and mean squared displacements (MSD) {\it vs.} accumulated difference in the steady state are shown in the Appendix, in Figs.  \ref{pe_gac}, \ref{SI_fullfycle_energy}, \ref{SI_fullcycle_stress}, and \ref{SI_msdavg}. The asymptotic values of the stroboscopic ($\gamma=0$)  energies $U$ are obtained by stretched exponential fits to the dependence on $\gamma_{acc}$. In most cases, the extracted asymptotic values correspond to values around which the energies fluctuate in the steady state. The maximum $\sigma_{xy}^{max}$ is obtained by averaging the stress at $\gamma_{\rm{max}}$ over $100-200$ cycles in the steady state. From the mean squared displacements of particles with respect to initial configurations in the steady state, we calculate diffusion coefficients $D$ using the $\gamma_{acc}$ as the time variable in the usual definition of $D$. Structural quantities, and the statistics of avalanches are likewise computed in the steady state, averaging over $100-200$ cycles.

%\sri{SS:Till here}

%(a) Plot of MSD, computed in the steady state and  averaged over different origin taken as the reference configuration,  against $\gamma_{acc}$ for different $\gamma_{max}$ for  $N=10^4$.

\section{ Results}
\subsection{Yielding transition}

\begin{figure*}[t]
  \centering{ 
  \includegraphics[width = 0.32\textwidth]{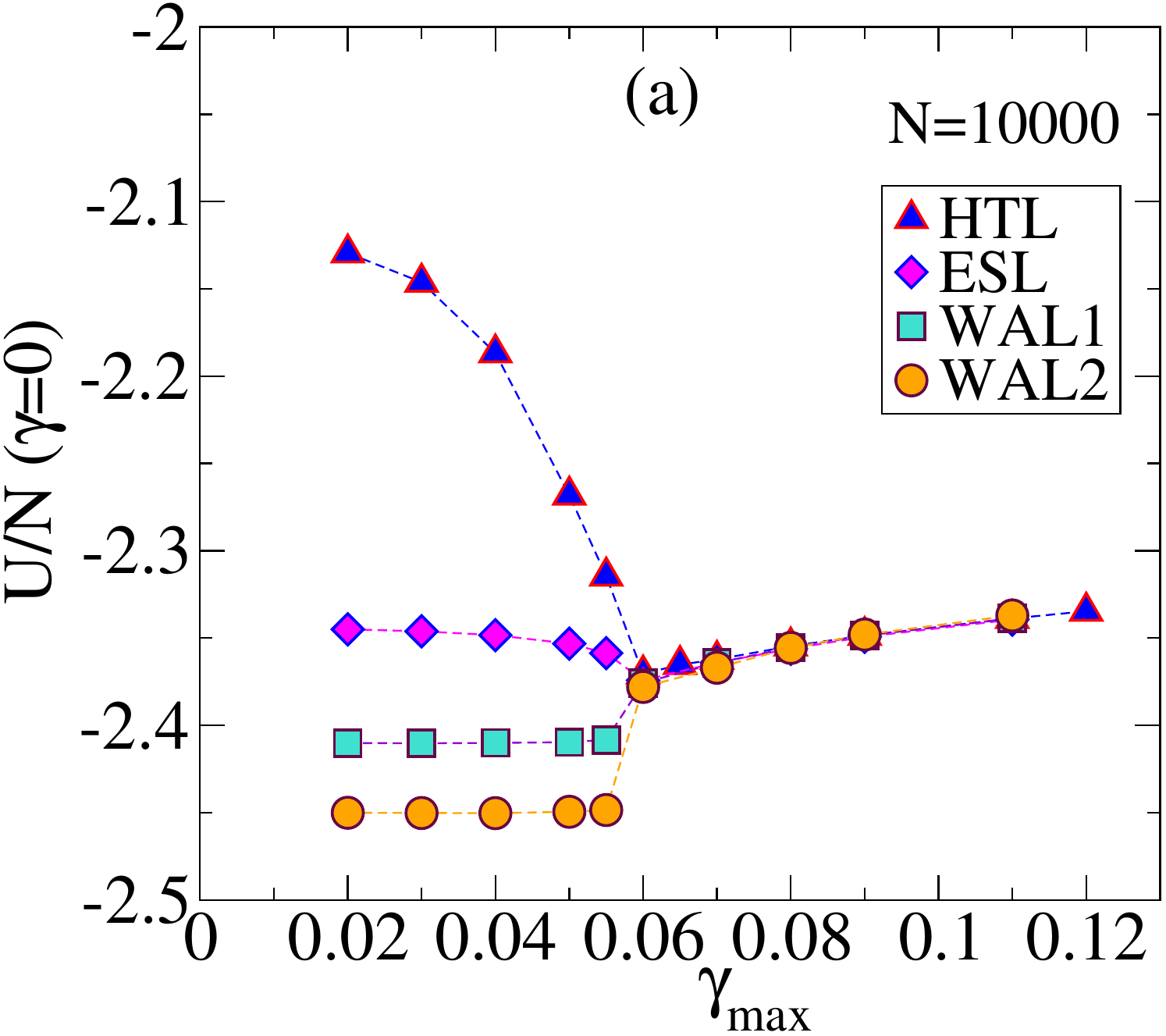}
  \includegraphics[width = 0.32\textwidth]{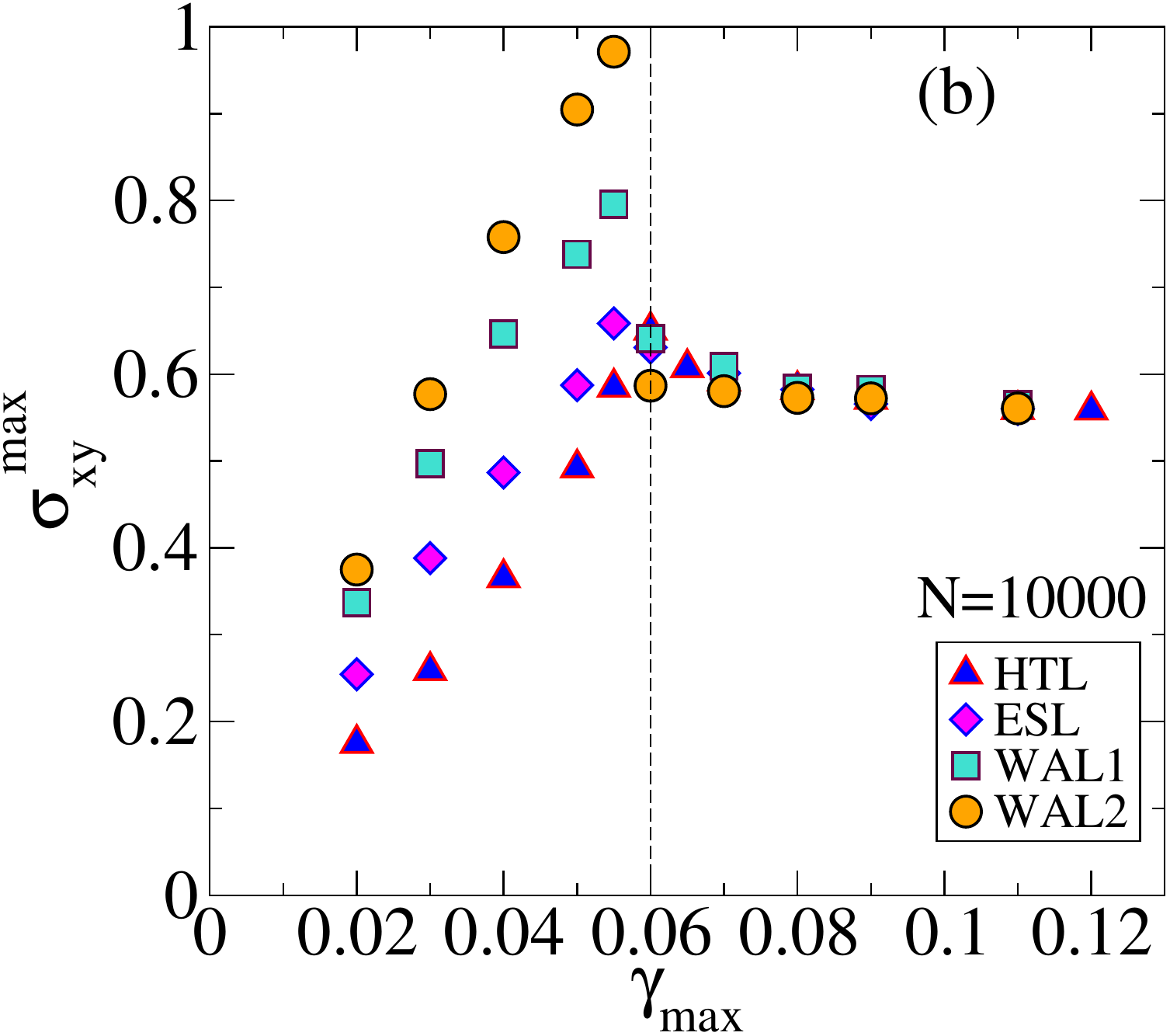}
  \includegraphics[width = 0.33\textwidth]{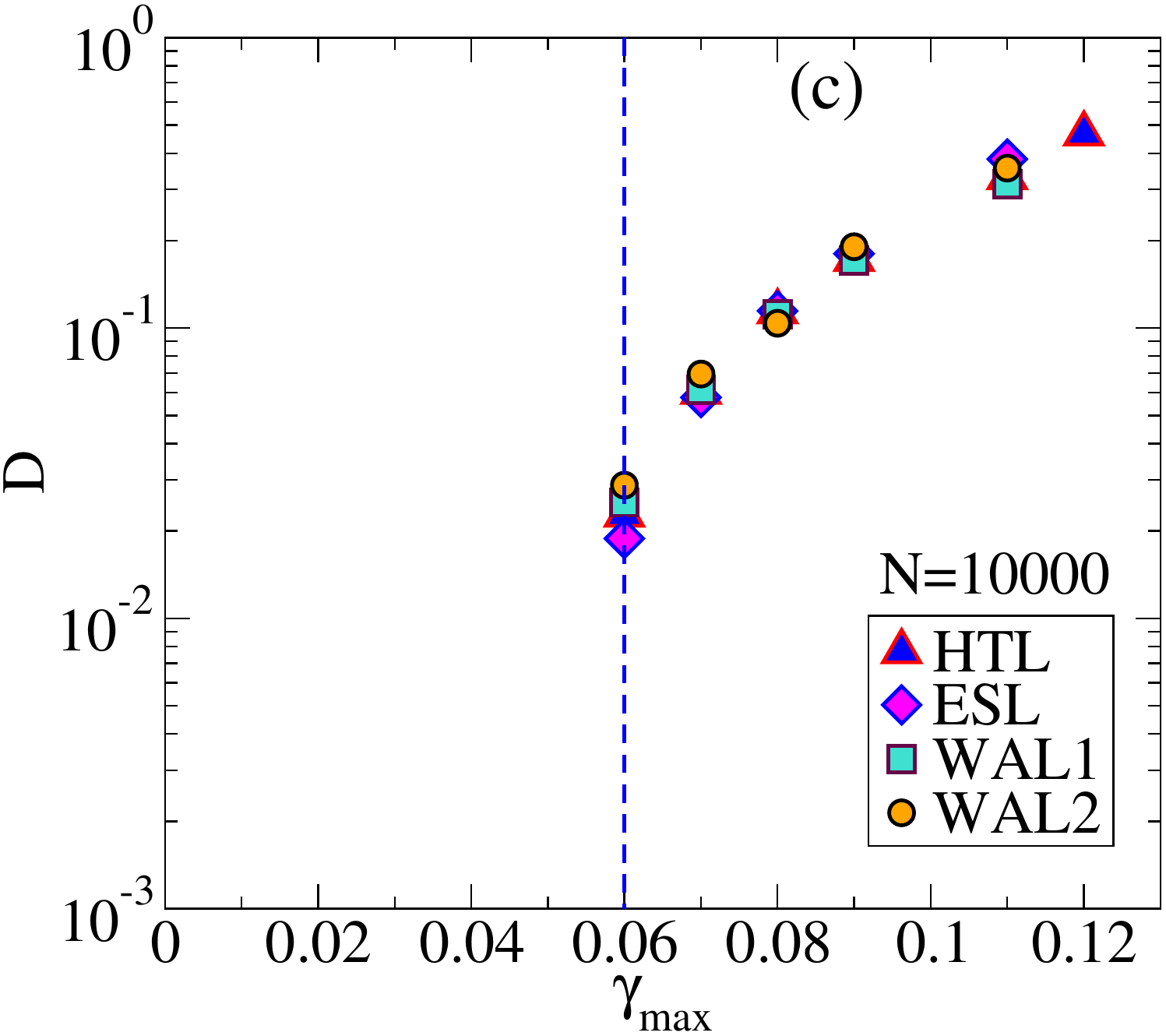}
  }
  \centering{ 
  \includegraphics[width = 0.32\textwidth]{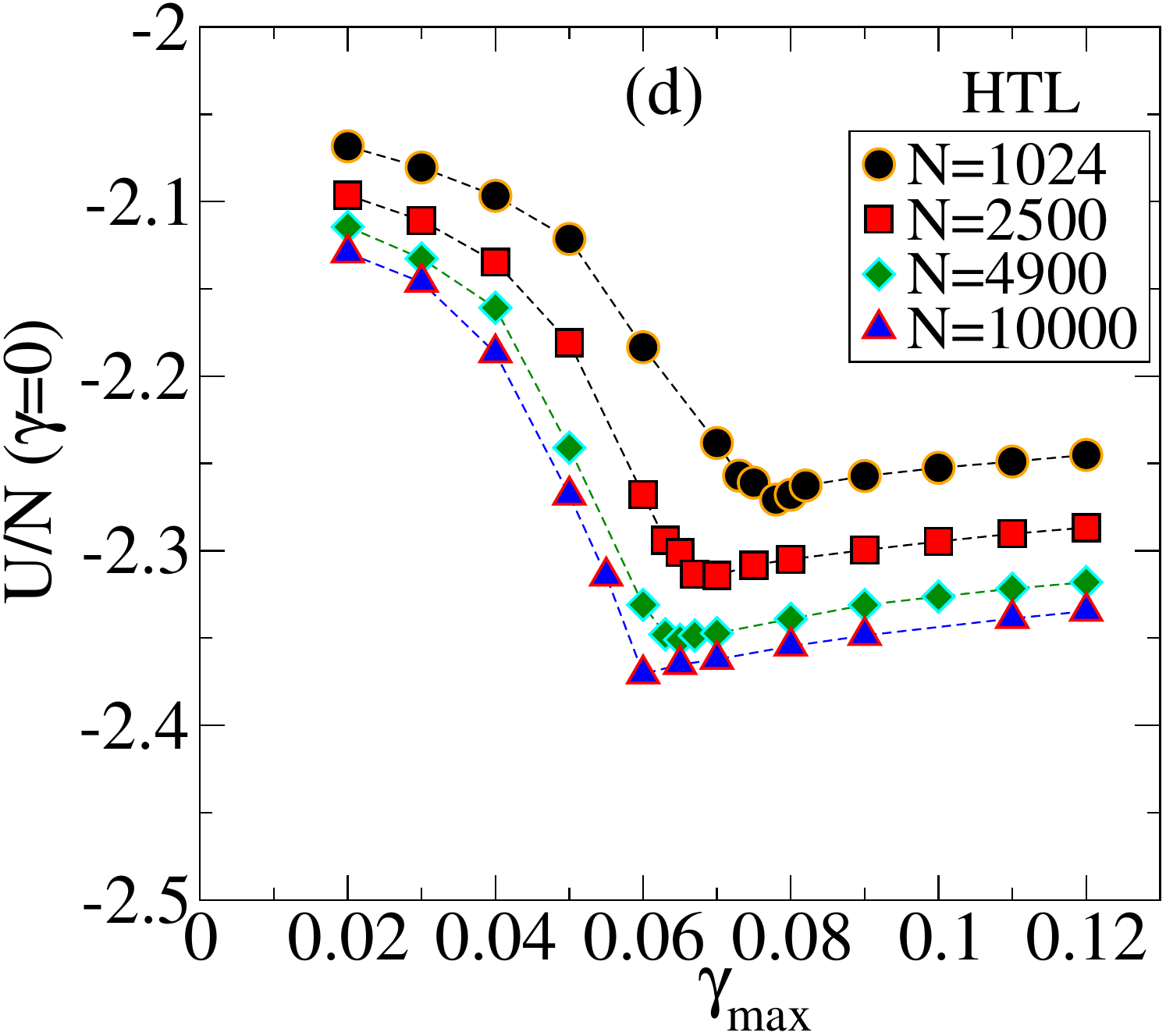}
  \includegraphics[width = 0.32\textwidth]{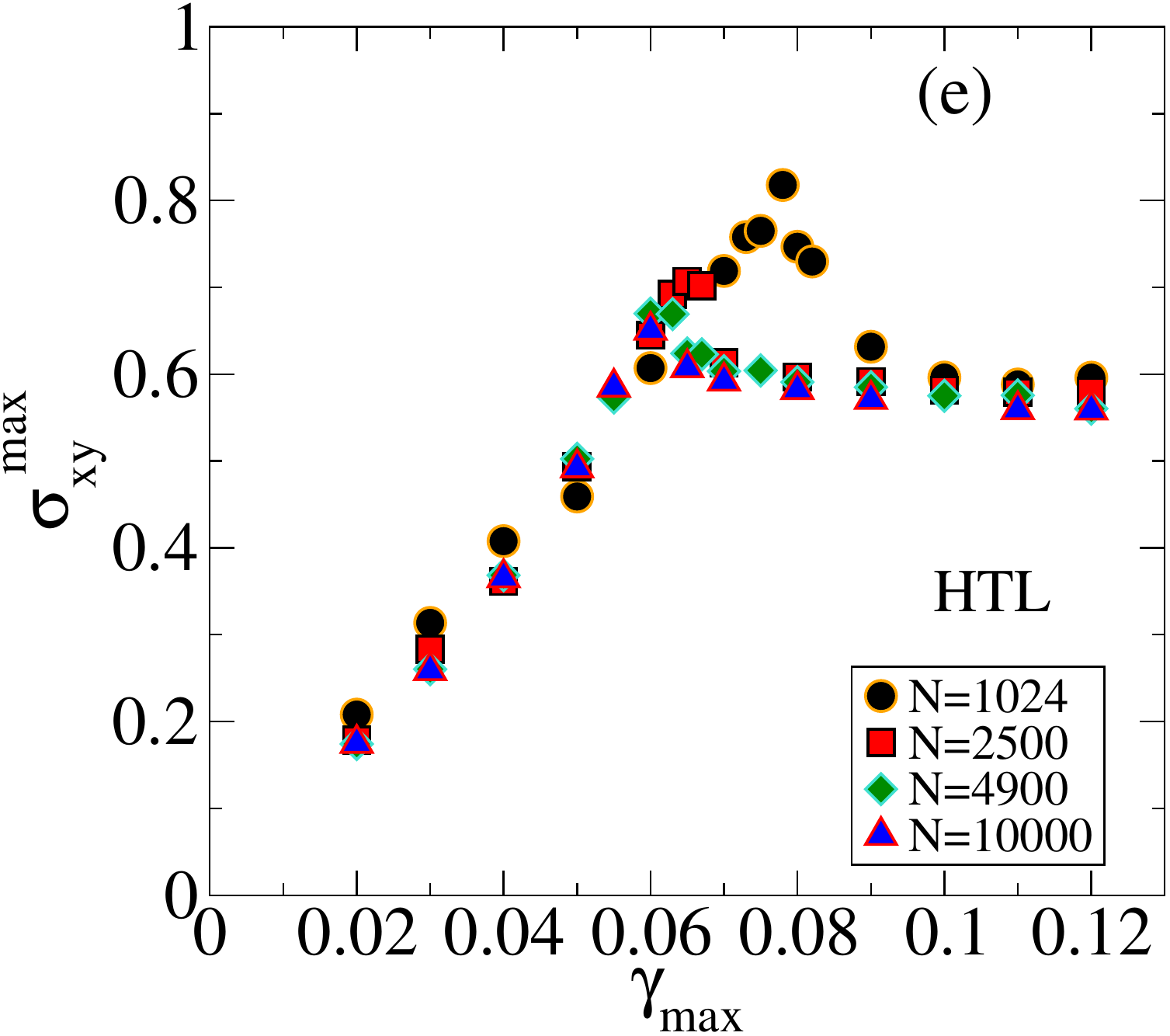}
  \includegraphics[width = 0.33\textwidth]{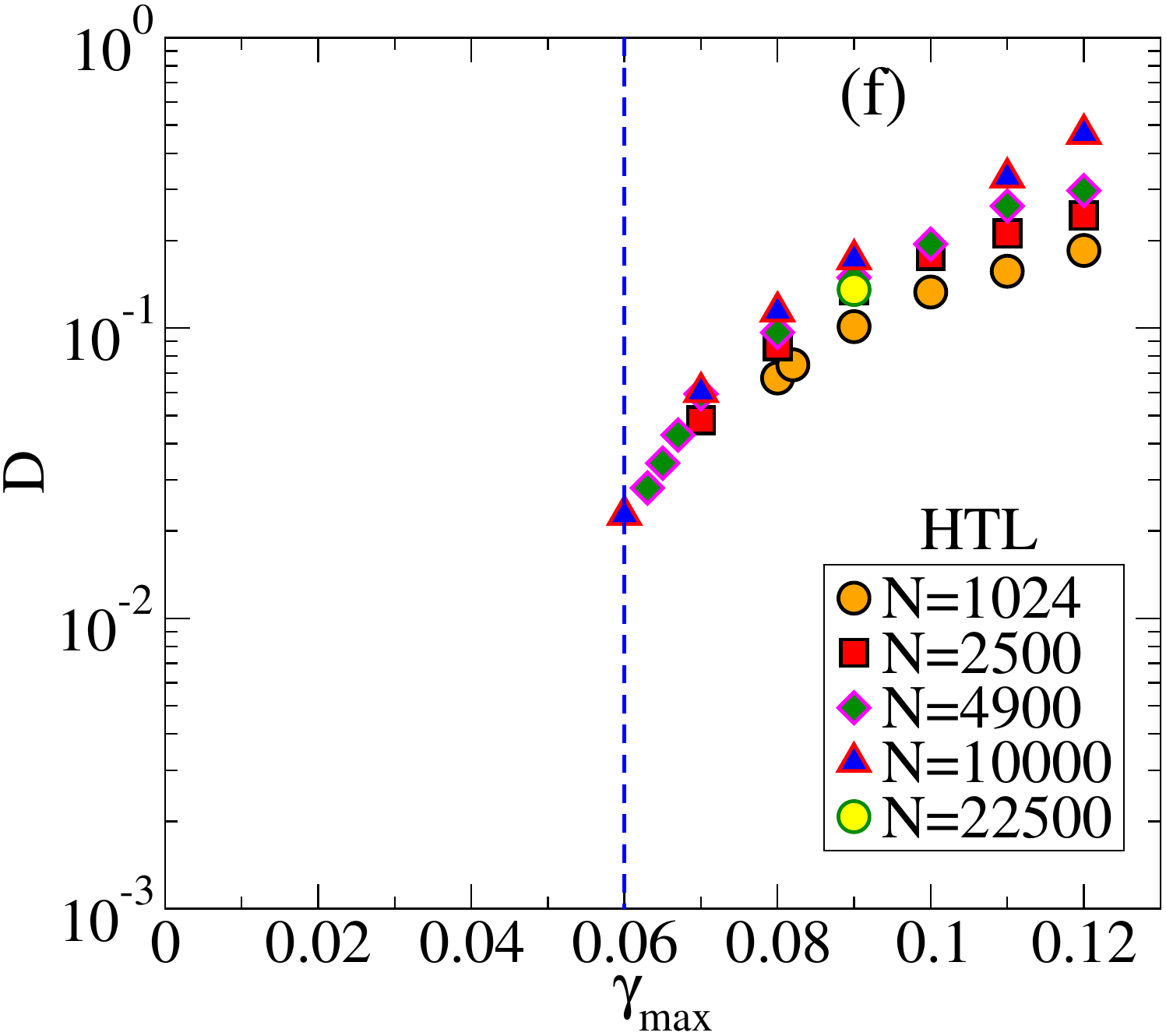}
  }
\caption{\label{fig_uss_sigma-max} Steady state potential energies (a, d) ($U/N$) of stroboscopic configurations,  (b,e) maximum value of stress $\sigma_{xy}^{max}$ and (c,f) diffusion coefficient $D$ against strain amplitude ($\gamma_{max}$). Upper panels are for different types of glasses: HTL, ESL, WAL1, and WAL2 with system size $N=10000$. Lower panels are for the HTL glass for differnet system sizes $N=1024,2500,4900,10000$. Dashed vertical line in (c,f) indicates the yield strain amplitude $\gamma_{y}=0.06$.}
\end{figure*}

The steady state energies $U/N$ are shown in Figs. \ref{fig_uss_sigma-max}(a) and \ref{fig_uss_sigma-max}(d) as a function of strain amplitude $ \gamma_{max}$ for the four types of glasses as well as for different system sizes for the HTL. The HTL glass displays  significant finite size effects. In particular,  while the shape of the curves in Fig. \ref{fig_uss_sigma-max}(d) are very similar for different system sizes, the yield strain amplitude decreases with increasing  system size. Below yielding, all HTL glasses (also ESL) anneal to lower energies with increasing strain amplitude, and we find that lower energies are obtained for larger system sizes for HTL. 

As shown in Fig. \ref{fig_uss_sigma-max}(a), as the strain amplitude $\gamma_{max}$ increases, $U/N$ of HTL and ESL progressively decreases until $\gamma_{max}$ reaches the yield amplitude $\gamma_y=0.06$. At the yield strain the energy of both HTL and ESL glasses approach to the same energy state. Such an observation is consistent with the results of cyclic shear explored before \cite{fioccoPRE13,leishangthemNAT2017,BhaumikPNAS21}.  Well annealed glasses (WAL1 and WAL2), on the other hand,  display a different behavior. Unlike  poorly annealed glasses (HTL and ESL), well annealed glasses remain stable for amplitudes of strain below the yield amplitude and  retrace the same sequences of energies over the successive cycles of deformation (see Fig. \ref{SI_fullfycle_energy} in the Appendix), displaying the same behaviour observed in three dimensional systems ~\cite{BhaumikPNAS21,yehprl20}. Thus, the stroboscopic energies remain unaltered as the strain amplitude increases towards the yield strain amplitude $\gamma_y=0.06$. At the yield point, for well annealed glasses,  we observe a finite jump in energy whose size  increases with the degree of annealing. In previous work for a three dimensional glass \cite{BhaumikPNAS21}, a finite jump in energy is found even for poorly annealed glasses. Within the precision of sampling of $\gamma_{max}$ in this work, we cannot draw the same conclusion. However, very clearly, for well annealed samples, a finite jump in energy is observed, which appears to be inconsistent with a recent investigation of a two dimensional glass \cite{LembergPRE20}. On the other hand, for the three dimensional Lennard-Jones glass studied in Ref. \cite{BhaumikPNAS21} (as also that investigated in Ref. \cite{yehprl20}), the yield strain amplitude depends appreciably on the degree of annealing, which we do not find in the present case. The yield amplitude in all cases studied here is roughly the same. This may be a consequence of the limited range of annealing we investigate here, and requires further study to clarify. As in previous studies, however, all glasses investigated have the same energies above yielding, regardless of initial conditions. 

Next we study the variation of maximum stress ($\sigma_{xy}^{max}$) with strain amplitude for the different cases. As shown in Figs. \ref{fig_uss_sigma-max}(b) (different glasses) and  \ref{fig_uss_sigma-max}(e) (for different system sizes of HTL), the $\gamma_{max}$ dependence of $\sigma_{xy}^{max}$ below the yield point exhibits significant dependence of the annealing of the glasses,  and a jump in the case of the well annealed glasses whose size increases with increasing degree of annealing. Fig. \ref{fig_uss_sigma-max}(e) displays substantial finite size effects both in the location of the yielding point and the value of the stress at yielding. As remarked earlier in the context of energy change, we do not attempt here to resolve whether a finite jump is present in the case of poorly annealed glasses. Similar to energy values, $\sigma_{xy}^{max}$ also shows the same $\gamma_{max}$ dependence in post yield regime for all the glasses. 

\begin{figure}[t]
\centering{ 
\includegraphics[width = 0.4\textwidth]{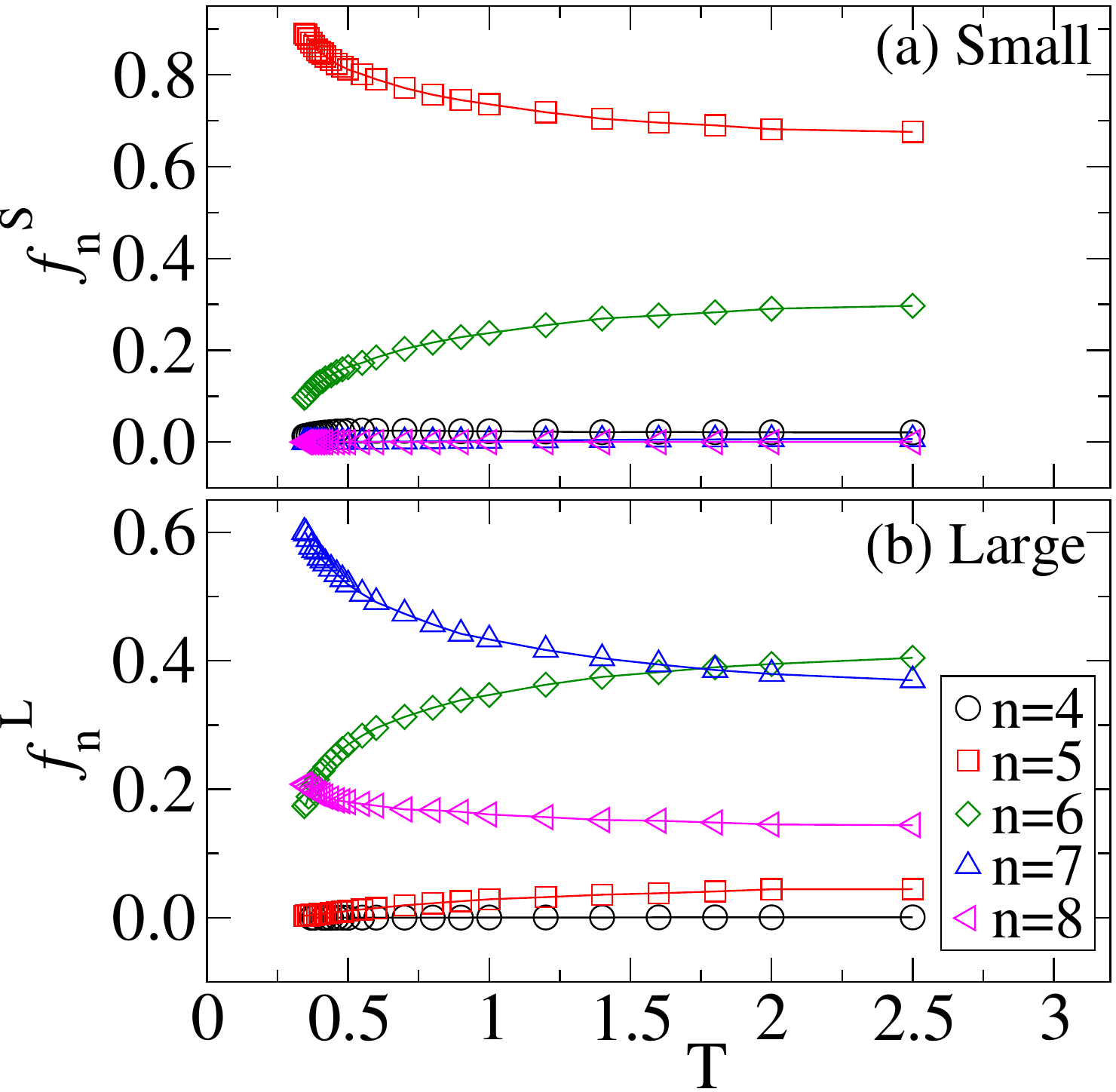}
}
\caption{\label{fnLSLiq}  (a) Fraction of small particles $f_n^S$ and (b) Fraction of large particles $f_n^L$ with $n$ nearest neighbours against $T$ for inherent structure configurations for $N=4900$. }
\end{figure}

% (a) Plot of distribution of local strain $\epsilon_d$ for various $\gamma_{max}$ for $N=10000$. The straight line has the power-law with exponent $-2$. (b) Distribution of displacement of the constituent particles of the triangles whose local strain $\epsilon_d$ is greater-than $0.22$ shows a maxima at $\delta r=0.25$. We have chosen $\delta r_c=0.25$ as cutoff value to extract active particles in a plastic event. (c) Plot of distribution cluster size of the avalanche for different $\gamma_{max}$ for system size $N=10000$. Two distinct set can be observed, the set corresponds to $\gamma_{max}\ge \gamma_{y}$ have the power-law $-5/3$ which is greater than mean field value.

We next consider the presence or absence of diffusive motion of particles to identify the  yielding transition for the four types of glasses. We find that for $\gamma_{max}<\gamma_y$, the MSD values are constant, indicating the absence of cumulative movement from one cycle to the next,  whereas for $\gamma_{max}> \gamma_y$, the particle motion is diffusive (see Fig. \ref{SI_msdavg} of the Appendix). We characterize such diffusive motion as a function of accumulated strain (relative to a reference point in the steady state) with $MSD =  4 D \Delta \gamma_{acc}$ where $D$ is the diffusion coefficient
and $\Delta \gamma_{acc}$ is the difference in $\gamma_{acc}$ between the configurations used to compute the MSD. The values of $D$ as function of $\gamma_{max}$ is  presented in Fig. \ref{fig_uss_sigma-max}(c) for the the different glasses for $N = 10000$ and for different system sizes for HTL in Fig. \ref{fig_uss_sigma-max}(f). The diffusion coefficients are zero below the yielding transition and exhibit a finite value at the yielding transition, and an increase with $\gamma_{max}$ above. Fig. \ref{fig_uss_sigma-max}(f) illustrates the finite size effects present for HTL, with the diffusion coefficients being bigger for larger systems. 

Thus, the change in the  $\gamma_{max}$ dependence of the energy, maximum stress, and the diffusion coefficient, consistently mark the discontinuous yielding transition, with the observed discontinuities of energy and stress being clearly dependent on the degree of annealing, and the jump in the diffusion coefficient exhibiting no such dependence.

\subsection{Structural change across the yielding transition}

Although some attention has  been paid in the past to the nature of structural changes associated with shear deformation and yielding in glasses \cite{shiprb06,parmarPRX2019,adhikari2018,bhaumiksilica2021,kawasakiPRE16,DenisovSR15,VasishtPRE20,VasishtPRE2020b}, such investigations have been relatively rare. Also, some of the conclusions drawn by such studies are at odds with each other. For example, it has been argued that a
sharp change in the symmetry associated with anisotropy is observed when crossing the yielding transition in a colloidal glass subjected to cyclic shear experimentally~\cite{DenisovSR15}, but no such anisotropy has been observed in a computer simulation investigation \cite{adhikari2018}. It has been claimed that in a system of jammed repulsive  non-Brownian particles, the static structure does not change in the pre- and post-yield regime \cite{kawasakiPRE16}. However, clear indications of structural change have been observed in other simulation studies \cite{shiprb06,parmarPRX2019,VasishtPRE20,VasishtPRE2020b}. In order to understand the nature of structural change in the pre-yield and post-yield regimes for the model glass we study, we perform a detailed analysis of structure.

\begin{figure}[t]
\centering{ 
\includegraphics[width = 0.49\textwidth]{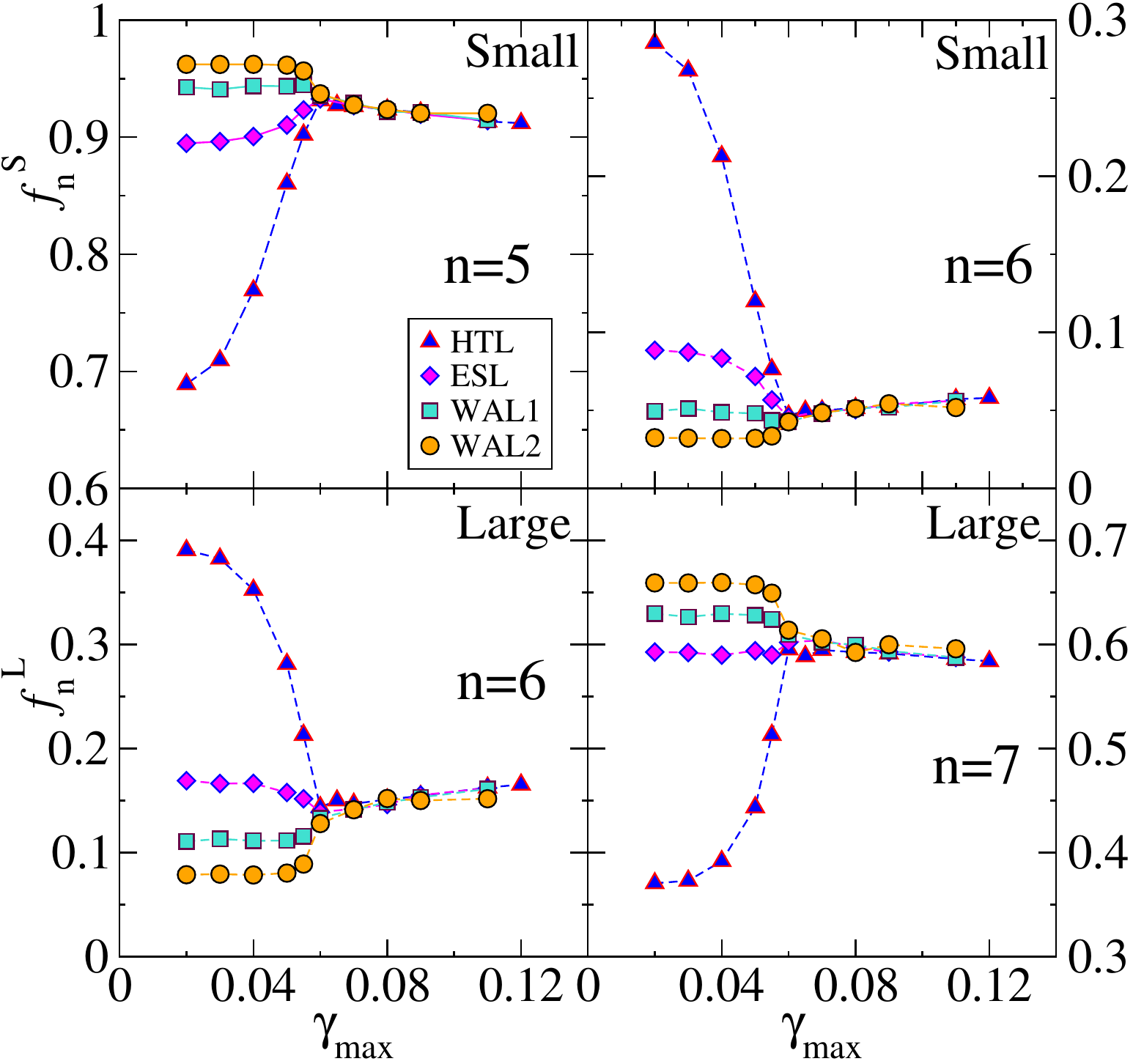}
}
\caption{\label{fLSGlass} {\bf } Fractions $f_n^S$ for $n=5,6$ and $f_n^L$ for $n=6,7$ against strain amplitude $\gamma_{max}$ for different types of glasses with $N=10000$ in the  steady state.}
\end{figure}

\begin{figure*}[t]
\centering{ \includegraphics[width = 0.99\textwidth]{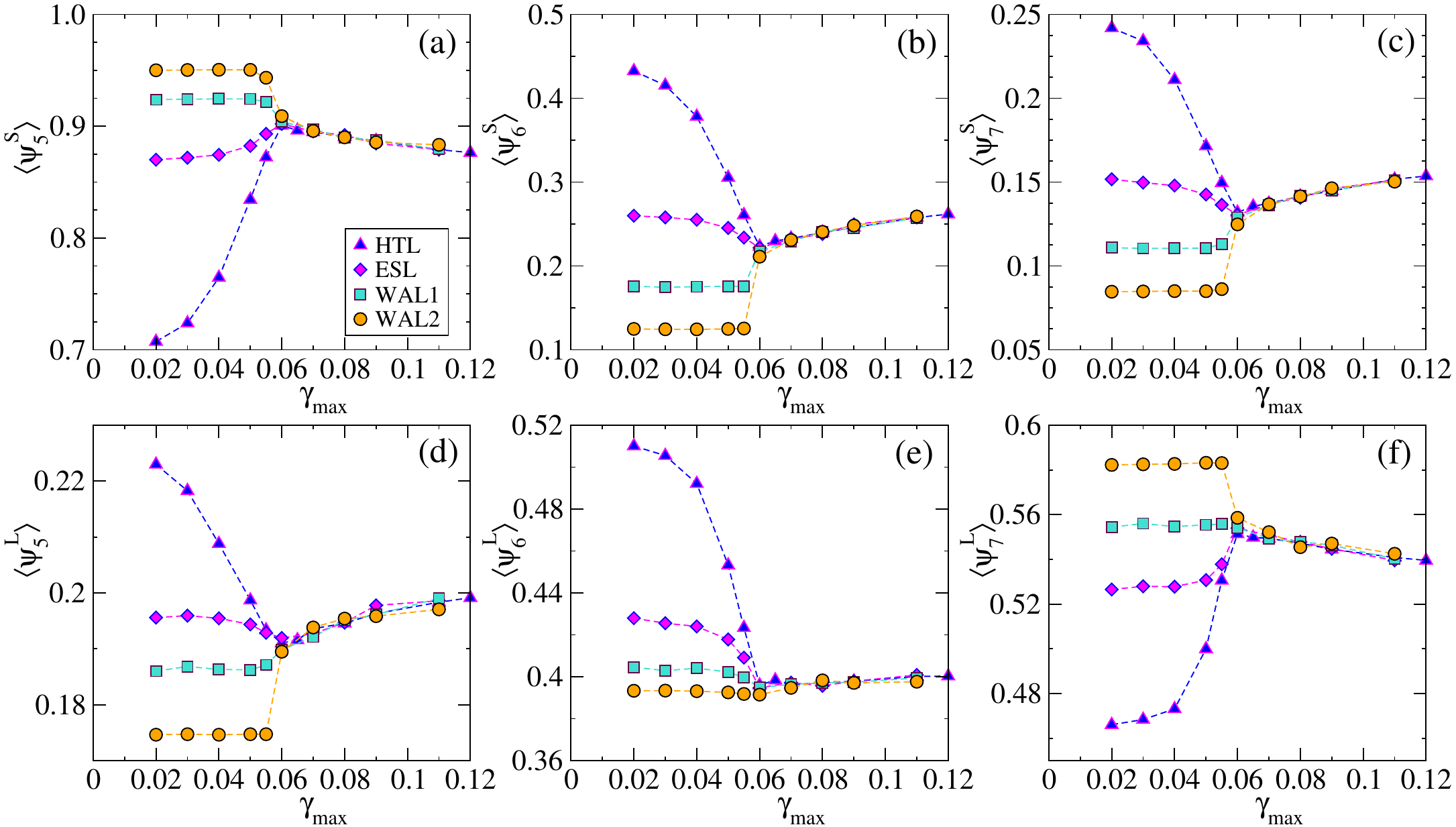}
}
\caption{\label{psiall} {\bf } Average orientational order parameter for (a-c) small particles ($\langle \psi_n^S\rangle $) and for (d-f) large particles ($\langle \psi_n^L\rangle $) against strain amplitude $\gamma_{max}$ for different types of glasses. Different panels are for different values of $n=5,6,7$. System size $N=10000$. }
\end{figure*}

\begin{figure*}[t]
    \centerline{
   \includegraphics[width=.35\linewidth]{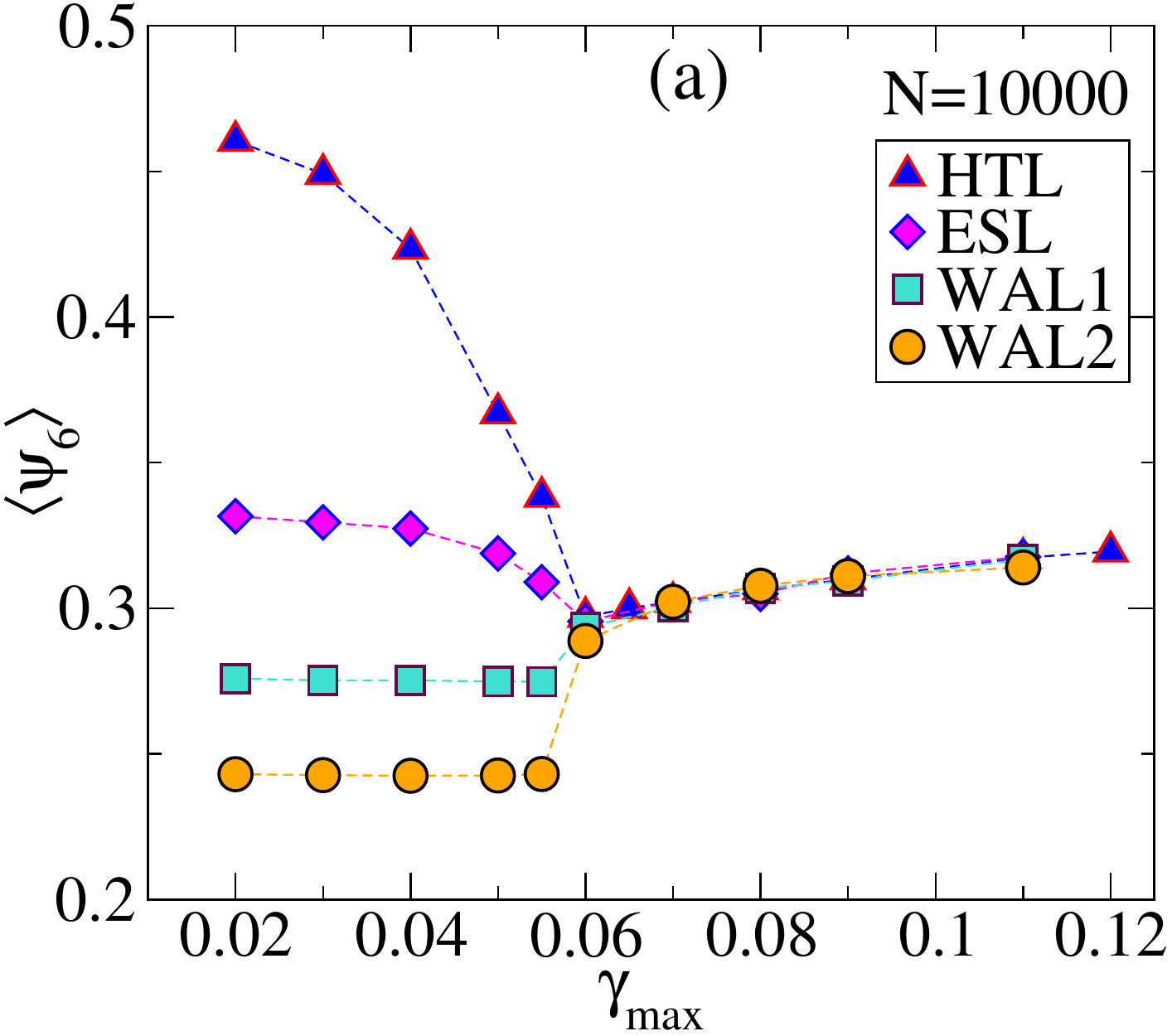} \quad
    \includegraphics[width=.35\linewidth]{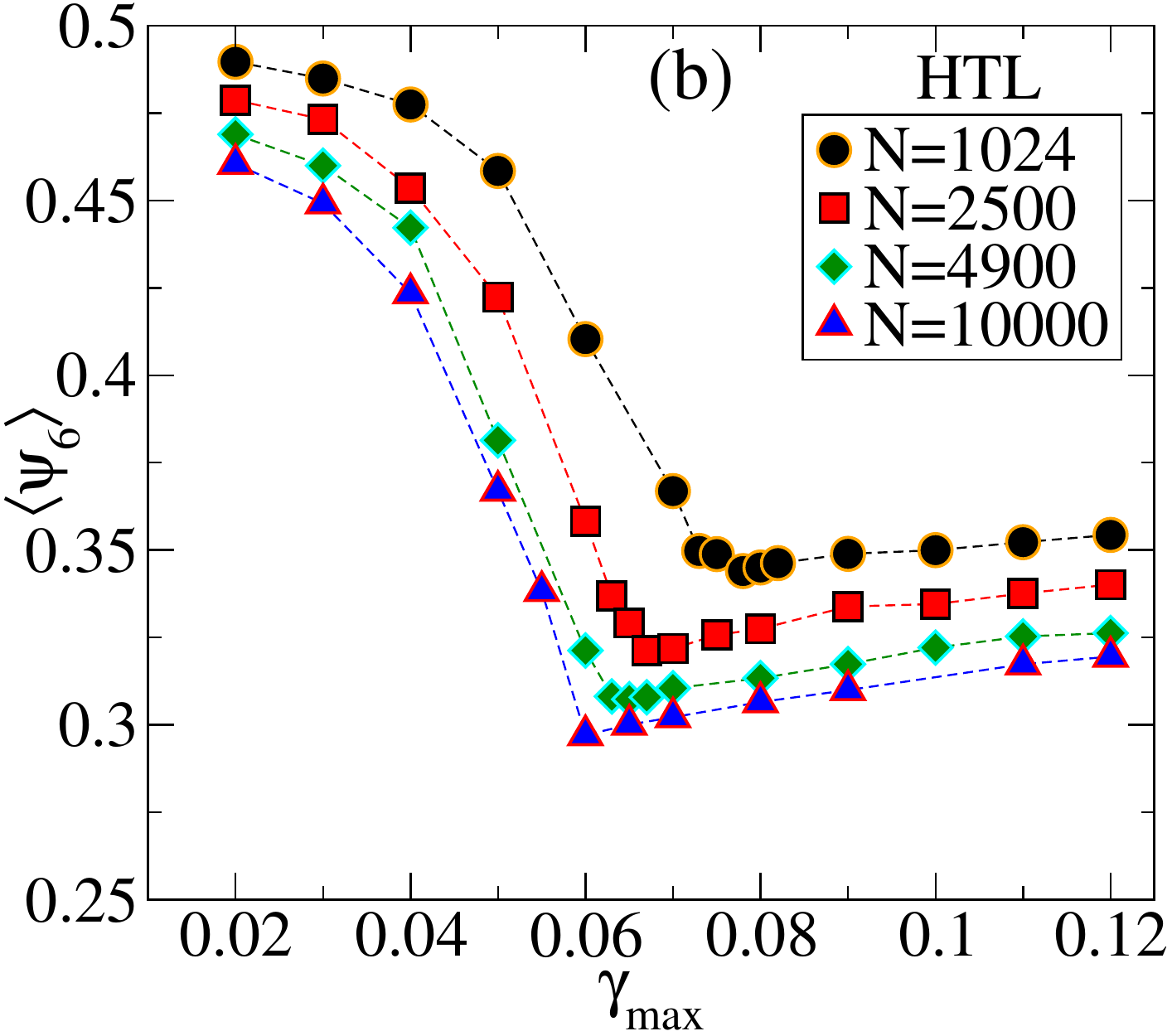}}
\caption{\label{fig_psi6GlassSystem}  Average orientational order parameter $\langle \psi_6\rangle=N^{-1}\sum_{i=0}^N\psi_6^i$ averaged over all the particles against strain amplitude $\gamma_{max}$ for (a) different types of glasses for a given system size $N=10000$ and (b) for the HTL glass with different system sizes $N=1024,2500,4900,10000$.}
\end{figure*}

We begin by analysing the structural change in the inherent structures corresponding to equilibrium liquid configurations, to benchmark our subsequent analysis of the sheared glasses. Since our glasses are composed of two types of particles (large, "L", and small, "S"), we investigate the short range order around each type of particle by considering the number of neighbors around each type of particle. Fig. \ref{fnLSLiq} shows the fraction of large and small particles with different numbers of neighbors in the first shell (defined as the geometric neighbors of a particles in a generalised Delaunay tessellation for configurations of bidisperse discs \cite{sastryPRE1997a,maitiEPJE2013}) as a function of temperature, $f^L_n$, $f^S_n$, where $L$ and $S$ refer to the type of particle and $n$ the number of neighbors. The data clearly reveal that $f^S_{n=5}$, $f^L_{n=7}$ and $f^L_{n=8}$ grow as temperature is lowered, and all other fractions, in particular $f^S_{n=6}$ and $f^L_{n=6}$, decrease. This indicates that the preferred order is for small particles to be surrounded by $5$ neighbours, and large particles to be surrounded by $7$  or $8$ neighbors. Local structure with $6$ neighbours, for either large or small particles, corresponds to unfavorable arrangements at low temperatures. 

With these results as reference, we now consider the fractions $f^L_{n=7}$, $f^S_{n=5}$, corresponding to favorable motifs, and   $f^L_{n=6}$, $f^S_{n=6}$, corresponding to unfavorable motifs, for the different glasses we study, as a function of strain amplitude $\gamma_{max}$. As shown in Fig. \ref{fLSGlass}, $f^L_{n=7}$ and $f^S_{n=5}$ are large for the well annealed glasses and small for the poorly annealed glasses, and they remain constant for well annealed glasses as $\gamma_{max}$ increases (for $\gamma_{max} < \gamma_y$, whereas they increase with increasing $\gamma_{max}$ for the poorly annealed glasses, indicating an increase in the fraction of favourable structural motifs. In contrast, $f^L_{n=6}$ and $f^S_{n=6}$ show exactly the opposite behaviour. Above the yield point, $f^L_{n=7}$ and $f^S_{n=5}$ decrease, whereas $f^L_{n=6}$ and $f^S_{n=6}$ increase, and the values do not depend on the initial annealing degree of the glasses.

\begin{figure*}[t]
\centering{ 
\includegraphics[width =
    0.32\textwidth]{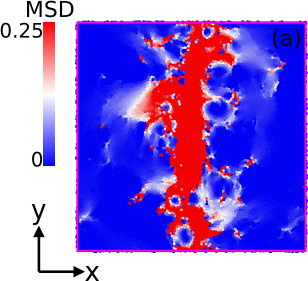}
\includegraphics[width =
    0.31\textwidth]{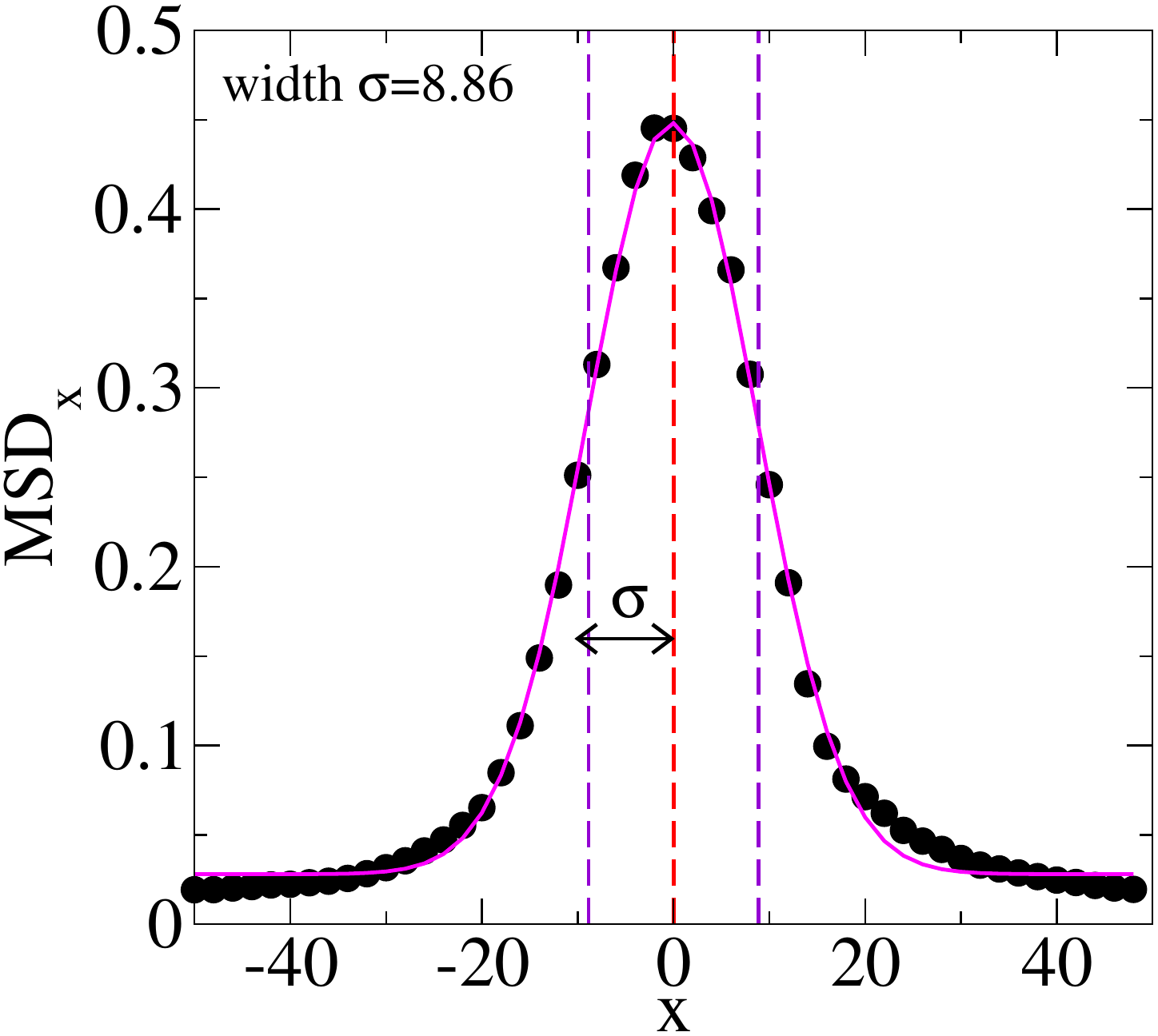}
\includegraphics[width =
    0.34\textwidth]{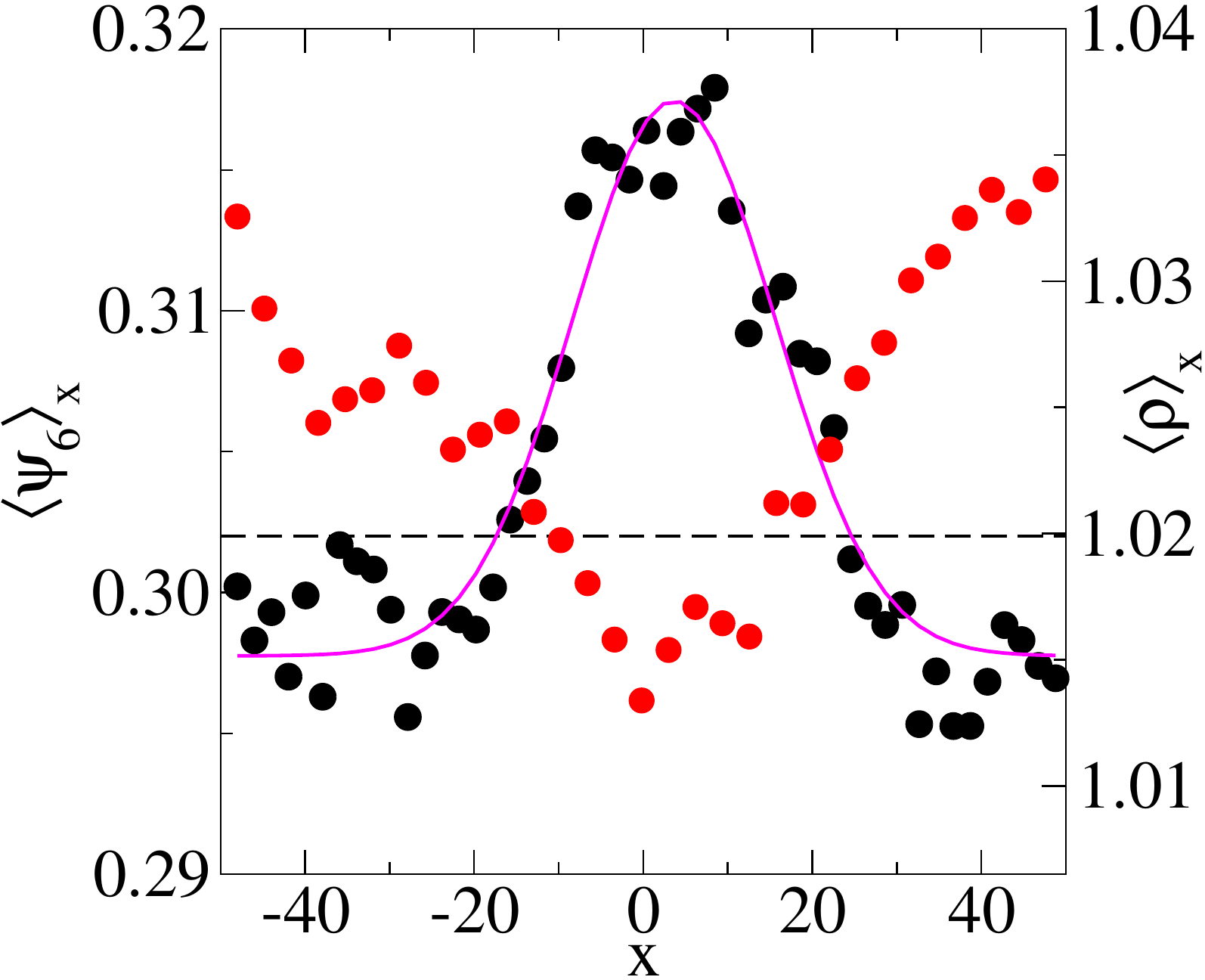}
}
\caption{\label{shearband} (a) Colour map of particle displacements for a zero strain configuration from the steady state for strain amplitude $\gamma_{max}=0.07$. Particles are coloured according to their mean square displacement (MSD) computed over a strain cycle. Particles with $MSD>0.25\sigma^2_{SL}$ are shown in red and the others according to the colour bar shown. (b) Slab-wise averaged $MSD_x$ in the x-direction showing a Gaussian form with width $w=8.86 \sigma_{SL}$ within which the particles are highly mobile. (c) Slab-wise averaged $\langle \psi_6\rangle_x$ (averaged over all particles) (black dots) and slab-wise density (red dots) against the x-coordinate. Horizontal dashed line indicates values of $\langle \psi_6\rangle=0.302$ for the whole system and overall system density $\rho=1.02$. All the data are for an HTL glass configuration with $N=10000$. }
\end{figure*}

In order to capture the overall  structural changes in a small number of order parameters, we compute the two dimensional orientational order parameters 
\begin{equation}
\label{eq_psin}
\psi^i_n=\frac{1}{N_i}\sum_{j=1}^{N_i}\exp(i n\theta_{ij})
\end{equation}
where $N_i$ is the number of nearest neighbours of a particle $i$ obtained by Delaunay  tessellation as mentioned before, and $\theta_{ij}$ is the angle made by the vector from particle $i$ to its neighbor $j$ with the $x$-axis. We compute the average values $\langle \psi^L_n \rangle$ and $\langle \psi^S_n \rangle$ for the large and small particles, as averages over the respective single particle values, for $n = 5, 6, 7$, which are shown in Fig. \ref{psiall}. We see that $\langle \psi^L_n \rangle$ and $\langle \psi^S_n \rangle$ capture the same trends as the fractions $f^L_{n}$, $f^S_{n}$, with the exception that $\langle \psi^L_{n=6} \rangle$ exhibits roughly the same values for the well annealed glasses as for the yielded glasses. Nevertheless, $\langle \psi_{n=6} \rangle$, averaged over all the particles, shown in Fig. \ref{fig_psi6GlassSystem} is a good indicator of overall structural change, and correlates strikingly well with the changes in the energy for all the different glasses and for different system sizes. In investigating the spatial variation of structure in the presence of strain localisation, therefore, we employ $\langle \psi_{n=6} \rangle $ as the single indicator of structural change. Fig. \ref{sllod} (b) shows the temperature dependence of $\langle \psi_{n=6} \rangle $ from MD simulations, for reference, for different system sizes.

\subsection{Strain Localisation}

The localisation of strain above the yielding point under cyclic shear has recently been investigated in three dimensional glasses \cite{parmarPRX2019,Mitra_2021}. The shear bands, defined as regions with greater mobility from one cycle of strain to the next, have been shown to have comparatively higher energy as well as lower density. We thus consider whether similar strain localisation is observed for the two dimensional system we study, and what aspects of structural change it is accompanied by. In Fig. \ref{shearband}(a) we show a snap shot of the stroboscopic configuration for  $\gamma_{max}=0.07$, in the steady state, for the HTL glass. Particles are coloured according to their mean square displacement(MSD) computed between two successive stroboscopic configurations and all the particles that have $MSD>0.25\sigma^2_{SL}$ are coloured in red. We see that the most mobile particles are highly correlated in space and form system spanning band like structures. In Fig. \ref{shearband}(b) we show the slab wise averaged (along x-direction) mean square displacement ($MSD_x$), which clearly demonstrates the existence of shear band, within which particles move substantially between successive cycles. The $MSD_x$ profile can be described, to a good approximation, by a Gaussian with width $w$ roughly equal to $8.86\sigma_{SL}$. We see that most mobile particles are highly localized within the width of the shear band, with $MSD_x \approx 0.25\sigma_{SL}^2$ or greater within the shear band. We next investigate the structural change with the shear band. As defined in Eq. (\ref{eq_psin}) we  compute $\psi_{n=6}$ for each particle. The slab wise averaged $\langle \psi_6 \rangle_x$ along with the slab-wise density are presented in \ref{shearband}(c). We see a reduction of structural order accompanied by a reduction of density inside the shear band. Such an observation is consistent with observations for the three dimension model glasses investigated earlier \cite{parmarPRX2019,Mitra_2021}.

\subsection{Avalanches}
\begin{figure*}[t]
\centering{ 
\includegraphics[width =
    0.3\textwidth]{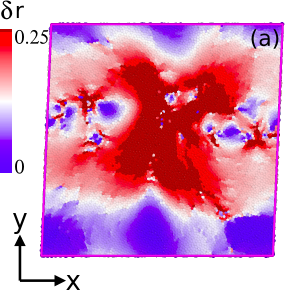} %\includegraphics[width =0.3\textwidth]{./avalanche_cluster_set1_cycle2502_naval49.png}
    \includegraphics[width =0.32\textwidth]{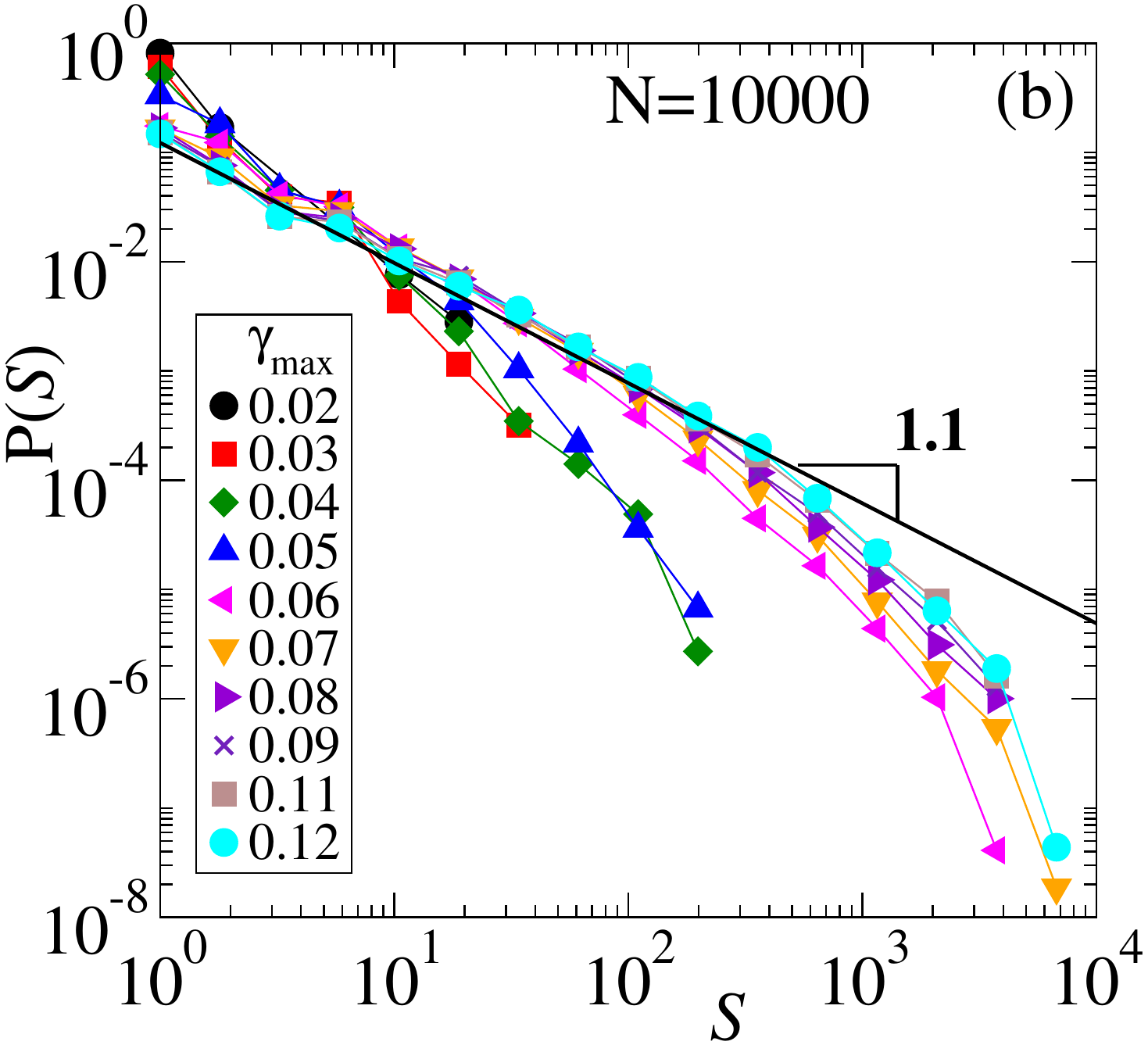}
    \includegraphics[width =
    0.32\textwidth]{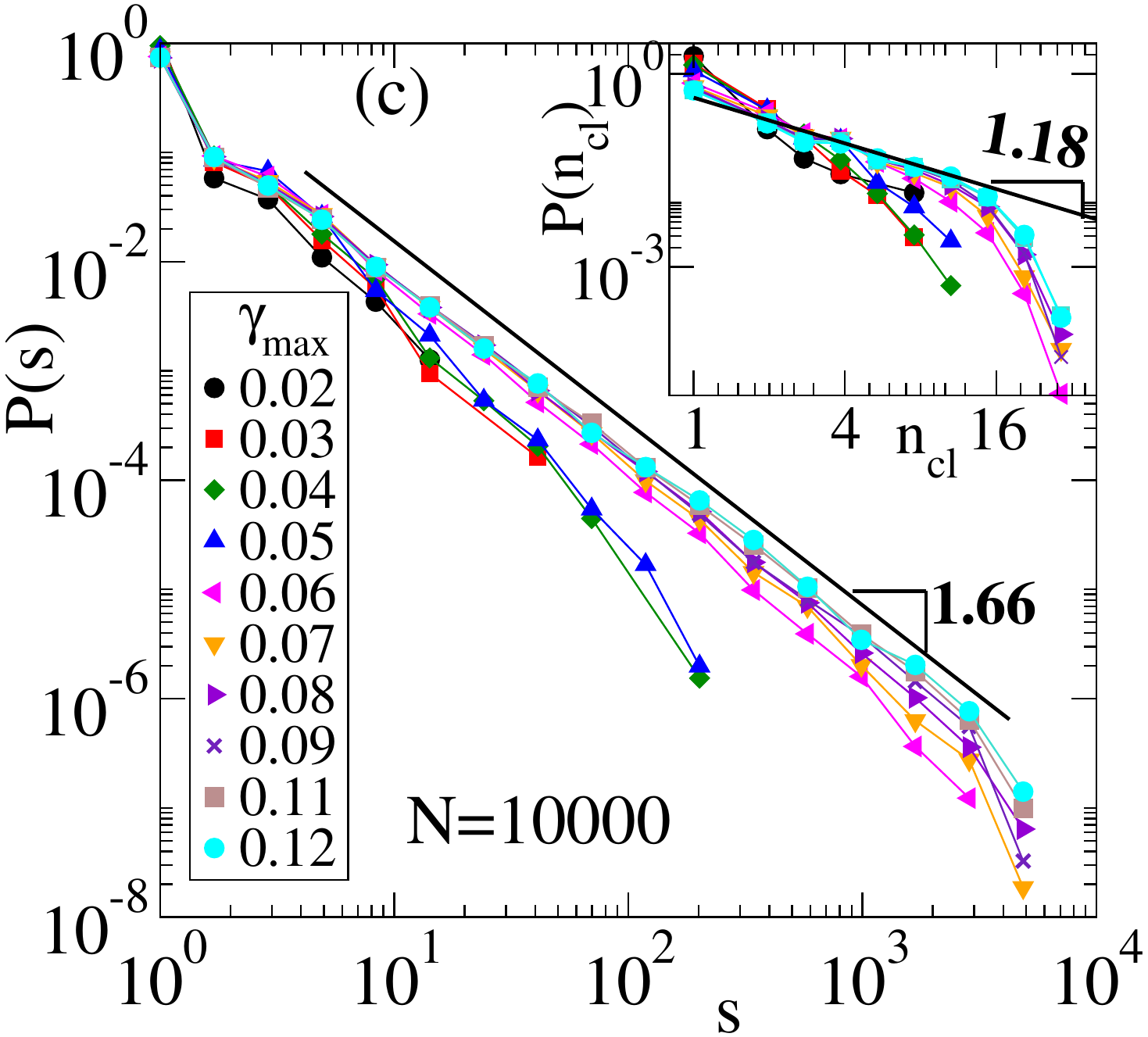} }
\caption{\label{fig_P_ed_s} (a) Colour map of particle displacements during a plastic rearrangement at $\gamma \approx 0.042$ during a strain cycle of amplitude $\gamma_{max}=0.06$. Particles with the lowest displacement $\delta r$ are shown in blue and those with the largest $\delta r$ are shown in red. Shown in maroon are the active particles with $\delta r>0.25$ which form a spatially correlated cluster with a cluster size $s=1359$ at the center. Distribution of (b) avalanche size $P(S)$ and (c) cluster size $P(s)$ for different $\gamma_{max}$ for system size $N=10000$. For both the cases two distinct sets can be observed, one corresponding to $\gamma_{max}\ge \gamma_{y}$ exhibiting a power law 
with exponent $\tau_a=1.1$ and $\tau_c=1.66$, for avalanche and cluster size distributions   respectively. Inset of (c) shows the distribution of the number of clusters in an avalanche, $P(n_{cl})$, with a power law with exponent $\kappa \approx 1.18$. }
\end{figure*}
Next we study the statistics of avalanches accompanying plastic rearrangements of particles during strain deformation of the system. Such events are accompanied by a discontinuous drop in stress and in energy. In our analysis here, we consider only the HTL glass.  We consider  steady state stroboscopic configurations and consider plastic rearrangement events in the first quadrant of the cycle, with $\gamma$ varying from $0\to \gamma_{max}$. To obtain the configurations just before and just after the plastic events precisely, we vary $\gamma$ with $d\gamma=10^{-6}$. With a given $\gamma_{max}$, for each sample we consider $20$ cycles to collect information on non-affine displacements corresponding to each drop event, identified following procedures in Refs. \cite{leishangthemNAT2017,parmarPRX2019}, by comparing the change in energy with the expected change if the change is elastic. Even in the presence of a plastic rearrangement, the long range elastic strain field leads to a continuous range of single particle displacements that are power law distributed
\cite{Eshelby1957,Dyre1999}, and one needs to identify particles that are displaced as part of the plastic rearrangement {\it vs.} those that are elastically displaced in response. In Refs. \cite{SchroderJCP2000,fioccoprl14,leishangthemNAT2017}, it was observed that the distribution of single particle displacements exhibits a power law regime, followed by an exponential cutoff. A value of $\delta r$ that separates the two regimes was chosen as a cutoff, and particles with displacements larger than the cutoff were identified as being part of the plastic core of the event. It was, however, also noted in \cite{leishangthemNAT2017} that the power law regime follows the expected scaling ($p(\delta r) \sim (\delta r)^{-(2D-1)/(D-1)}$) only for small amplitudes of shear. Thus, we follow the procedure of identifying a cutoff which falls within the exponential tail, but verifying that the distribution of avalanches and clusters is not sensitive to the choice. We also investigated a similar prescription in Ref. \cite{salernoPRE13} that employs the same reasoning but employs local deviatoric strain instead of displacements. We find, however, that the resulting avalanche and cluster distributions are sensitive to the choice of the cutoff and therefore, pending further investigation, we do not employ the deviatoric strain method in this work. In Fig. \ref{SI_distdelr_ps} of the Appendix, we show the distributions of particle displacements $\delta r$, along with different choice of the cutoff.  Fig. \ref{SI_distdelr_ps} also shows that the power law part of the cluster size distribution is insensitive to a choice of cutoff $\delta r_c$ from $0.15$ to $0.3$. In what follows, we employ the choice $\delta r_c = 0.25$ to identify what we term {\it active} particles.

In Fig. \ref{fig_P_ed_s}(a) we show a colour map of particle displacements for a plastic rearrangement. Imposing a cut-off value $\delta r_c=0.25$ on the displacement field we identify all the active particles which are coloured in maroon. It can be observed that such active particles are highly correlated in space and form a cluster. In general, the set of all active particles (whose number yields the avalanche size) may be composed of disconnected clusters of rearranging particles. We identify such clusters, as well as the number of such clusters in a given avalanche, and investigate their properties. If an avalanche of size $(S)$ (total number of active particles) consists $n_{cl}$ number of clusters and $i$th cluster has size $s_i$, then one has 
\begin{equation}
S=\sum_{i=0}^{n_{cl}}s_i.
\end{equation}

We compute the cluster statistics of the active particles, considering two active particles (or type A, B) to be in the same cluster if they are separated by less than a distance $r_{AB}^{min}$, the first minimum of the relevant partial pair correlation function ($r_{AB}^{min}=1.03,1.40,1.70$ for $SS$,$LL$, and $SL$ pairs respectively). The distribution of the investigated quantities, namely the avalanche size ($S$),  cluster size ($s$), and the number of clusters in an avalanche, $n_{cl}$, are expected to exhibit power law regimes \cite{PriolPRL21}, characterised by exponents identified below: 

\begin{equation}
    P(S)\sim S^{-\tau_a},\ \ \   P(s)\sim s^{-\tau_c},\ \ \ P(n_{cl})\sim n_{cl}^{-\kappa}.
    \label{pSpspncl}
\end{equation}
 We study the size distribution of avalanches and clusters for various strain amplitudes as shown in Fig. \ref{fig_P_ed_s}(b) and \ref{fig_P_ed_s}(c), respectively. The  distribution of the number of clusters is shown in the inset of Fig. \ref{fig_P_ed_s}(c). For all the three quantities, one can identify two distinct sets of distributions that correspond respectively to strain amplitudes $\gamma_{max}$ below, and above, the yield strain $\gamma_y$ as also observed in  a previous study for a three dimensional glass for cluster sizes \cite{leishangthemNAT2017}. For $\gamma_{max}>\gamma_y$ in particular, the distributions exhibit clear power-law regimes, with exponents $\tau_a=1.1\pm 0.1$, $\tau_c=1.66 \pm 0.06$ and $\kappa=1.18 \pm0.07$. The errors indicated represent the range of variation of the exponent values obtained by fitting the different data sets and ranges of sizes. Results are found to be similar over a range of choice of the cutoff $\delta r_c$ as already mentioned (and shown in the Appendix, Fig. \ref{SI_distdelr_ps}). We discuss the relationship between $\tau_a$, $\tau_c$ and $\kappa$, following the arguments in Ref. \cite{PriolPRL21} for models of crack propagation, in the presence of long range interactions\cite{Laurson2010}, and as discussed for three dimensional glasses in Ref. \cite{bhaumiksilica2021}.  
 
 Considering an avalanche of size $S$, we represent the number of the size of the clusters present as $n(s|S)$. Since the sum of the sizes of the clusters must equal the size of the avalanche, we have, by definition, 
 \begin{equation}\label{eq:scond}
 \int_{1}^{S} ~ s~ n(s|S) ~ds = S.
 \end{equation}\label{eq:clustsize} 
 Likewise, by definition, the total number of clusters present is given by 
  \begin{equation}
 \int_{1}^{S} ~  n(s|S)~ ds = n_{cl}(S).
 \end{equation}
 The distribution of cluster sizes for a given $S$ exhibits a power law form up to the maximum size of $S$, as also noted in \cite{PriolPRL21} (The distributions, however, are noisy within the statistics we have and thus we do not rely on them directly in the discussion below, but instead on average quantities as described later.)  We thus assume $n(s|S) = A s^\tau$, and the condition Eq. \ref{eq:scond} requires $A = (2-\tau) S^{\tau-1}$ (for large $S$). Note that we do not assume $\tau = \tau_c$ for reasons that will be discussed further later.  With this choice, and assuming that $\tau > 1$, we obtain for the number of clusters 
   \begin{equation} \label{nclS}
n_{cl}(S) \sim S^{\tau - 1}. 
 \end{equation}
 Defining the exponent $\gamma_{ns}$ by $\langle n_{cl}\rangle_S\sim S^{\gamma_{ns}  }$, we have  $\gamma_{ns} = \tau - 1$, or  
 
\begin{figure*}[t]
\centering{
 \includegraphics[width =
    0.35\textwidth]{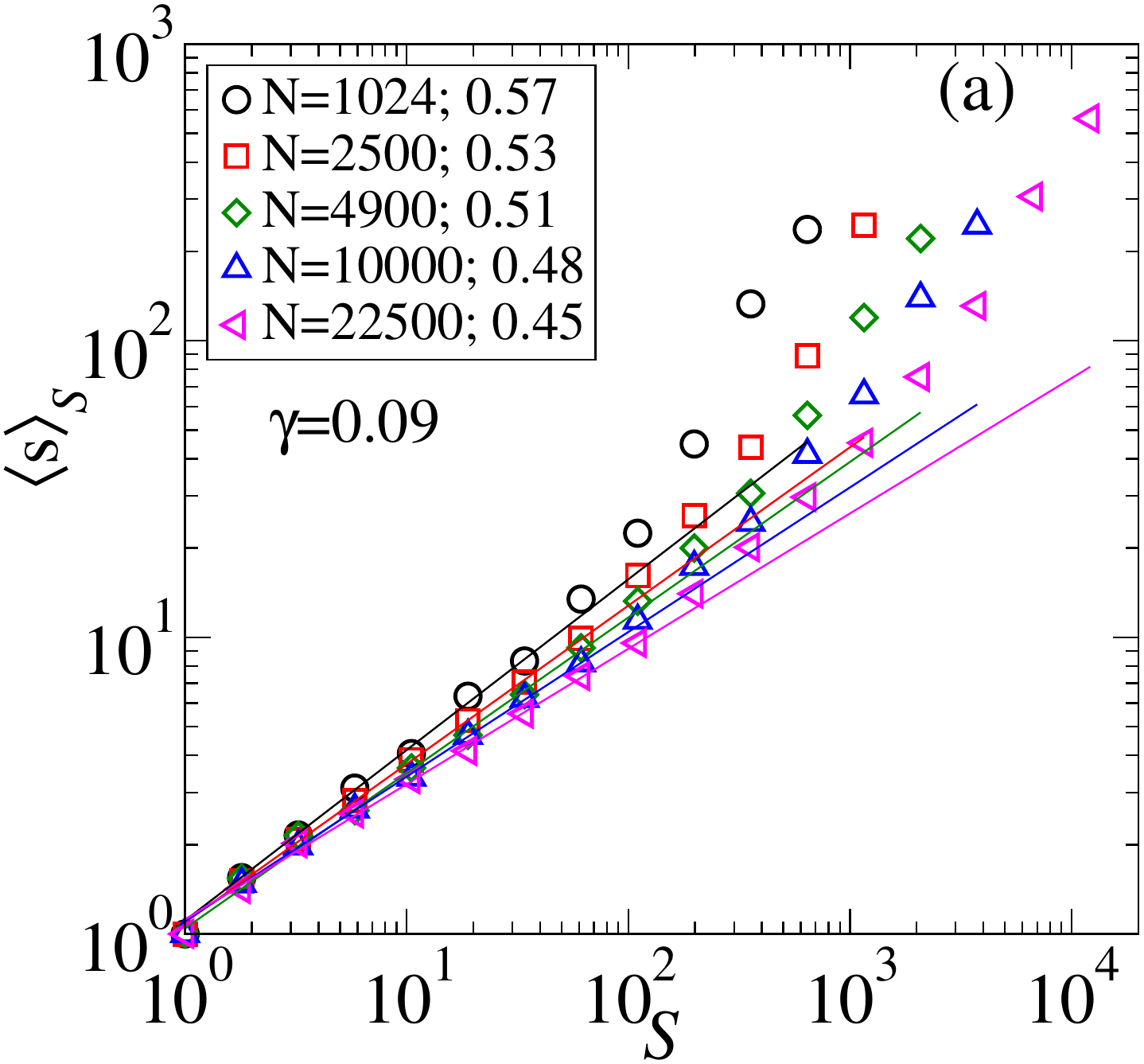}
  \includegraphics[width =
    0.35\textwidth]{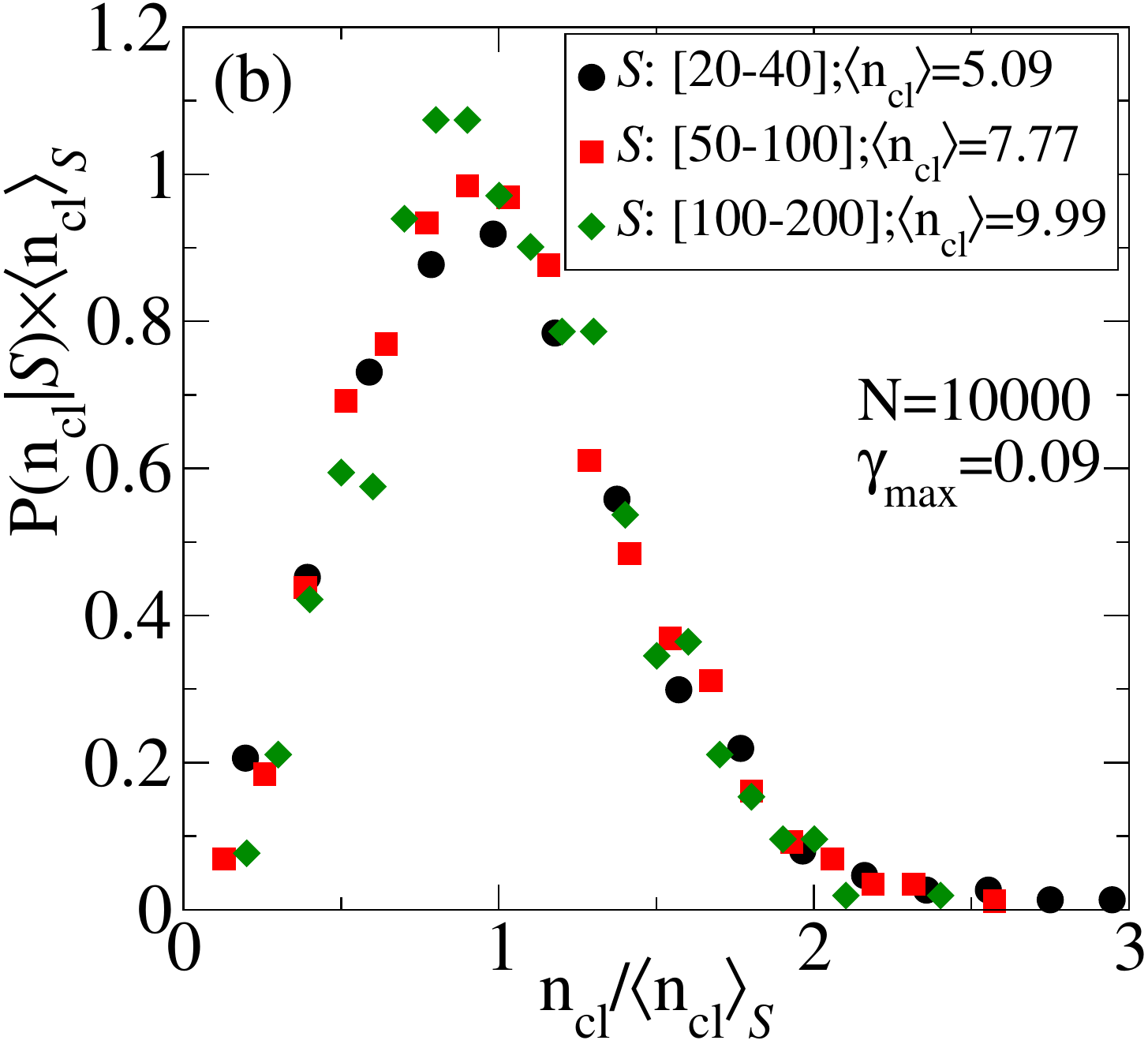}
    }
\caption{\label{Sconditional} (a) The mean cluster size $\langle s\rangle_S$ for a given avalanche size $S$ for different system sizes at $\gamma_{max}=0.09$. For the small $S$ range, $\langle s\rangle_S$ increases as a power law with exponent converging to $0.45$ for the  largest system size. Legends indicate the system size and the exponent value. (b) Scaled conditional probability distribution of the number of clusters for given avalanche size $S$ for a system of size $N=10000$. Legends indicate the range of $S$ values and the mean number of clusters.}
\end{figure*}

 %The average cluster numbers $\langle n_{cl}\rangle_S$ of an avalanche size $S$ should scale as 
%\begin{equation}
%\langle n_{cl}\rangle_S\sim S^{\gamma_{ns}  } .
%\label{nclS}
%\end{equation}
%We note that clusters belonging to an avalanche of size $S$ do not have a typical size, as they follow power-law distribution $P(s|S)\sim s^{-\tau}$ which claims the average cluster size $\langle s\rangle_S$ up to a given avalanche size $S$ to scale as $\langle s\rangle_S = \int sP(s|S)ds \sim S^{2-\tau}$. Since $\langle s\rangle_S = S/\langle n_{cl}\rangle_S$, we obtain a scaling relation
\begin{equation}\label{gtau} 
\tau=\gamma_{ns}+1
\end{equation}
We note that the above relations also imply that the mean cluster size 
\begin{equation}
\langle s\rangle_S\sim S^{2 - \tau}
\label{mclust}
\end{equation}
as along noted in \cite{PriolPRL21}. Fig. \ref{Sconditional}(a) shows that indeed, the mean cluster size scales for large system sizes with exponent $2 - \tau = 0.45$, corresponding to $\tau = 1.55$.

Normalising $n(s|S)$ we obtain the distribution $P(s|S) \sim s^{-\tau}$ and with $P(S) \sim S^{-\tau_{a}}$, we obtain, the full distribution of $s$ to be 
\begin{equation}
P(s) = \int_{s}^{\infty} P(s|S) P(S) dS \sim s^{-(\tau + (\tau_a -1))}.
\label{sdist}
\end{equation}
Thus, the exponent describing the cluster sizes is 
\begin{equation}
\tau_c = \tau + (\tau_a - 1)
\label{sexp}
\end{equation}
which, proposed in \cite{bhaumiksilica2021}  has not, to our knowledge, previously been discussed. We discuss the validity of this relation in the next section. 

We next consider the distribution of the number of clusters. Since the number of clusters $n_{cl}$ for a given avalanche size $S$ are distributed with an average value $\langle n_{cl}\rangle_S \sim S^{\gamma_{ns}}$, the conditional probability $P(n_{cl}|S)$ should have the functional form $P(n_{cl}|S)\sim S^{-\gamma_{ns}}g(n_{cl}/ S^{\gamma_{ns}})$, which results in $\langle n_{cl}\rangle_S \sim S^{\gamma_{ns}}$, through $\langle n_{cl}\rangle_S=\int n_{cl}P(n_{cl}|S) dn_{cl}$. The scaling function $g$, shown in Fig. \ref{Sconditional}, is peaked around the mean value $\langle n_{cl}\rangle_S$, as also seen in \cite{PriolPRL21}. The full distribution of $n_{cl}$ can be obtained as 
\begin{eqnarray}
P(n_{cl})&=&\int P(n_{cl}|S)P(S)dS \nonumber \\ 
&\sim& n_{cl}^{-(1+(\tau_a-1)/\gamma_{ns})}.
\label{pncl_n}
\end{eqnarray}
Using Eq. \ref{pSpspncl}, \ref{gtau}, and \ref{pncl_n} we obtain the scaling relation
\begin{equation}
\kappa =1+\frac{\tau_a-1}{\tau-1}
\label{kappatau_relation}
\end{equation}
This relation is well satisfied with the measured values of the exponents described above, namely $\tau_a=1.1\pm 0.1$, $\tau=1.55 \pm 0.06$ and $\kappa=1.18 \pm0.07$. To further characterize the scaling behaviour we  perform a system size analysis in the next section. 

\subsection{Finite size scaling}

\begin{figure*}[t]
\centering{ \includegraphics[width =
    0.32\textwidth]{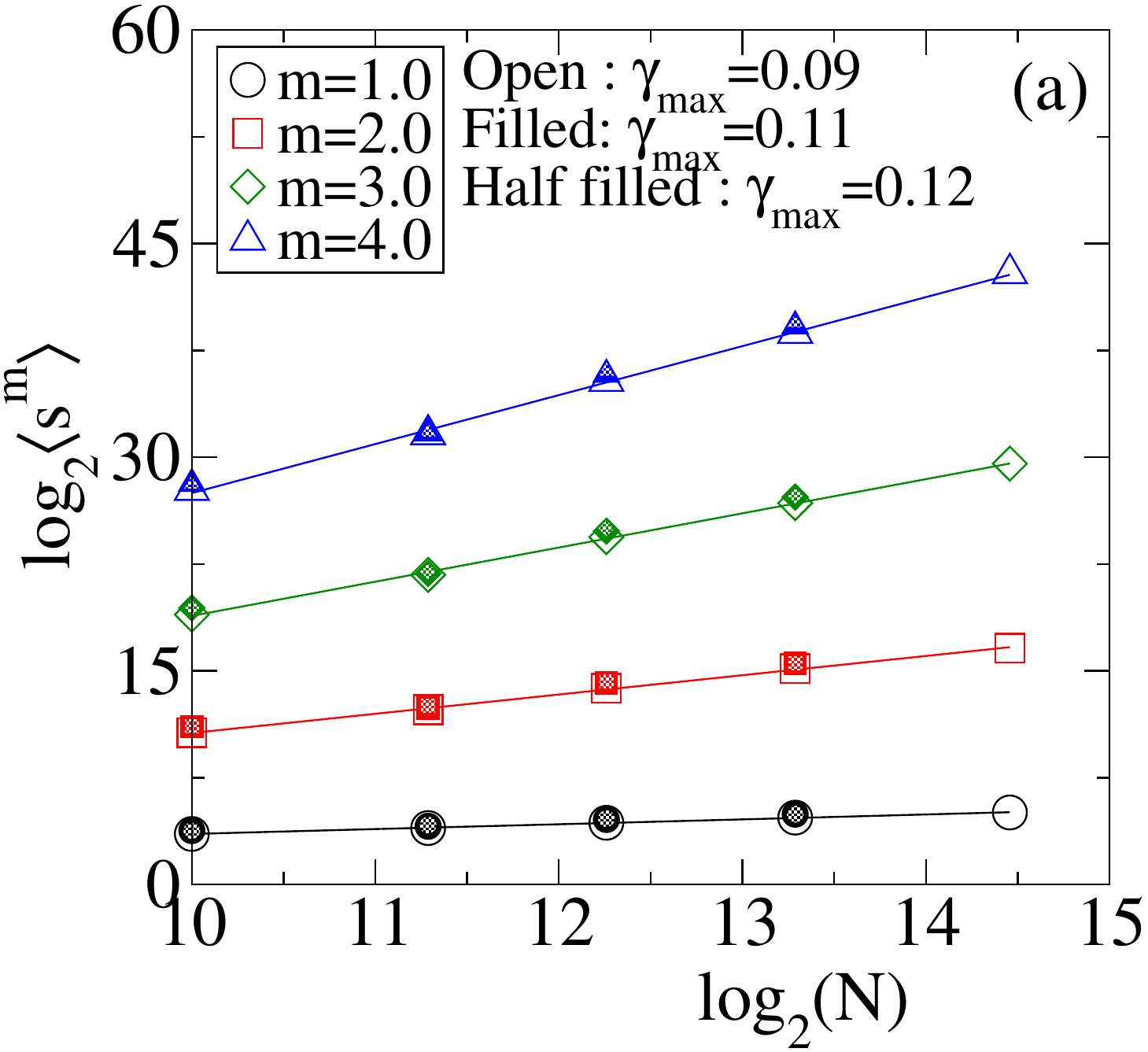} 
\includegraphics[width =
    0.32\textwidth]{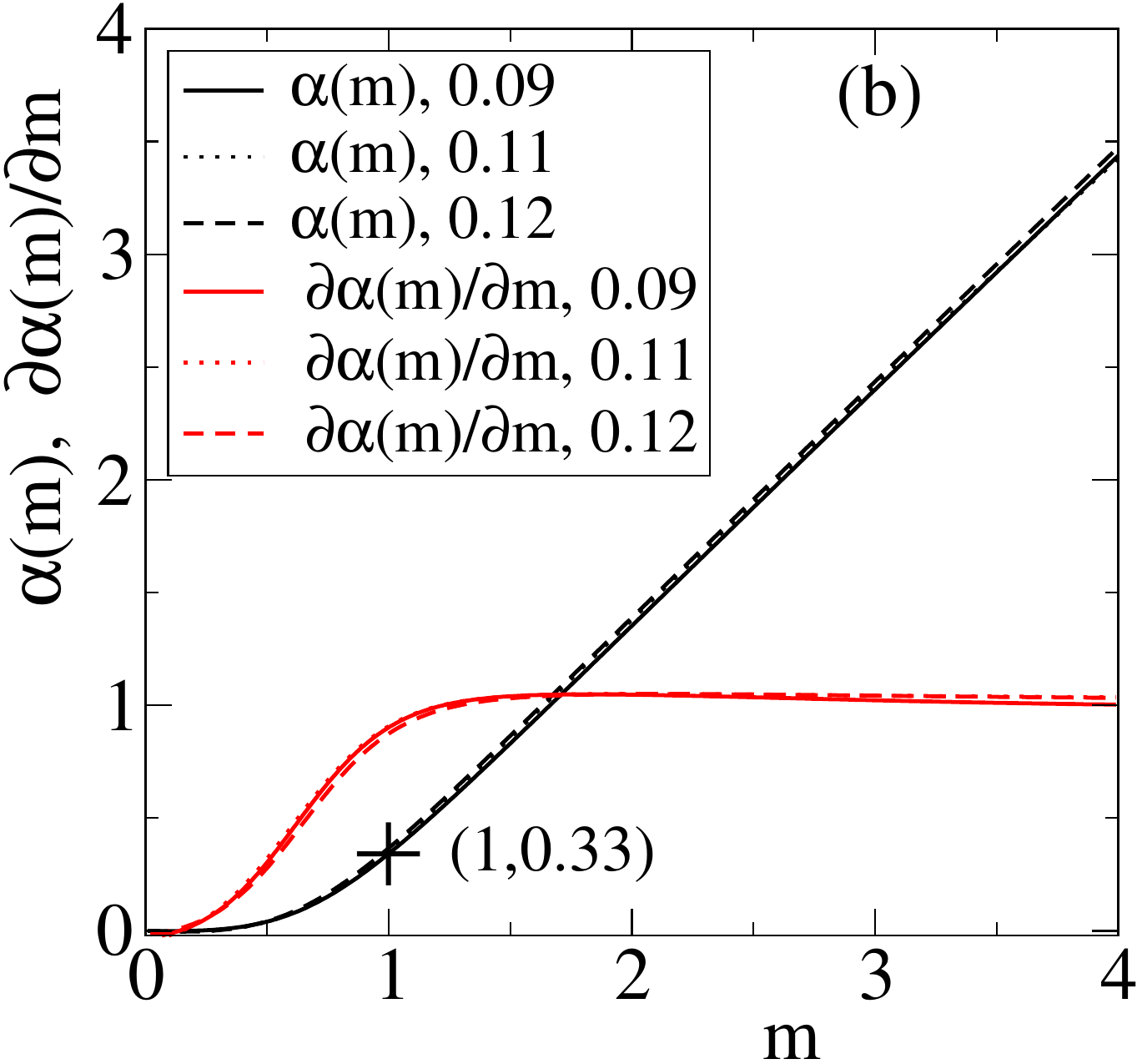}
\includegraphics[width =
    0.34\textwidth]{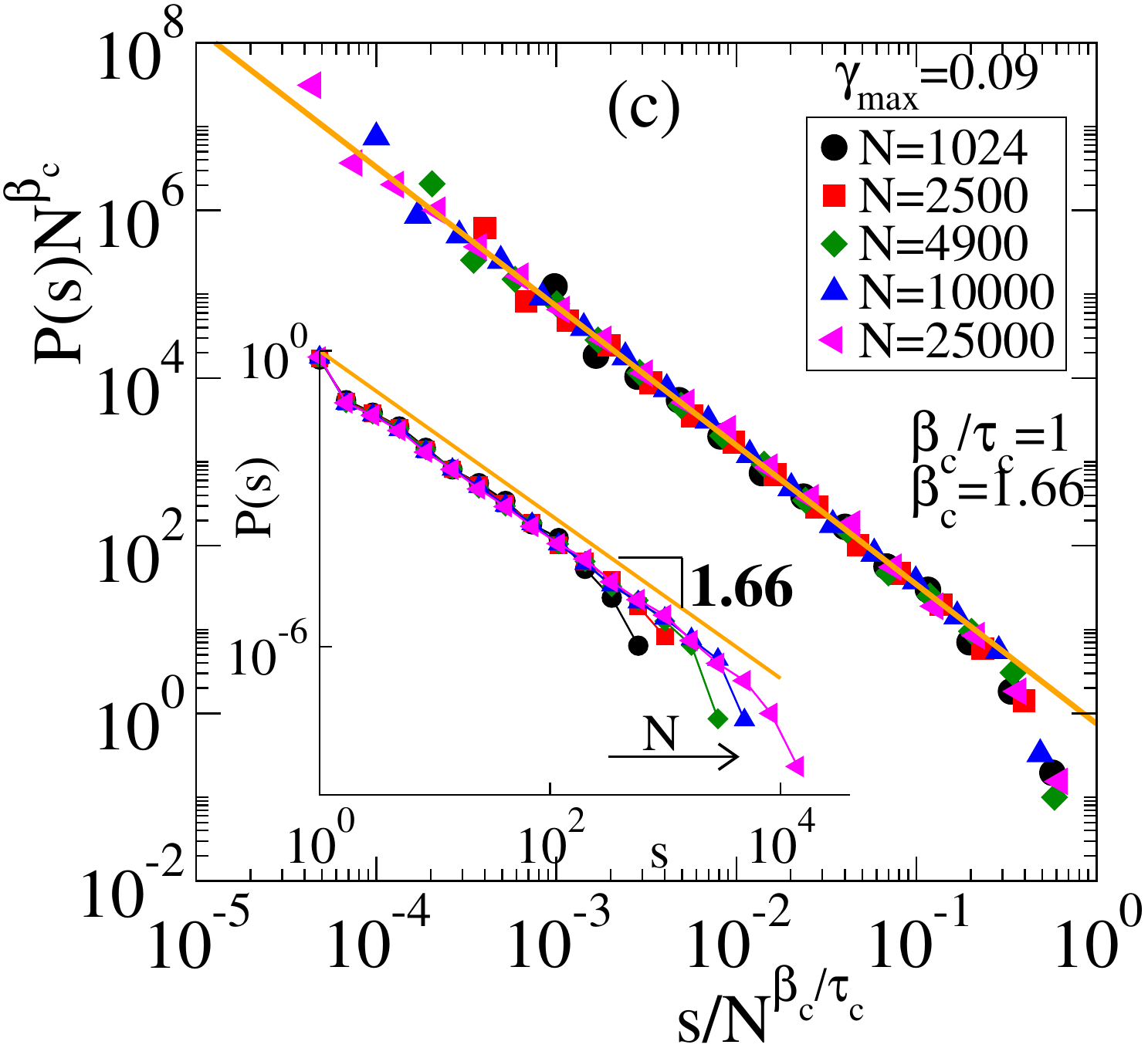}}
\caption{\label{fig_fssPs} (a) Moments of the cluster size
  $\langle s^m \rangle$ against system size $N$  for
  different values of $m$. Lines are the fitted straight lines obtained
  from linear regression. (b) Moment exponent  $\sigma(m)$ and its derivative with respect to $m$. (c) Scaled distribution of cluster size against scaled cluster size for different system sizes for $\gamma_{max}=0.09$. Inset shows the unscaled data.}
\end{figure*}
\begin{figure*}[t]
  \centering{ 
    \includegraphics[width = 0.32\textwidth]{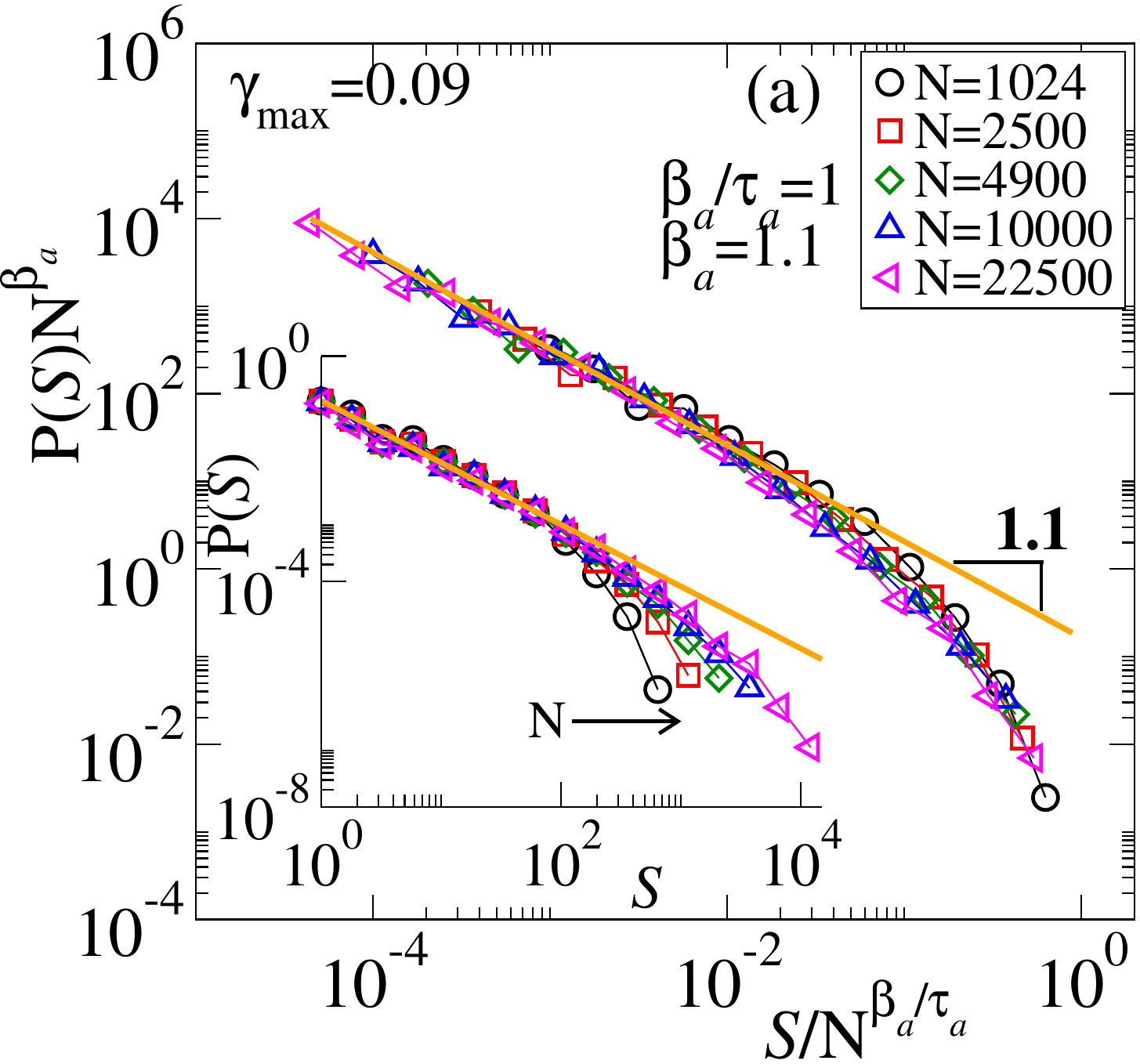}
    \includegraphics[width = 0.32\textwidth]{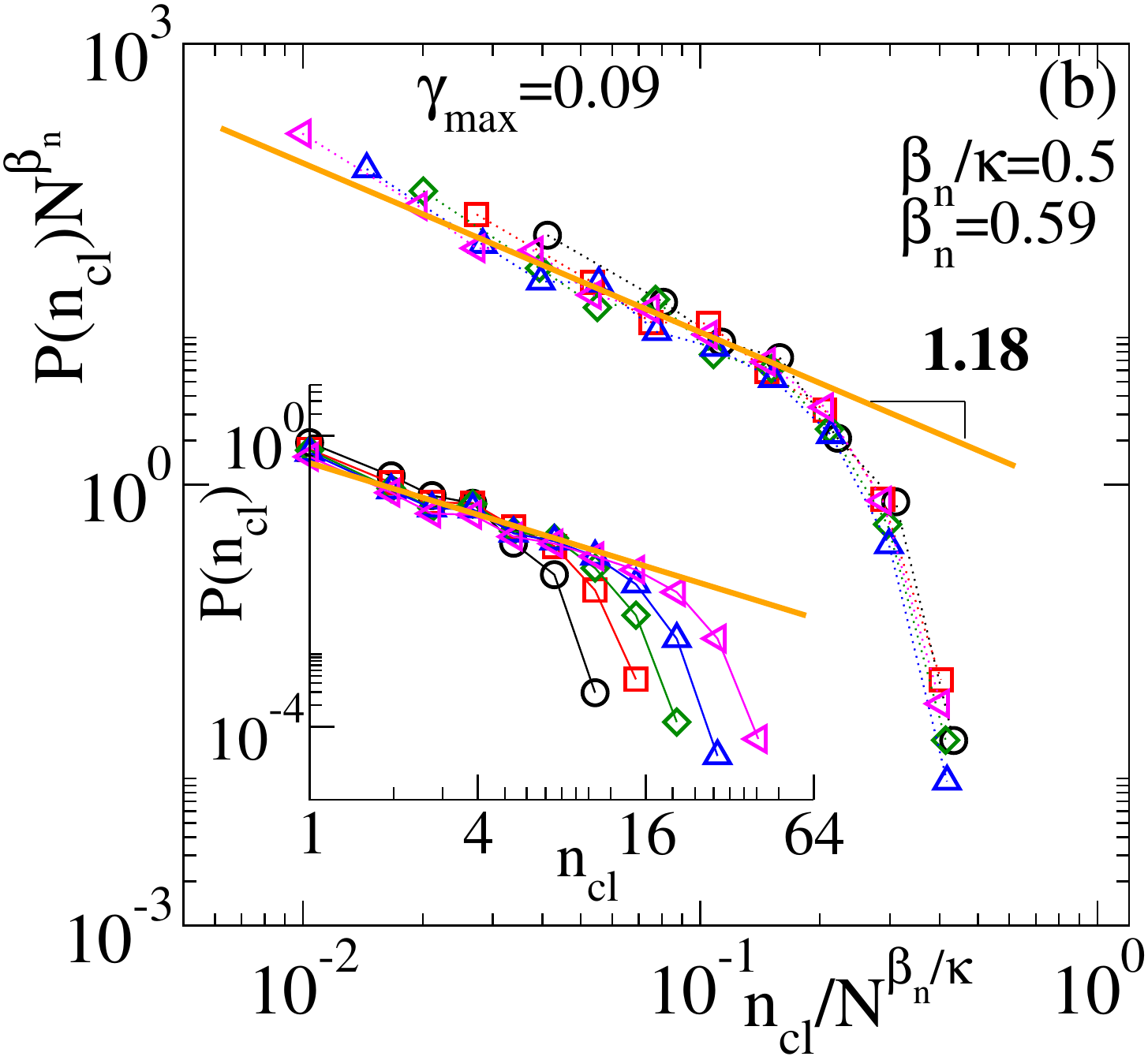}
         \includegraphics[width = 0.31\textwidth]{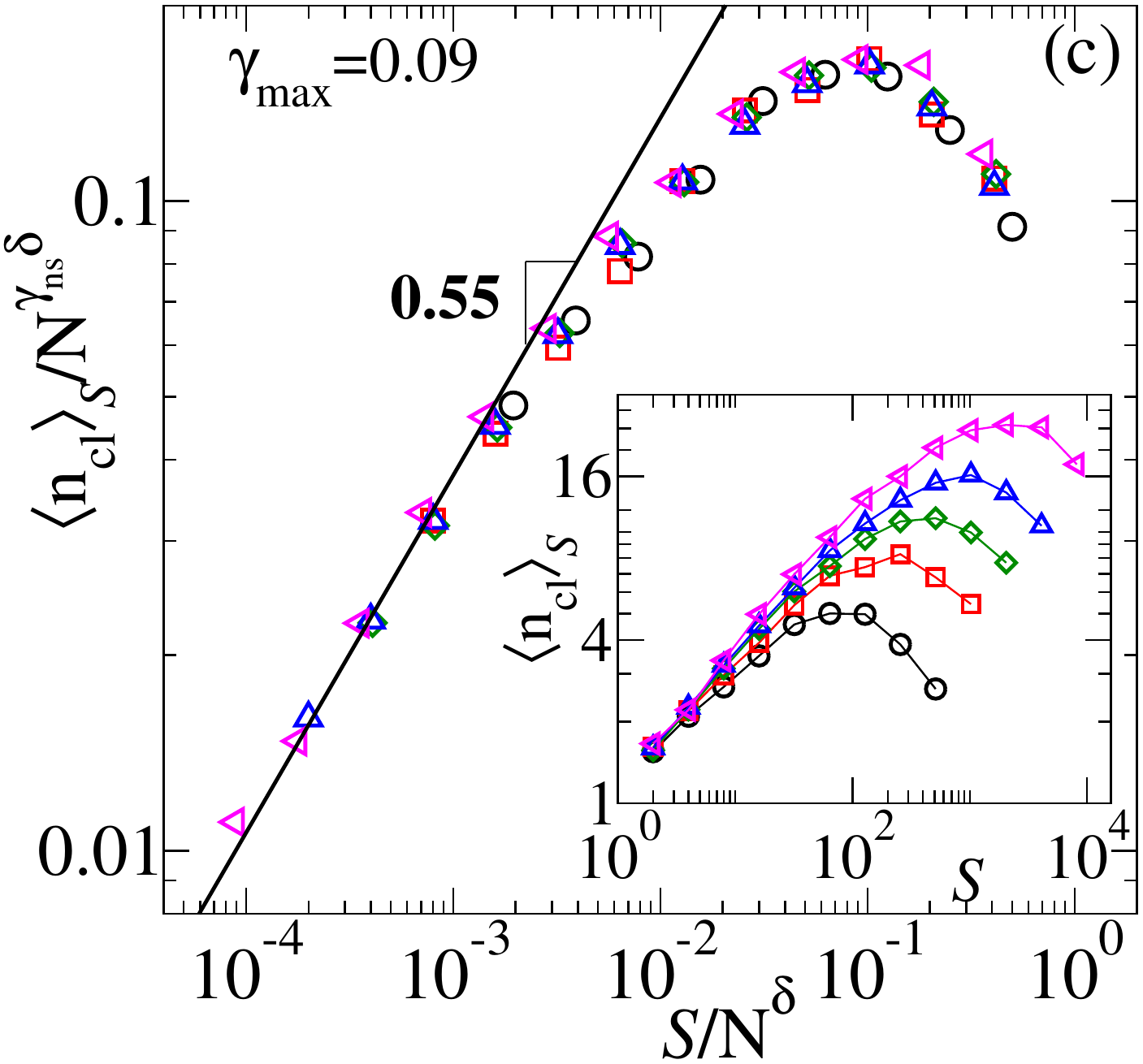}
     }
  \caption{\label{fig_fssSncl} (a) Scaled avalanche size distributions against scaled avalanche size for different system sizes with $\tau_a=1.1$ and $\beta_a/\tau_a=1.0$. Inset shows the unscaled data which can be well described by the power-law $P(S)\sim S^{-1.1}$. (b) Scaled cluster number distributions for different system sizes. Best data collapse is obtained using $\beta_n/\kappa=0.5$ and $\kappa=1.18$. Inset shows the unscaled data with increasing cutoff as $N$ increases. The straight line through the data points indicates the  power law $P(n_{cl})\sim n_{cl}^{-1.18}$. (c) Scaled average number of clusters $\langle n_{cl} \rangle_S N^{\gamma_{ns}\delta}$ against scaled avalanche size $S/N^{\delta}$ for different system sizes. From fitting of the data for the largest system sizes we obtain $\gamma_{ns}=0.55\pm 0.05$, as indicated.}
\end{figure*}

\begin{figure*}[t]
  \centering{ 
    \includegraphics[width = 0.35\textwidth]{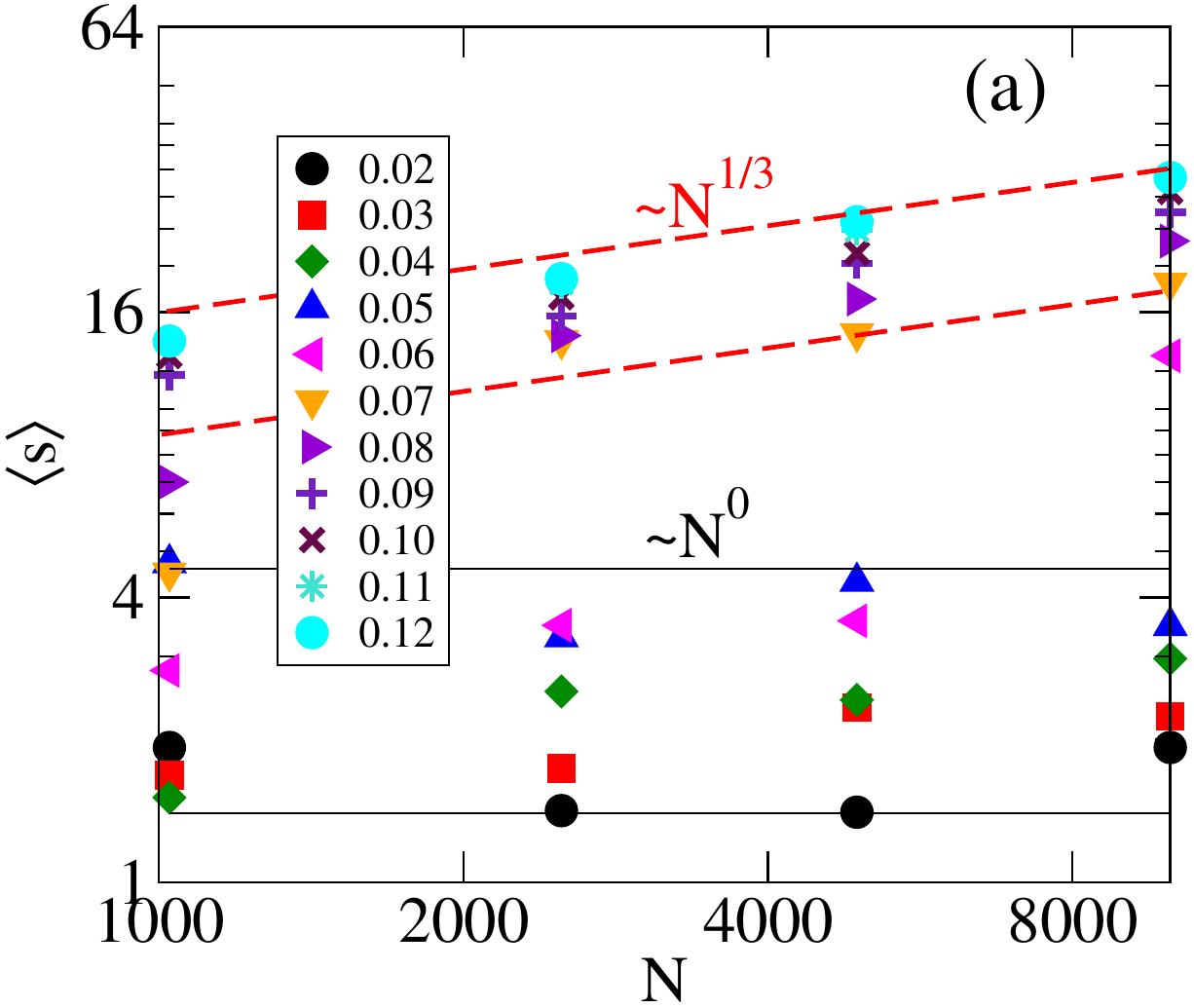}\quad
    \includegraphics[width = 0.35\textwidth]{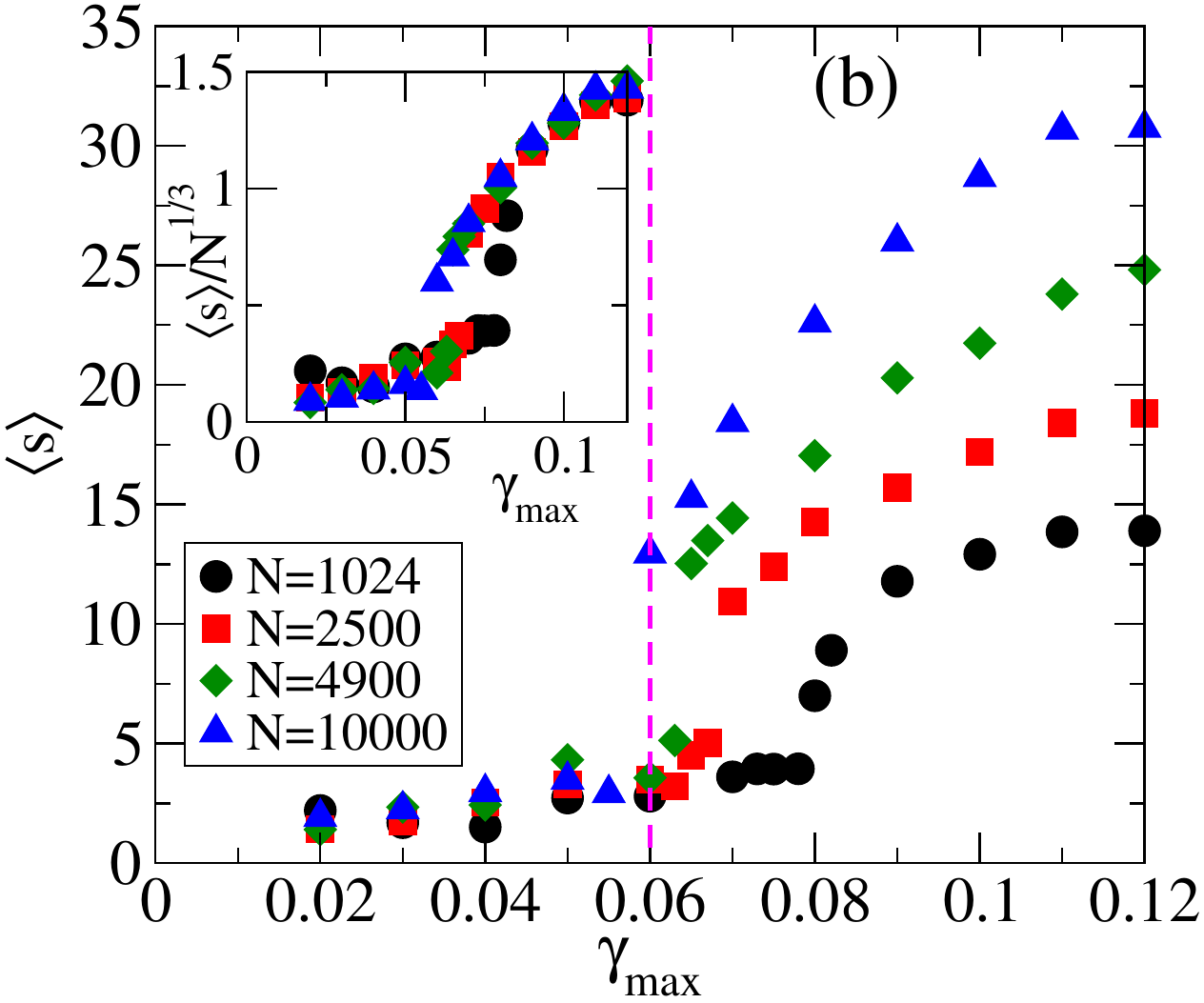}
     }
  \caption{\label{fig_avgs_g} (a) Average cluster size $\langle s \rangle $ against system size for different $\gamma_{max}$. (b) Average cluster size against strain amplitude $\gamma_{max}$ for different system sizes. Inset: scaled cluster size against $\gamma_{max}$ to show the $\langle s \rangle \sim N^{1/3}$ scaling.}
\end{figure*}

To investigate the system size dependence, we collect the  statistics of avalanche sizes, cluster sizes and the number of clusters, for a range of system sizes $N=1024,2500,4900,10000,22500$ for the HTL glass. In the following we will describe in detail the finite size scaling analysis of various quantities. We begin by analysing the distribution of cluster sizes assuming a finite size scaling form
\begin{equation} 
P(s,N)=N^{-\beta_c}f\left[\frac{s}{N^{\beta_c/\tau_c}}\right]
\end{equation}
where the scaling function has the characteristics $f(x)\to x^{-\tau_c}$ for $x \to 0$, so that the power-law behaviour retrieved as $P(s)\sim s^{-\tau_c}$ and $f(x)$ decreases to zero rapidly when $x\to 1$. To extract the value of $\beta_c$ and $\tau_c$ we calculate the various moments of the distribution defined as: 
\begin{eqnarray}
\langle s^m \rangle &=& \int_0^\infty s^mP(s,N)ds=\int_0^\infty s^m
N^{-\beta_c}f(s/N^{\beta_c/\tau_c})ds \nonumber \\
&=&N^{\beta_c/\tau_c(m+1-\tau_c)}\int_0^\infty
z^m f(z)dz
\end{eqnarray}
where, in the last step we replace the variable $z=s/N^{\beta_c/\tau_c}$. Since the integration $\int z^mdz$ is a constant the moment $\langle s^m\rangle$ should vary with system size as 
\begin{equation}
\langle s^m\rangle\sim N^{\frac{\beta_c}{\tau_c}(m+1-\tau_c)}\sim N^{\alpha(m)},
\end{equation}
where  the moment exponent $\alpha(m)=\frac{\beta_c}{\tau_c}(m+1-\tau_c)$ is evaluated for different values of $m$ ranging from $0$ to $4$ in intervals of $0.01$. For some integer  values of $m$ the data of $\alpha(m)$ against $N$ are presented in Fig.\ref{fig_fssPs}(a) for different strain amplitudes, larger than the yield value. For $P(s)$ to obey finite size scaling (FSS) for a given $\gamma_{max}$, the moment exponent $\alpha(m)$ should have a constant gap between two successive values of $m$, {\em i.e.}, $\alpha(m+1) -\alpha(m) = \beta_c/\tau_c$ and a derivative of $\alpha(m)$ with respect to $m$ should converge to $\beta_c/\tau_c$ for higher values of $m$. In Fig. \ref{fig_fssPs}(b) we present the variation of $\alpha(m)$ and its derivative $\partial \alpha(m)/ \partial m$(obtained from the central difference) as a function of $m$ for several strain amplitudes. These 
data reveal that $\partial \alpha(m)/ \partial m$ converges satisfactorily for all $\gamma_{max}$ considered here in the post yield regime providing the value of $\beta_c/\tau_c=1$. By a linear fit of $\sigma(m)$ for large $m$  we determine $\beta_c/\tau_c=1.02\pm0.02$ and $\beta_c=1.66\pm 0.08$ which yield $\tau_c=1.69\pm 0.07$ consistent with the direct measurement described earlier. In Fig.\ref{fig_fssPs}(c) we present the scaling plot of the cluster size distribution that demonstrates the excellent quality of the data collapse for different system sizes. 

Similar to cluster sizes $s$, we perform finite size scaling for the avalanche sizes $S$  and for the number of clusters $n_{cl}$ by writing 
\begin{equation} 
P(S,N)=N^{-\beta_a}f_a\left[\frac{S}{N^{\beta_a/\tau_a}}\right]
\end{equation}
and 
\begin{equation} 
P(n_{cl},N)=N^{-\beta_n}f_n\left[\frac{n_{cl}}{N^{\beta_n/\kappa}}\right]
\end{equation}
where $f_a$ and $f_n$ are the scaling function and $\beta_a$ and $\beta_n$ are the associated scaling exponents. In Figs. \ref{fig_fssSncl}(a) and \ref{fig_fssSncl}(b) we present the scaled distribution for $S$ and $n_{cl}$ respectively. The best data collapse obtained for $\beta_a/\tau_a=1$ and $\tau_a=1.1$ for avalanche size $S$, whereas for $n_{cl}$ we use $\beta_n/\kappa=0.5$ and $\kappa=1.18$. Given the limited range of the power laws, we find that the moment analysis along the lines performed for cluster sizes is less satisfactory, and we thus do not use such a procedure in these cases.

We next investigate the system size scaling of average number of clusters $\langle n_{cl}\rangle_S$ for a given avalanche size $S$. Keeping in mind Eq. \ref{nclS} and Eq. \ref{gtau}, we assume the finite size scaling form as 
\begin{equation}
\langle n_{cl}\rangle_S =N^{\gamma_{ns}\delta}g_n(S/N^{\delta})
\label{fss_nclS}
\end{equation}
where the scaling function $g_n(x)$ grows as $x^{\gamma_{ns}}$ for small $x$. In Fig. \ref{fig_fssSncl}(c) we show the scaled data of average cluster number according to Eq. (\ref{fss_nclS}). A reasonable data collapse obtained for different system sizes using $\gamma_{ns}=0.55$ and $\delta=1$. The $\gamma_{ns}$ value is consistent with the estimated exponent $\tau = 1.55$. Note that the exponent value we estimate, of $\tau_c \approx 1.66$, with $\gamma_{ns}=0.55$  requires Eq. \ref{sexp} to be true, and thus, the we conclude that the exponents we estimate
\\
\begin{equation}
\tau = 1.55 ,\ \  \gamma_{ns} = 0.55  ,\ \   \tau_c = 1.66 ,\ \ \tau_a = 1.1 ,\ \  \kappa = 1.18
    \label{allexp}
\end{equation}
are all satisfactorily and consistently explained by the analysis we presented in the previous section.

%slightly smaller value than that predicted from the scaling relation Eq. (\ref{nclS}) using $\tau=1.66$. Such a difference we attribute to the finite size of the system. We note in this connection that the avalanche distributions do not show convincing power law regimes, and it may be necessary to go to larger system sizes. The avalanche distribution will clear affect the analysis of the statistics of the number of clusters. 

Next we consider the variation of average cluster size $\langle s \rangle$ with system size across the yield strain amplitude. In Fig. \ref{fig_avgs_g}(a) we show $\langle s \rangle$ against $N$, which  exhibits different scaling behaviour in the post- and pre-yield regimes: $\langle s \rangle \sim N^{1/3}$ when $\gamma_{max}>\gamma_y$ and the $N$ dependence is negligible below $\gamma_{y}$. Fig. \ref{fig_avgs_g}(b) shows the data of $\langle s \rangle$ against strain amplitude for different $N$. It can be seen that $\langle s \rangle$ does not depend on system size for $\gamma_{max}<\gamma_{y}=0.06$, whereas it exhibits a $\langle s \rangle \sim N^{1/3}$ dependence for  $\gamma_{max} > \gamma_{y}$ as described before. The scaled averaged size is plotted against $\gamma_{max}$ in inset of Fig. \ref{fig_avgs_g}(b) in order to verify the scaling behaviour of $\langle s \rangle$ in post-yield regime. The observed data collapse confirms the $ N^{1/3}$ scaling, which is consistent with the first moment obtained, as shown in Fig. \ref{fig_fssPs}(b).

Finally, we study the statistics of energy drops during a plastic rearrangement, which has been investigated in the past as a quantifier of avalanche size.  The plastic component of energy drop ($\Delta U_{pl}$) (see Appendix Fig. \ref{SI_delE}) follows power-law scaling $P(\Delta U_{pl}) \sim \Delta U_{pl}^{-1.25}$ in agreement with observations in Ref. \cite{salernoPRE13,leishangthemNAT2017} for three dimensional glasses. The variation of  $\langle \Delta U_{pl} \rangle$ with system size for $\gamma_{max}>\gamma_y$ displays similar scaling behaviour, $\Delta U_{pl} \sim N^{1/3}$, as observed for the mean cluster size. 

\section{Summary and Discussion}

To summarize, we have studied the yielding behaviour of a two dimensional glass former under athermal quasi-static cyclic deformation, focusing in particular on the role of annealing of the glasses on the yielding behaviour. Upon repeated cycles of shear, the glasses reach steady states, which are either invariant from one cycle to the other (absorbing states) or whose properties fluctuate around a mean value, with the particles exhibiting diffusive behaviour from cycle to cycle. The former corresponds to strain amplitudes below yielding, whereas the latter to strain amplitudes above the yield strain amplitude. Our results demonstrate that the nature of yielding changes with the degree of annealing, consistently with results for three dimensional glasses, and that well annealed glasses exhibit a strongly discontinuous yielding transition, with the discontinuous changes in energy, stress and other properties increasing with the degree of annealing. We characterise the structural changes during shear deformation, and show that substantial changes in structure occur both above and below yielding. In particular, our results show that structural heterogeneities accompany the formation of shear bands above yielding. We have performed a detailed investigation of the statistics of avalanches, including an analysis of the decomposition of avalanches into clusters. We have analysed the relationship between exponents characterising the distribution of avalanches, cluster sizes and the number of clusters, and verify predictions from recent investigations \cite{PriolPRL21} of these exponents in the context of models for crack propagation with long range interactions. In addition, we propose (in an accompanying paper \cite{bhaumiksilica2021} and here) a new relationship between the exponents characterising avalanches and clusters and demonstrate the relationship to be valid. It will be interesting for this relationship to be investigated in other glasses. These analyses have been performed using detailed finite size scaling. We further verify for the two dimensional system studied that avalanche sizes do not grow with system size below yielding, whereas they do above, mirroring results for three dimensional glasses \cite{leishangthemNAT2017}. 

\section*{Appendix}

\renewcommand{\theequation}{A\arabic{equation}}
\renewcommand{\thefigure}{A\arabic{figure}}
\renewcommand{\thesection}{A-\arabic{section}} 
\renewcommand{\thesubsection}{A-\arabic{subsection}}
\setcounter{subsection}{0} 
\setcounter{figure}{0} 
\setcounter{equation}{0}

\begin{figure*}[t]
\centering{ 
\includegraphics[width = 0.32\textwidth]{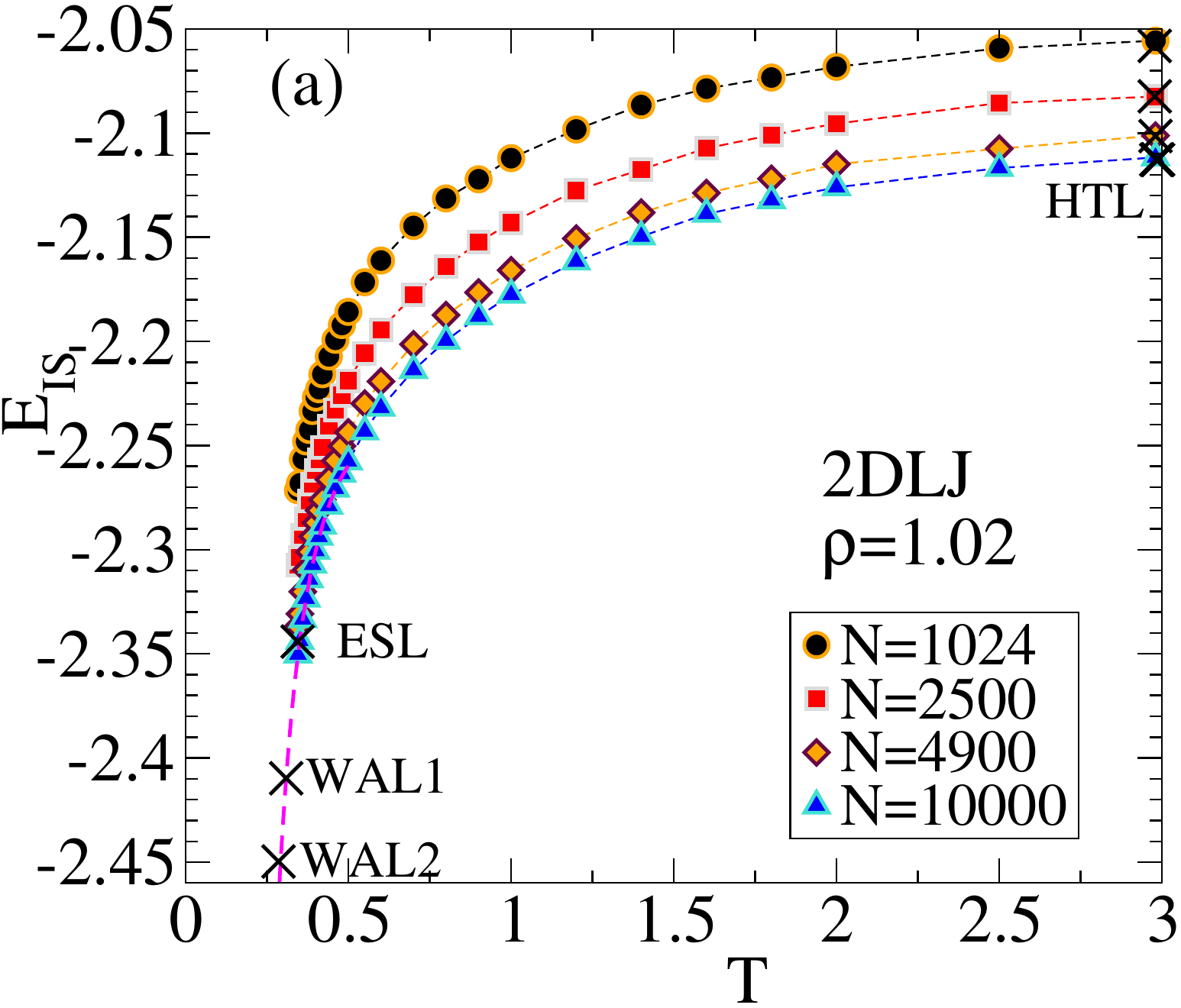}
\includegraphics[width = 0.32\textwidth]{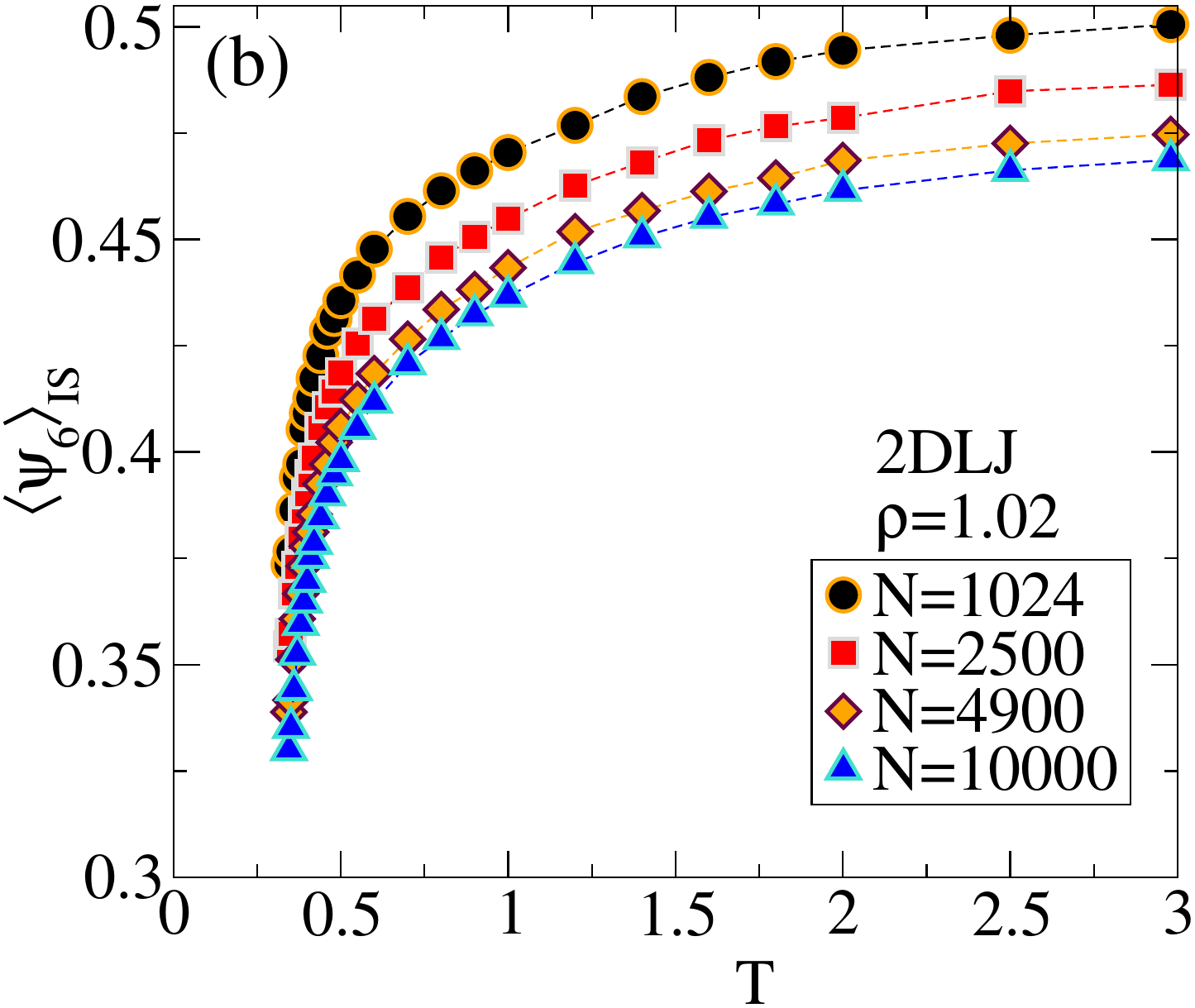}
\includegraphics[width = 0.32\textwidth]{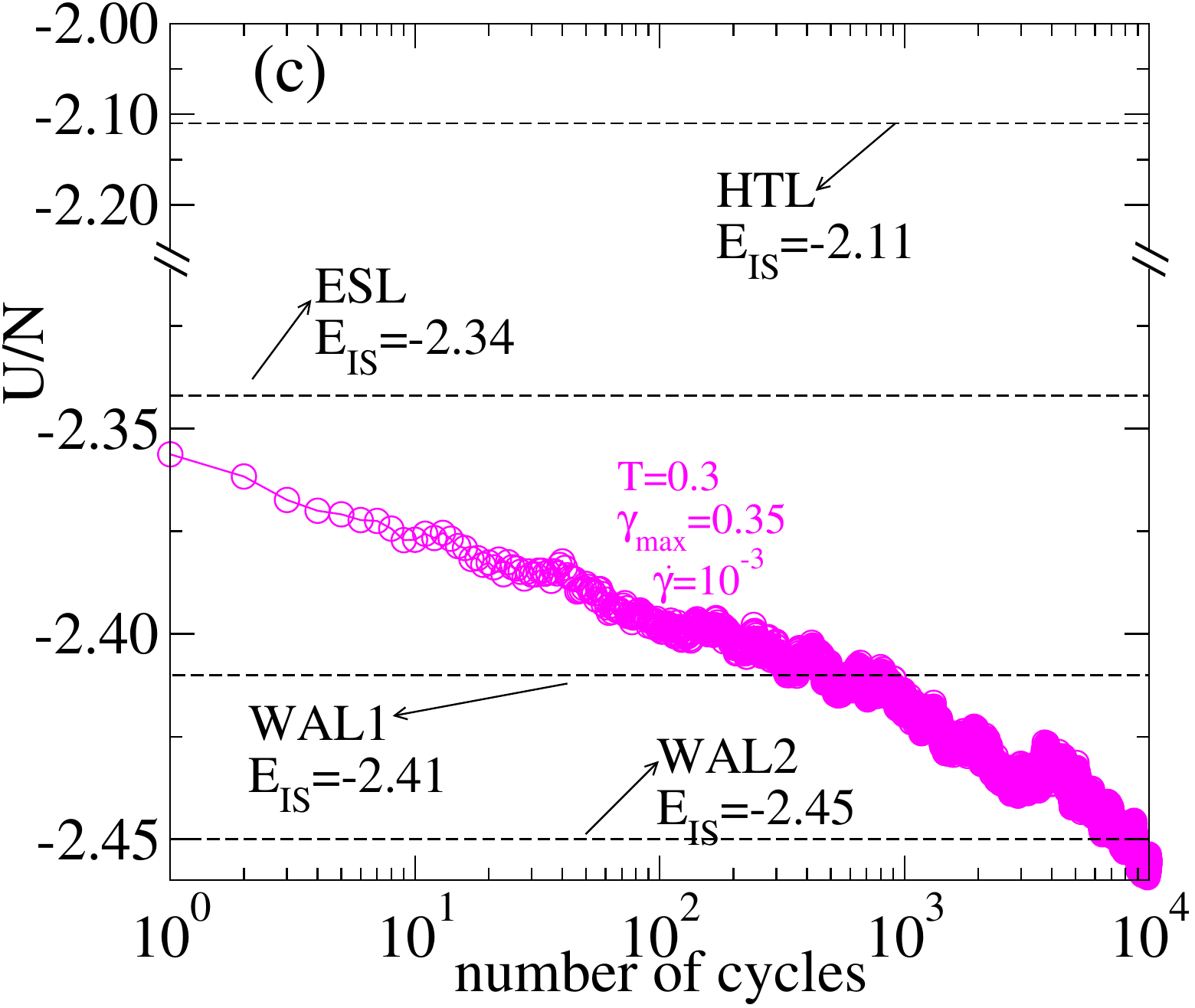}
 }
\caption{\label{sllod} (a) Variation of potential energy of the inherent structures obtained from NVT MD simulation against temperature for different system sizes. Cross marks indicate the energy of the four glasses  considered in this study. The dashed extrapolated line, obtained from fitting the data for $N=10000$ with $E_{IS}=a/T^{b}+c$ in the $T<1$ region, maps the $E_{IS}\ \  vs \ \  T$ in the low temperature regime. (b) Average orientational order parameter $\langle \psi_6\rangle $ of inherent structure configurations against temperature for different system sizes. (c) Potential energy of the inherent structure of cyclically sheared stroboscopic configurations as a function of the number of cycles. Dashed horizontal lines represent the energies of the four glasses  which have been investigated in this work. From the extrapolation of $E_{IS}\ \ vs\ \ T$ data we find the corresponding temperatures of WAL1 and WAL2 to be $T=0.271$ and $0.241$, respectively.
}
\end{figure*}

\subsection{Glass preparation through finite rate cyclic shear}
We explore the yielding transition as a function of the  degree of annealing of glasses subjected to cyclic shear deformation. To generate poorly annealed glasses (HTL and ESL) we use  conventional constant volume and temperature (NVT) molecular dynamics, while for well annealed glasses (WAL1 and WAL2) below the energy $-2.40$, we use a finite temperature, finite shear rate, cyclic shear protocol which we describe in this section. Taking  equilibrated configurations at $T = 0.35$ trajectory, the system evolved using the SLLOD equations of motion under repetitive cyclic shear for $T = 0.3$, shear rate $\dot{\gamma}=10^{-3}$ and amplitude $\gamma_{max} = 0.035$. The inherent structure energy of cyclically sheared stroboscopic configurations are shown in Fig. \ref{sllod}. With  increasing number of cycles, the system explores deeper and deeper energy configurations. We have collected two sets of configurations from these well annealed liquid configurations,  with energy corresponding to $-2.41$ (WAL1) and $-2.45$ (WAL2).

\subsection{Evolution of potential energy under oscillatory deformation}
\begin{figure*}[h]
\centering{ 
\includegraphics[width =0.3\textwidth]{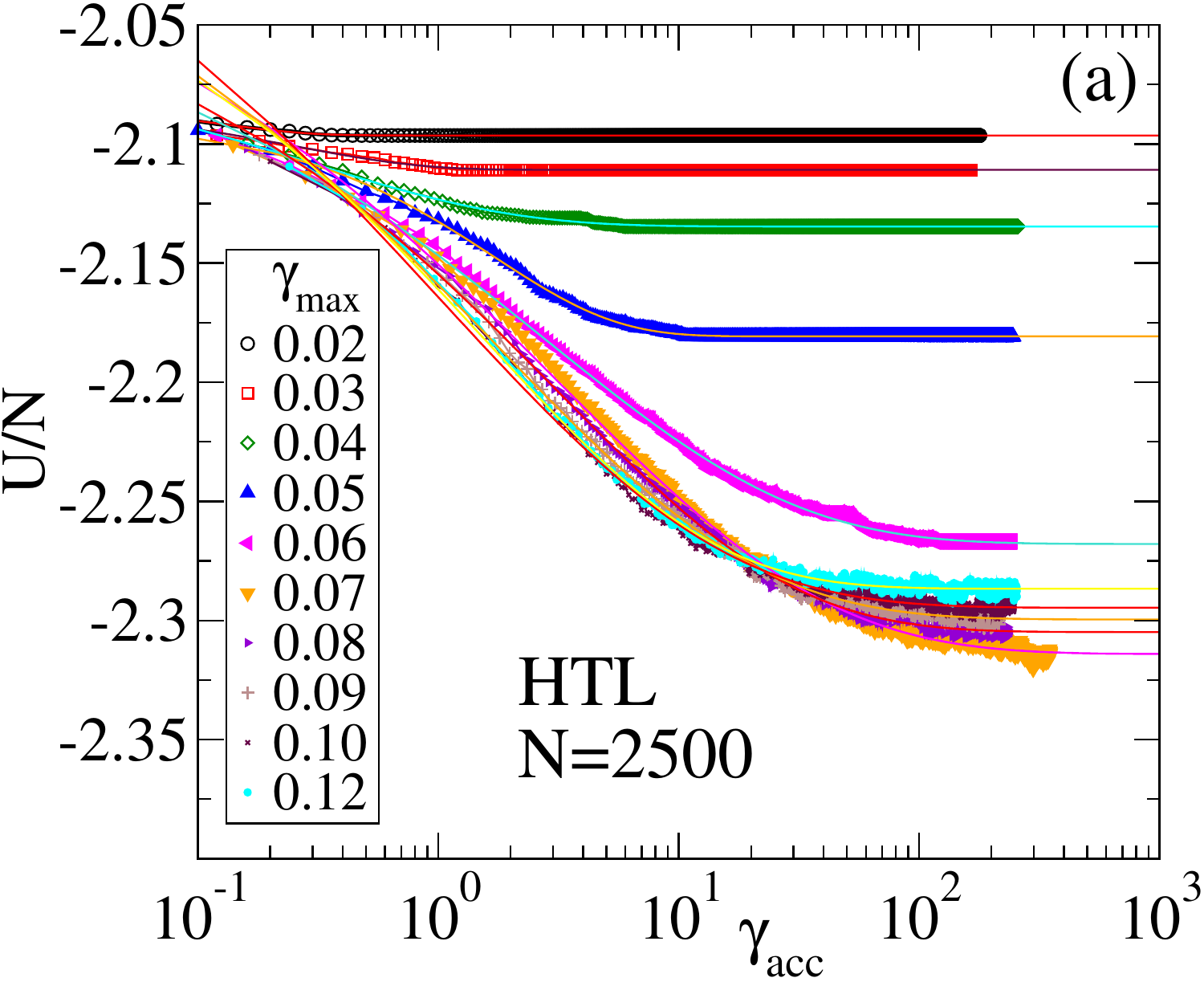}
\includegraphics[width =0.3\textwidth]{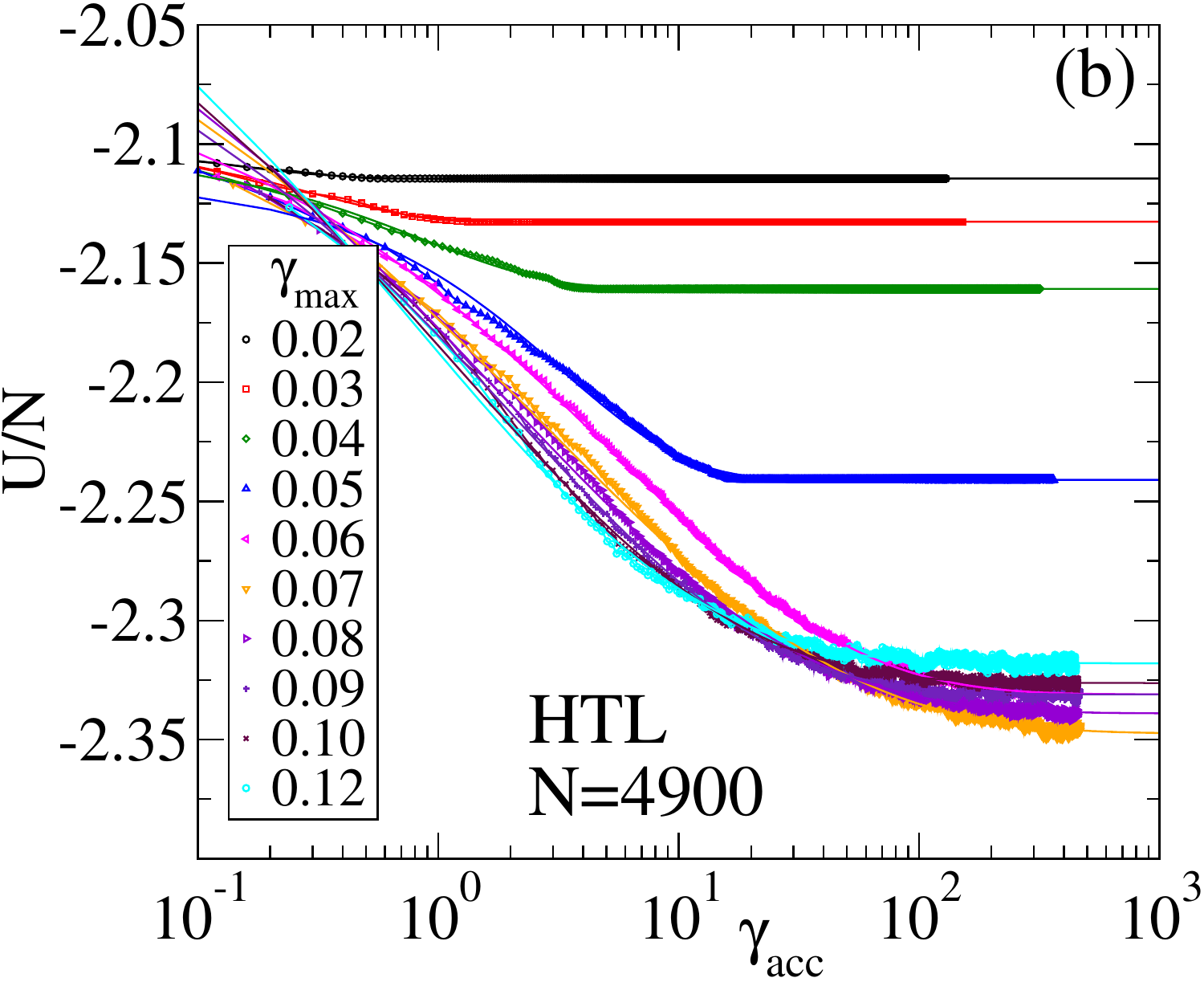}
\includegraphics[width =
    0.3\textwidth]{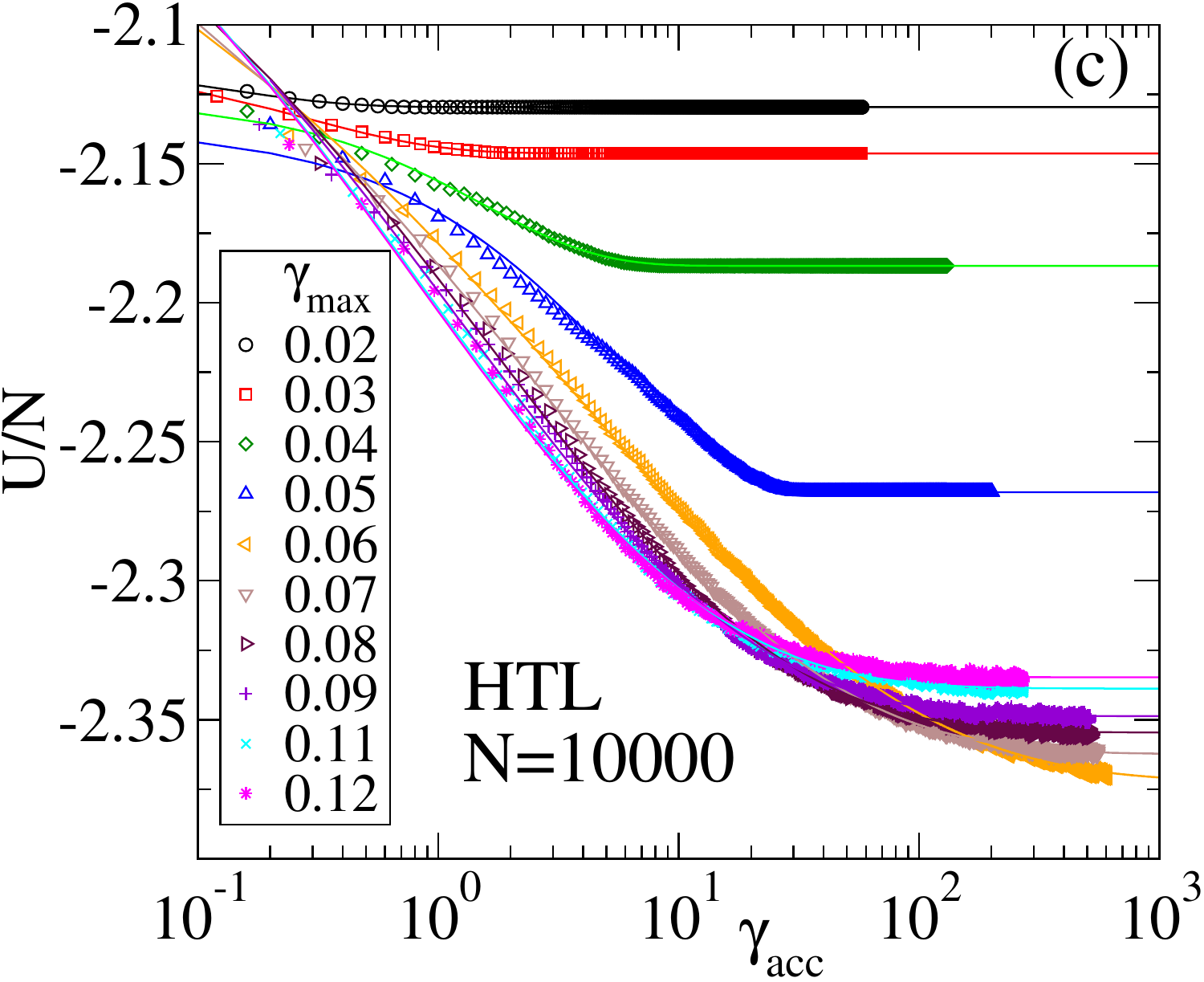}
}
\centering{ 
\includegraphics[width = 0.3\textwidth]{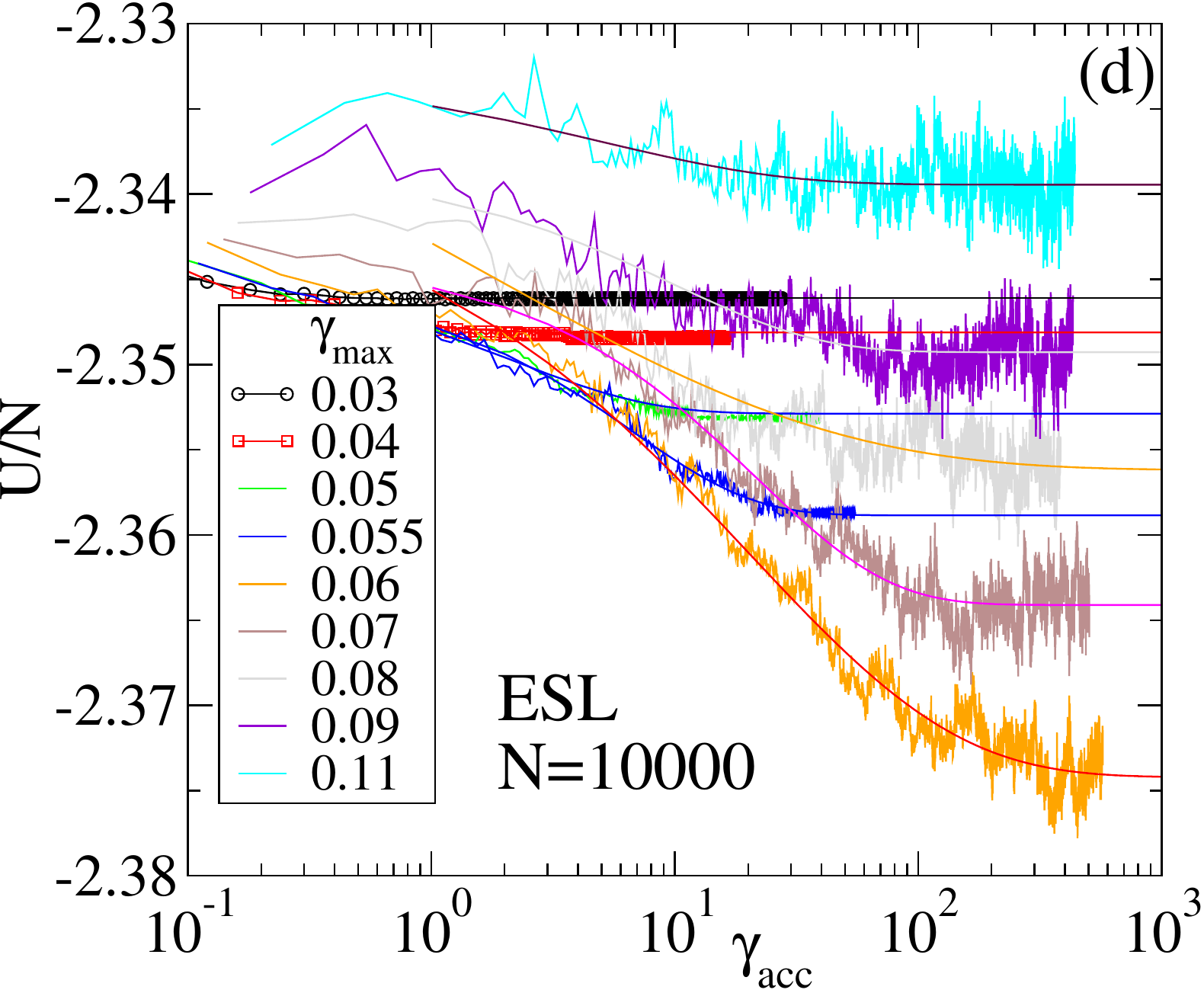}
\includegraphics[width = 0.3\textwidth]{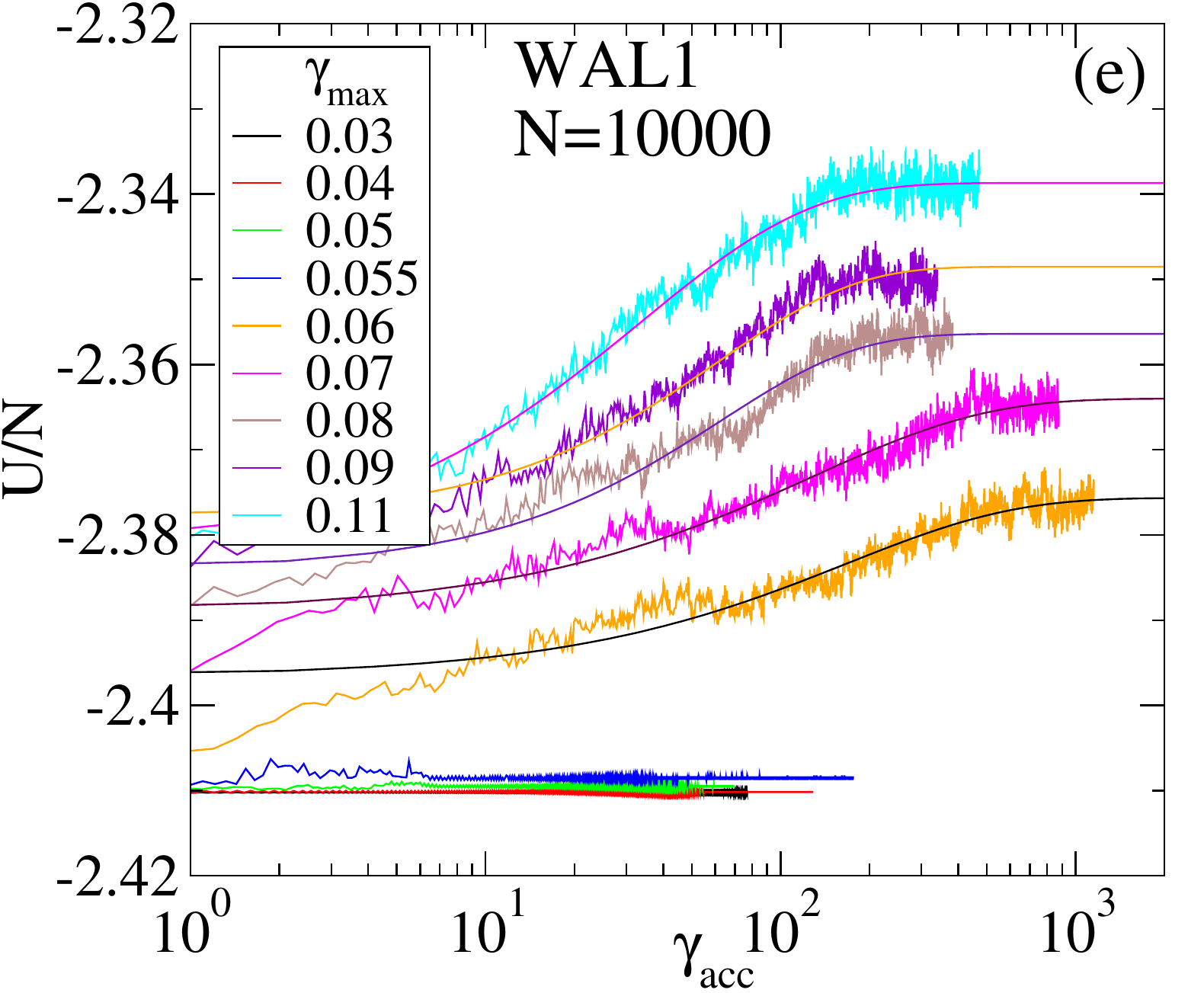}
\includegraphics[width = 0.3\textwidth]{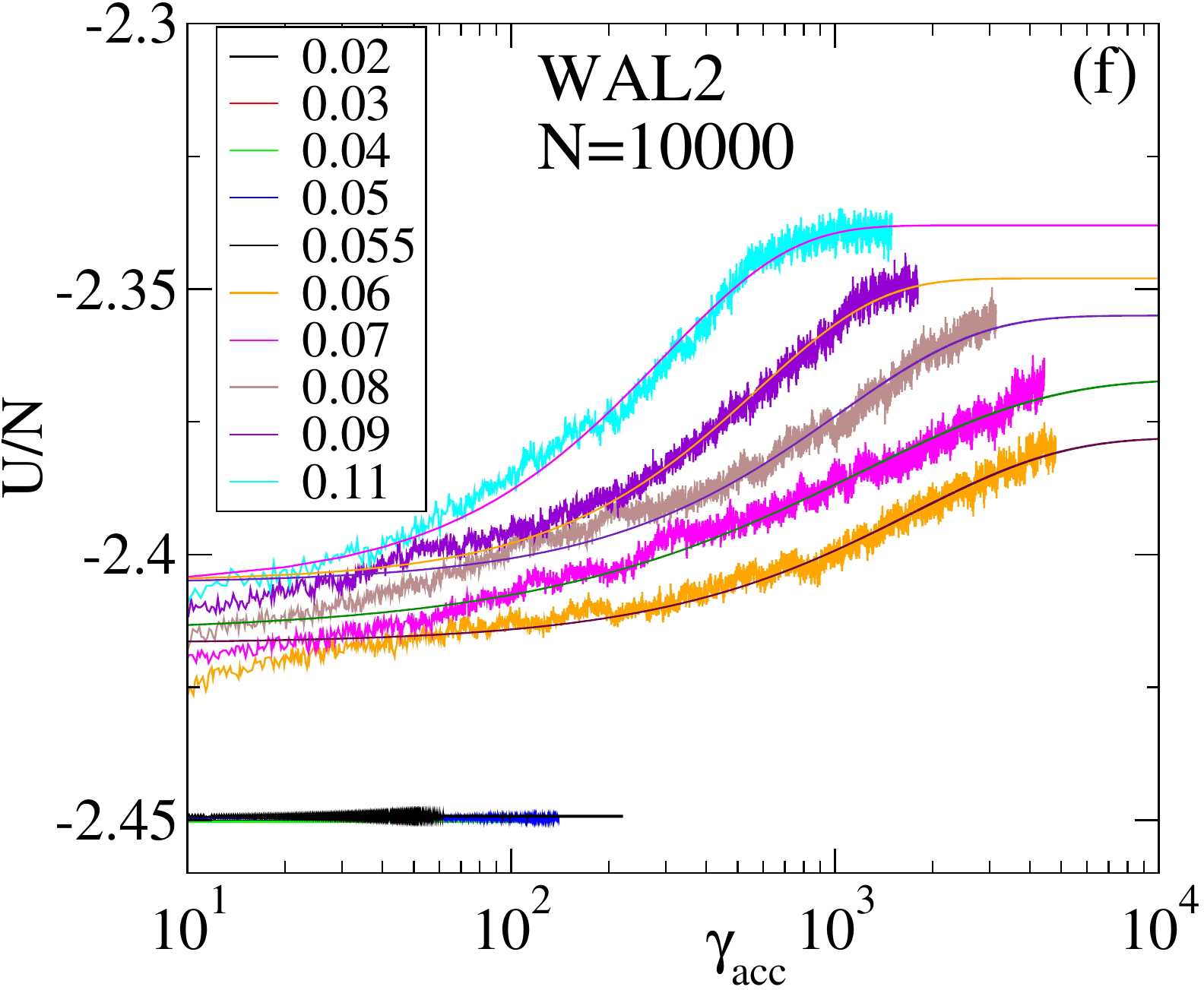}
 }
\caption{\label{pe_gac} Evolution of potential energy with accumulated strain ($\gamma_{acc}=4\times\gamma_{max}\times N_{cycle}$) for different system sizes for (a)-(c) HTL glasses for various strain amplitudes. (d)-(f) Evolution data of potential energy for ESL, WAL1, and WAL2 for a system of size $N=10000$ for different strain amplitudes. Solid lines through the data points are  fits to a stretched exponential form.}
%\end{figure*}
%\begin{figure*}[h]
\vspace{.5cm}
\centering{ \includegraphics[width = 0.85\textwidth]{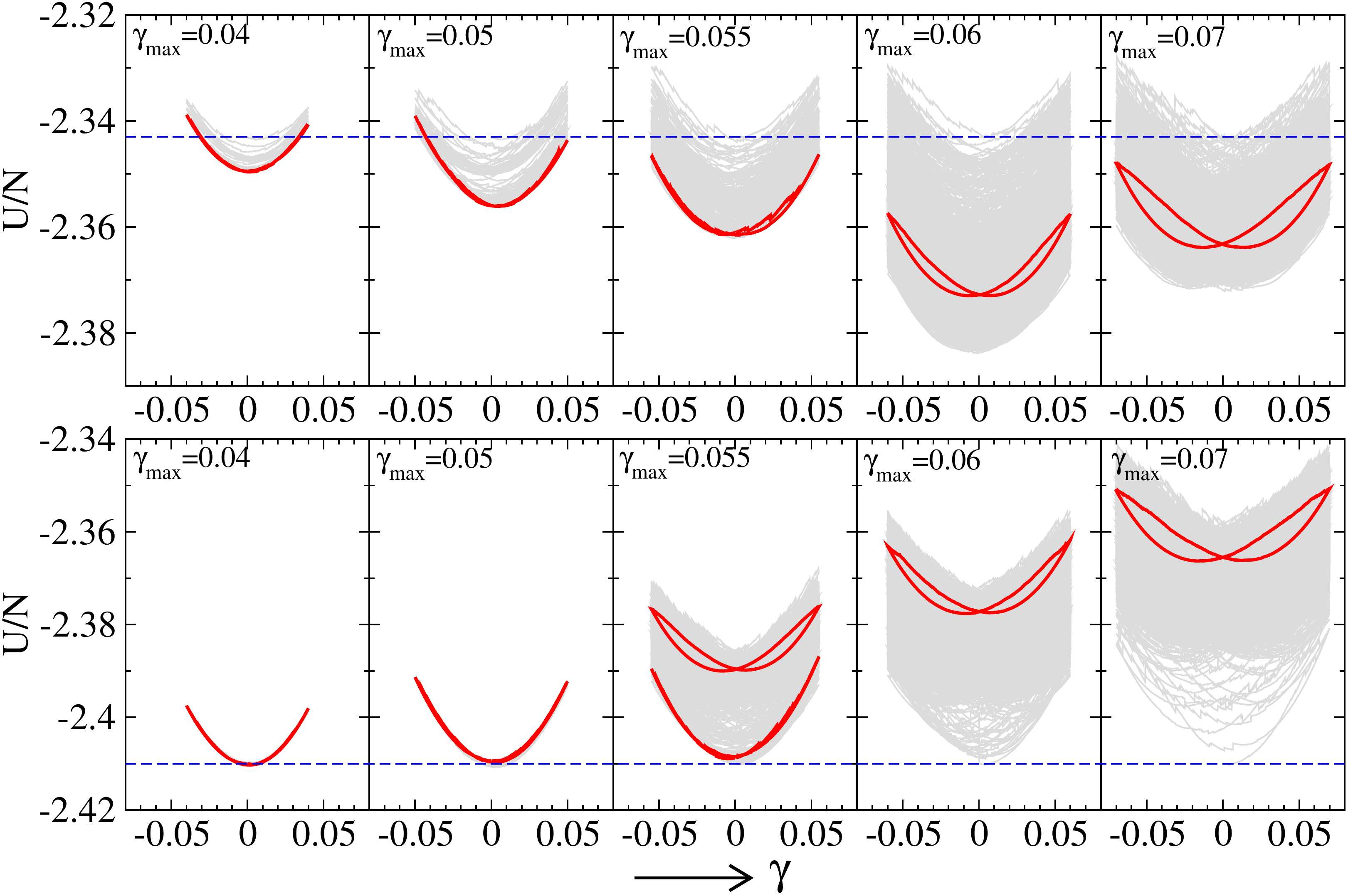}}
\caption{\label{SI_fullfycle_energy}  Potential energy against strain over the full cycle for  ESL (upper panels) and WAL1 (lower panels) and different strain amplitudes, for system size  $N=10000$. The energies from the early cycles are shown in grey while the averaged value of the energy in the steady state is shown in red. Dashed blue lines indicate the initial IS energy. For the case of ESL, the system gradually anneals with increasing of strain amplitude up to the critical amplitude $\gamma_{max}$, whereas for the case of WAL1, the system remains at the same energy  below the yield amplitude. Close to the yield strain amplitude, different samples either remain in the absorbing state or make a transition to the diffusive state. This bi-stable behaviour is shown for $\gamma_{max}=0.055$ where two distinct final states are displayed. }
\end{figure*}

In Fig.~\ref{pe_gac}, we present the evolution of the stroboscopic potential energy  configurations with accumulated strain for different cases (HTL for different system sizes,  along with ESL, WAL1, and WAL2 for $N = 10000$). The results display a trend similar to what have been observed in previous work~\cite{fioccoPRE13, BhaumikPNAS21} for three dimensional glasses. For HTL, we have considered $10-12$ samples for each system size for all the strain amplitudes, whereas for ESL, WAL1, and WAL2 we consider $3$ samples. Note that the evolution of energy in the post yield regime has a very slow convergence for WAL2 glass. For each cases we fit the energies  with a stretched exponential function to extract the asymptotic value of the steady state energy. 

In Fig.~\ref{SI_fullfycle_energy}, we present the evolution of the energy during the shear cycles leading to the steady state. In this case, the energy is presented as a function of strain, over the full cycles for a set of representative amplitudes below and above yielding. In the top panel, the evolution for the ESL is presented. The data illustrate the 
annealing that occurs with repeated cycles, with the energy becoming lower with successive cycles. When yielding occurs and the glass is in a diffusive state, the limit cycles present two minima in each cycle in agreement with previous observations~\cite{leishangthemNAT2017}. For the WAL1 case, presented in the lower panels of Fig.~\ref{SI_fullfycle_energy},  we observe no evolution  below yielding and, as in the previous case, the appearance of two minima above yielding. For WAL1 and $\gamma=0.55$, close to the yielding point, we observe two different final states that can be obtained depending on the initial configuration.

\subsection{Stress profile over the strain cycle}
\begin{figure*}[h]
\centering{
\includegraphics[width = 0.30\textwidth]{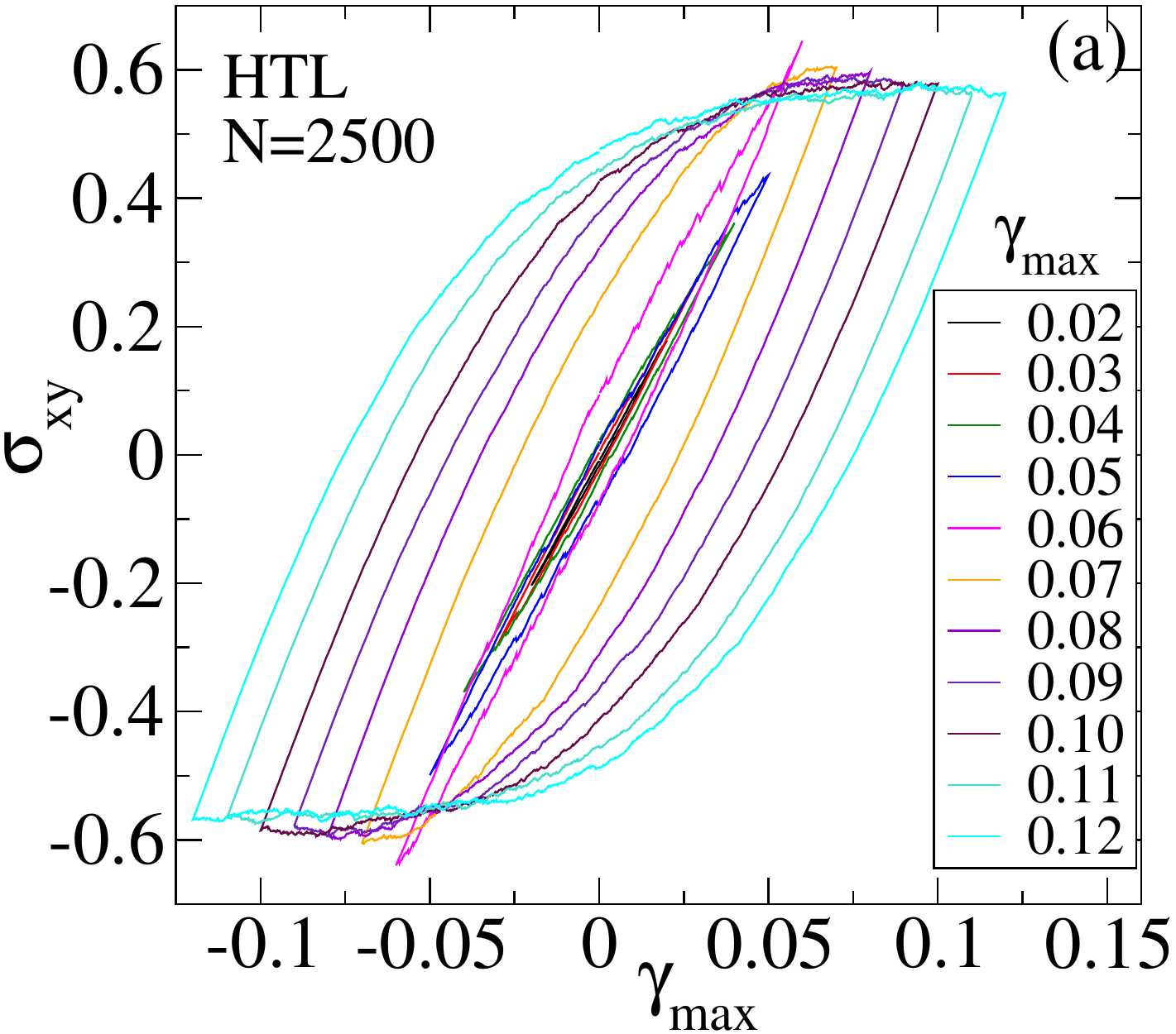}
\includegraphics[width = 0.30\textwidth]{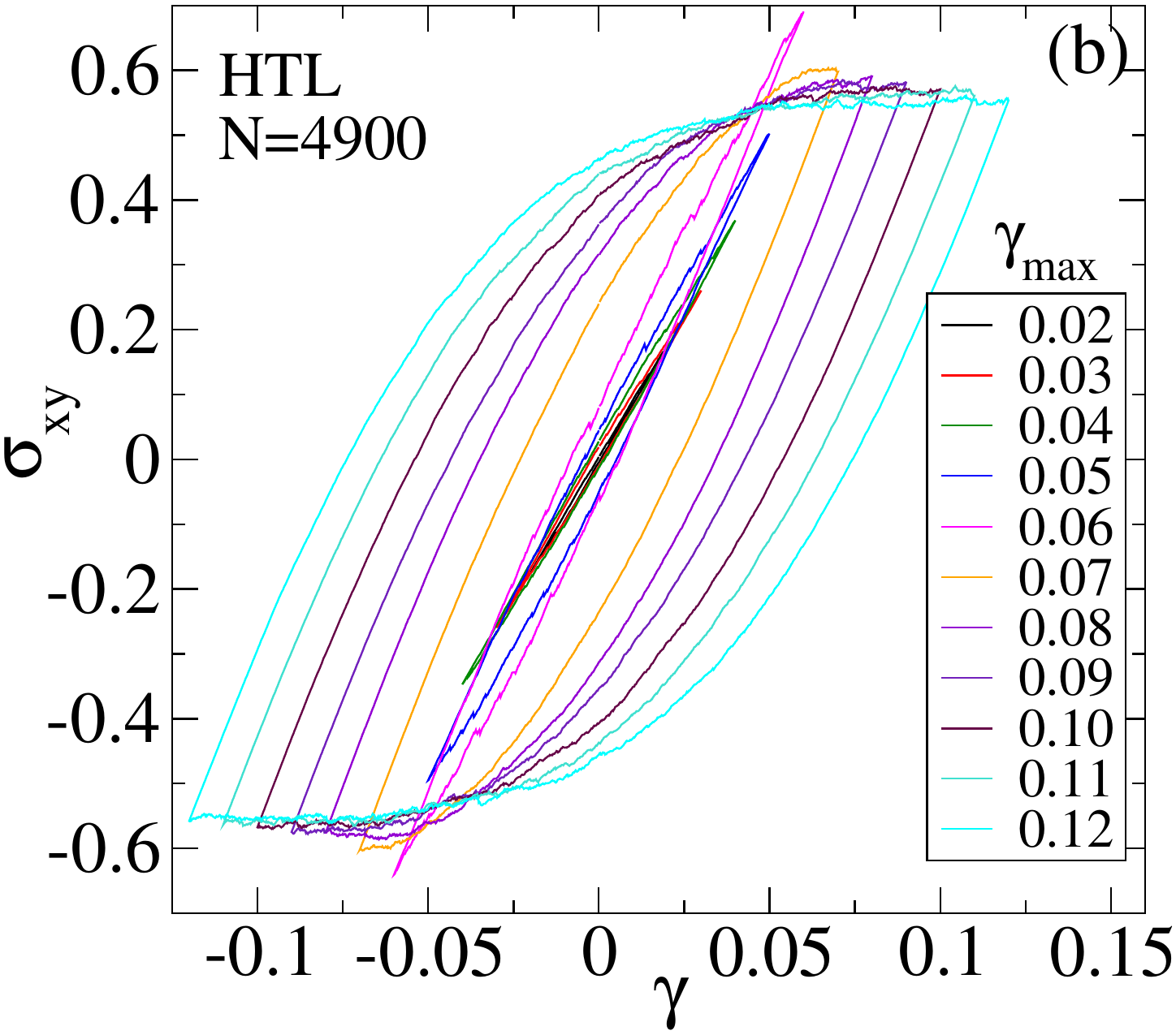}
\includegraphics[width = 0.30\textwidth]{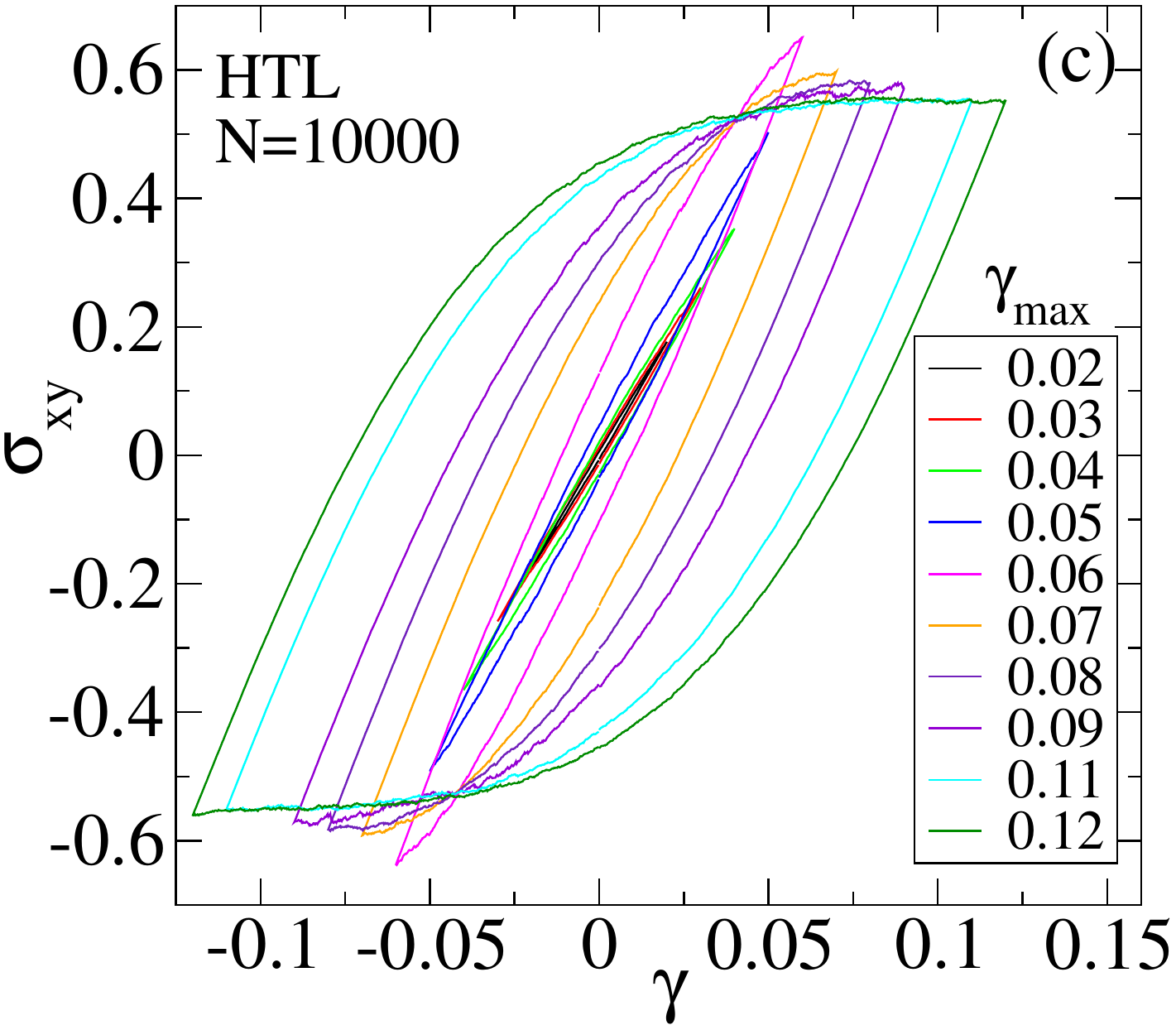}
}
\centering{
\includegraphics[width = 0.30\textwidth]{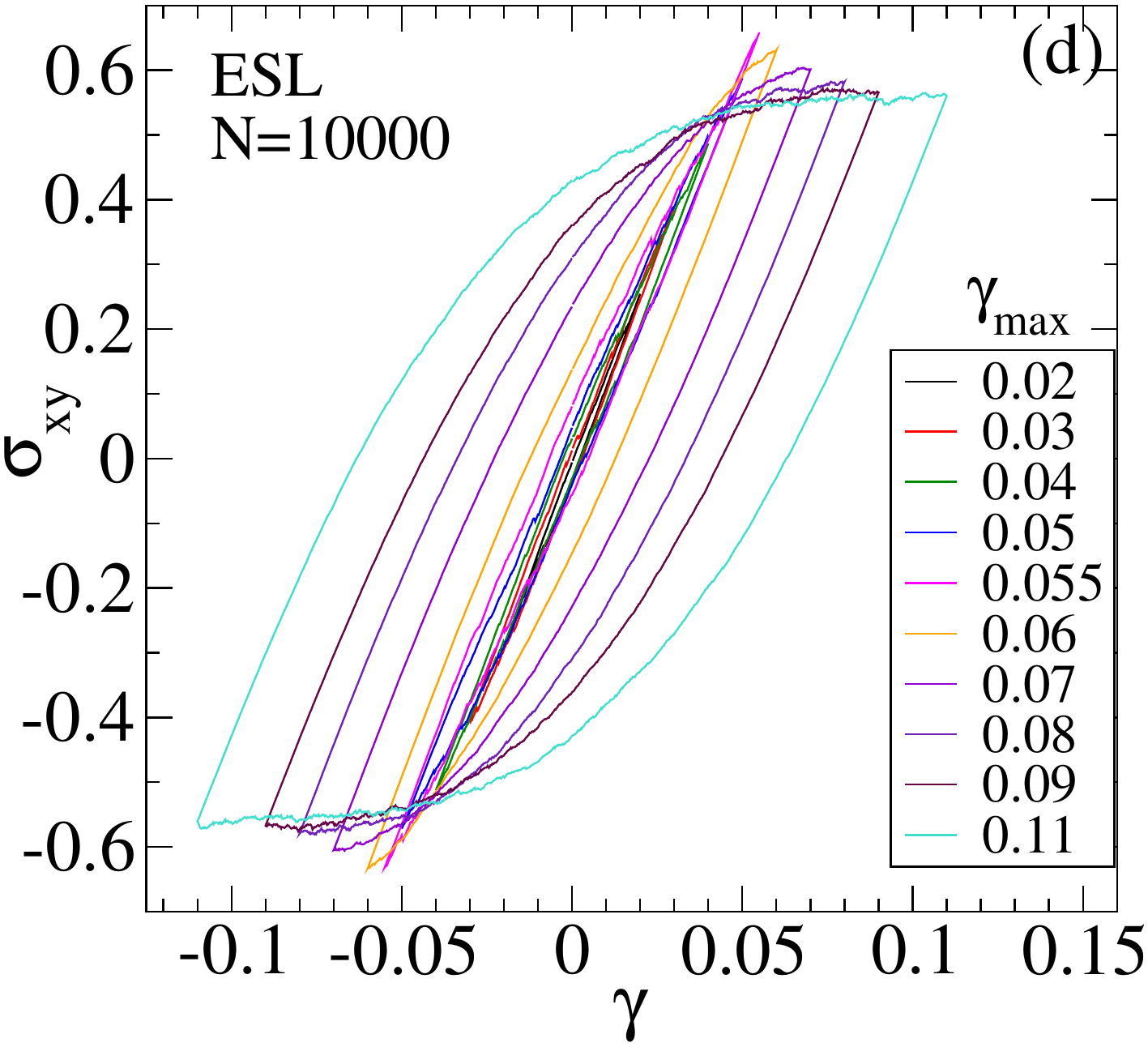}
\includegraphics[width = 0.30\textwidth]{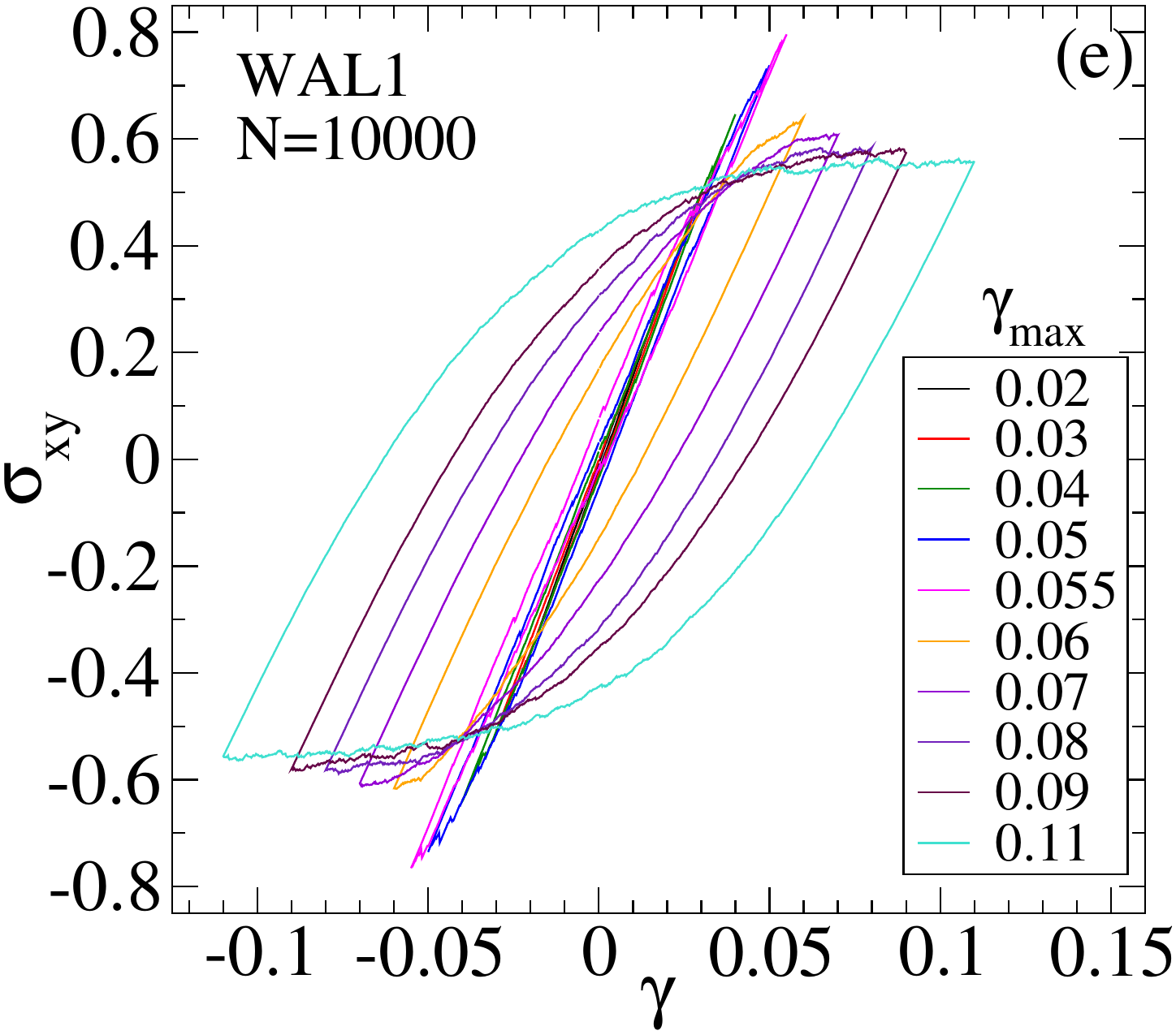}
\includegraphics[width = 0.30\textwidth]{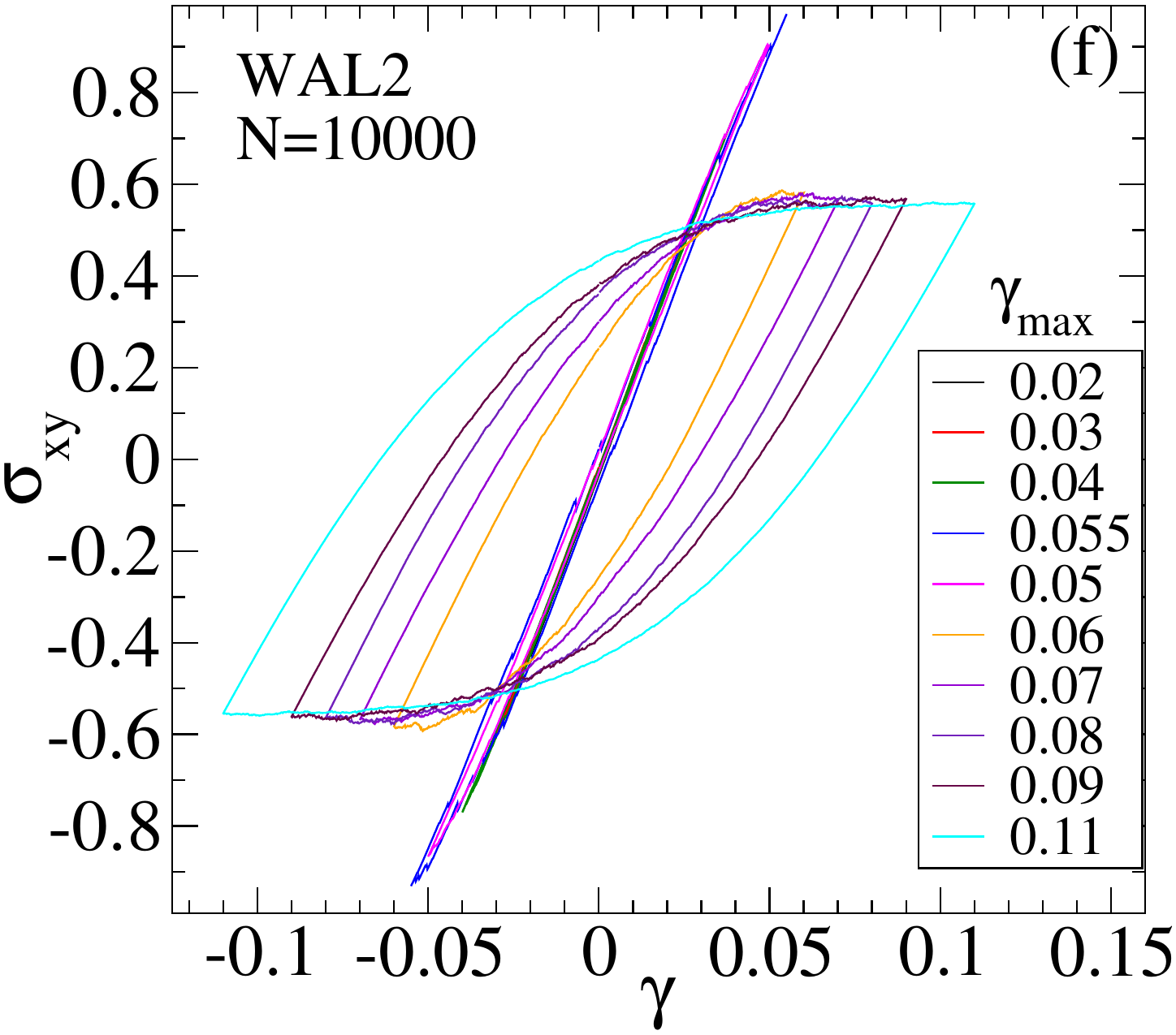} 
}
\caption{\label{SI_fullcycle_stress}  Stress-strain curves averaged over many cycles in the steady state for different $\gamma_{max}$, (a)-(c) for different system sizes ($N=2500,4900,10000$) of the HTL glass, and (d)-(f) for ESL, WAL1 and WAL2 respectively for $N=10000$.}
\end{figure*}

In Fig. \ref{SI_fullcycle_stress} we present averaged stress-strain curves over the complete deformation cycle in the steady state for different strain amplitudes. We average the data over 200-300 cycles in the steady state in each case. For low strain amplitude we see the stress-strain curves do not enclose a finite area, whereas for amplitudes larger than the yield strain they clearly do. For each strain amplitude $\gamma_{max}$, the maximum stress value $\sigma_{xy}^{max}$ are collected and presented in Fig. \ref{fig_uss_sigma-max}(b)-(c). We see that with increasing $\gamma_{max}$, $\sigma_{xy}^{max}$ increases and attains its maximum value at the yield strain amplitude beyond which it drops suddenly. Such behaviour is more prominent for well annealed glasses.

\subsection{Mean Square Displacement in the steady state}
\begin{figure*}[h]
\centering{ 
\includegraphics[width = 0.30\textwidth]{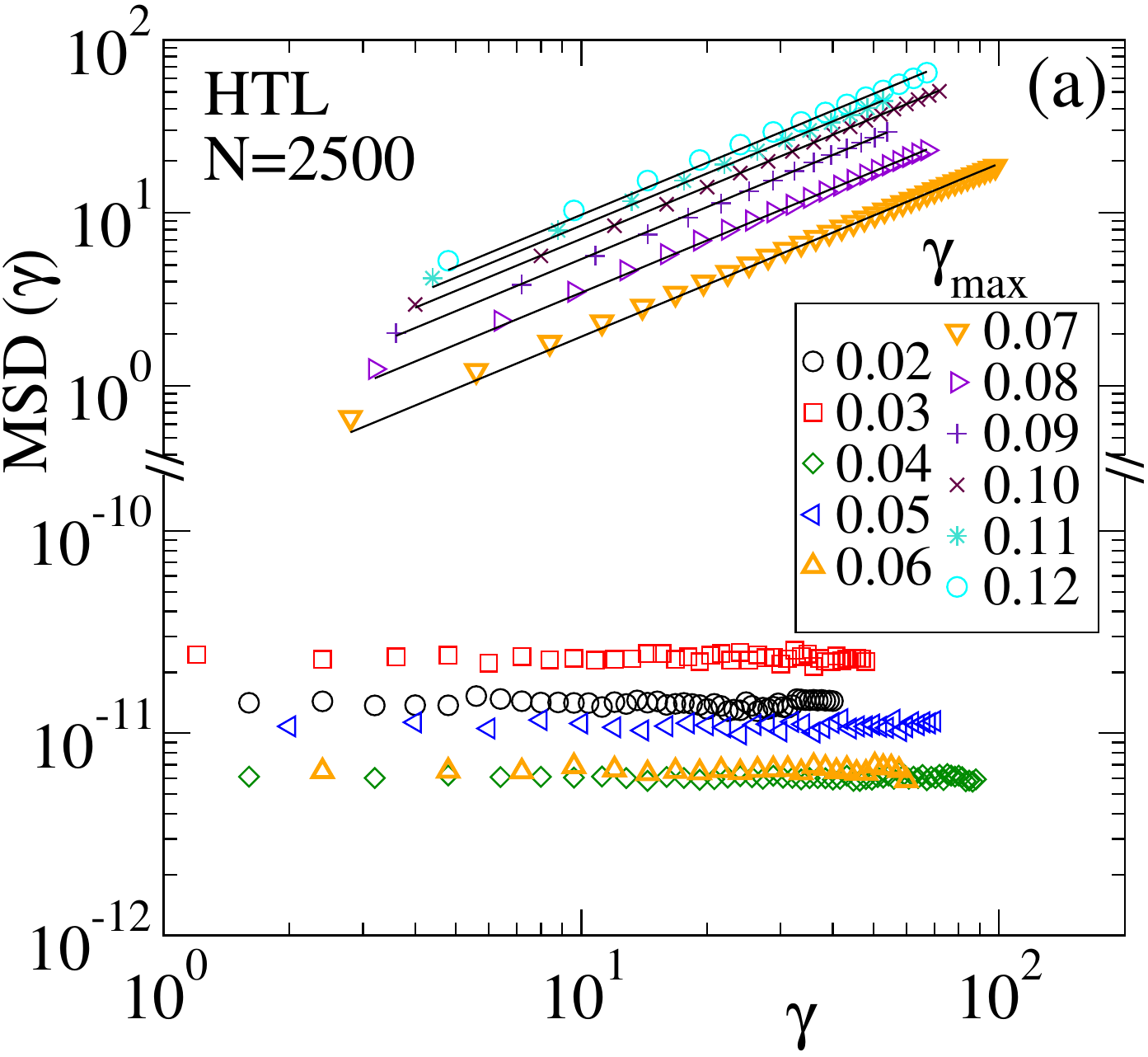}
\includegraphics[width = 0.30\textwidth]{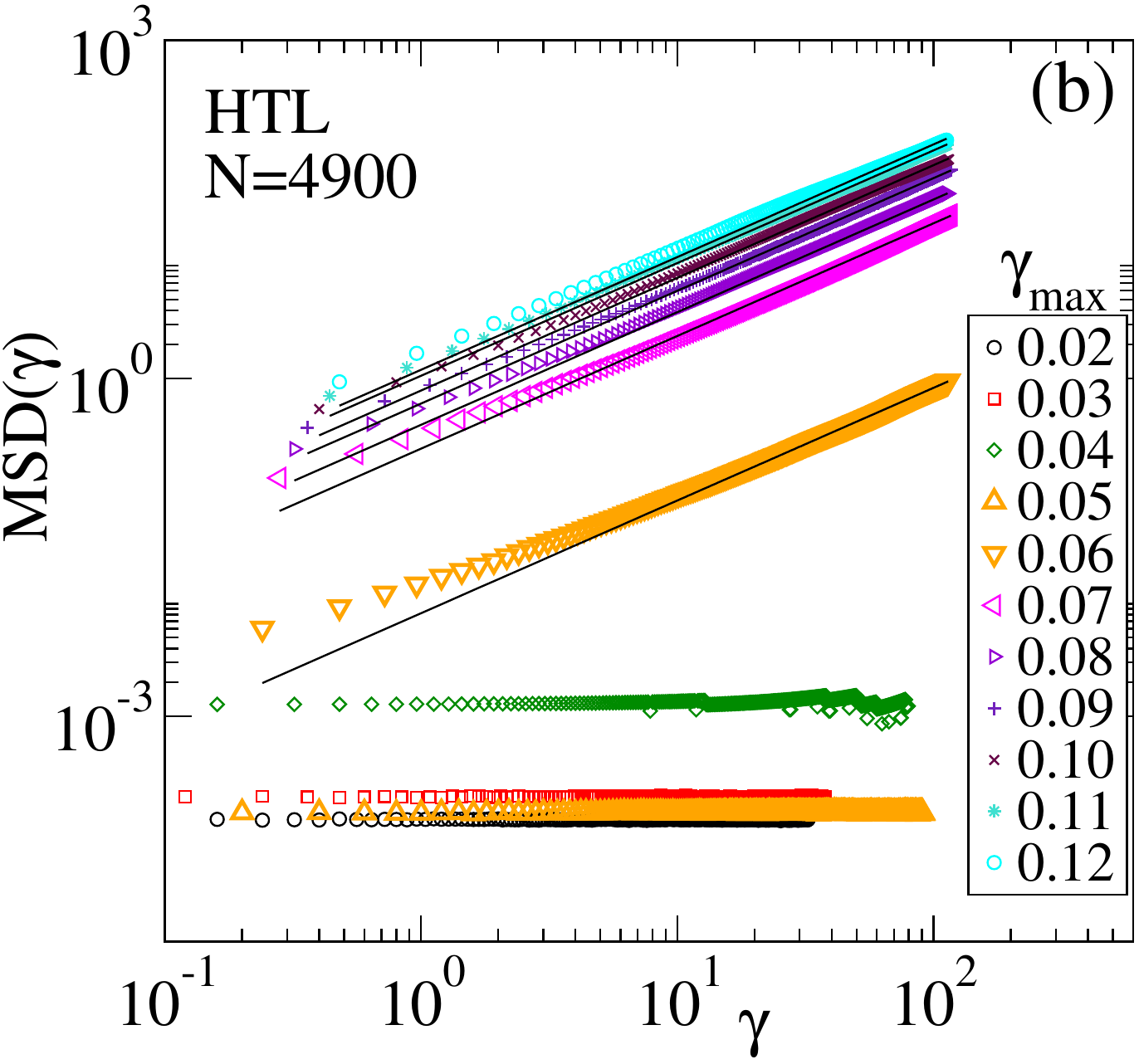}
\includegraphics[width = 0.30\textwidth]{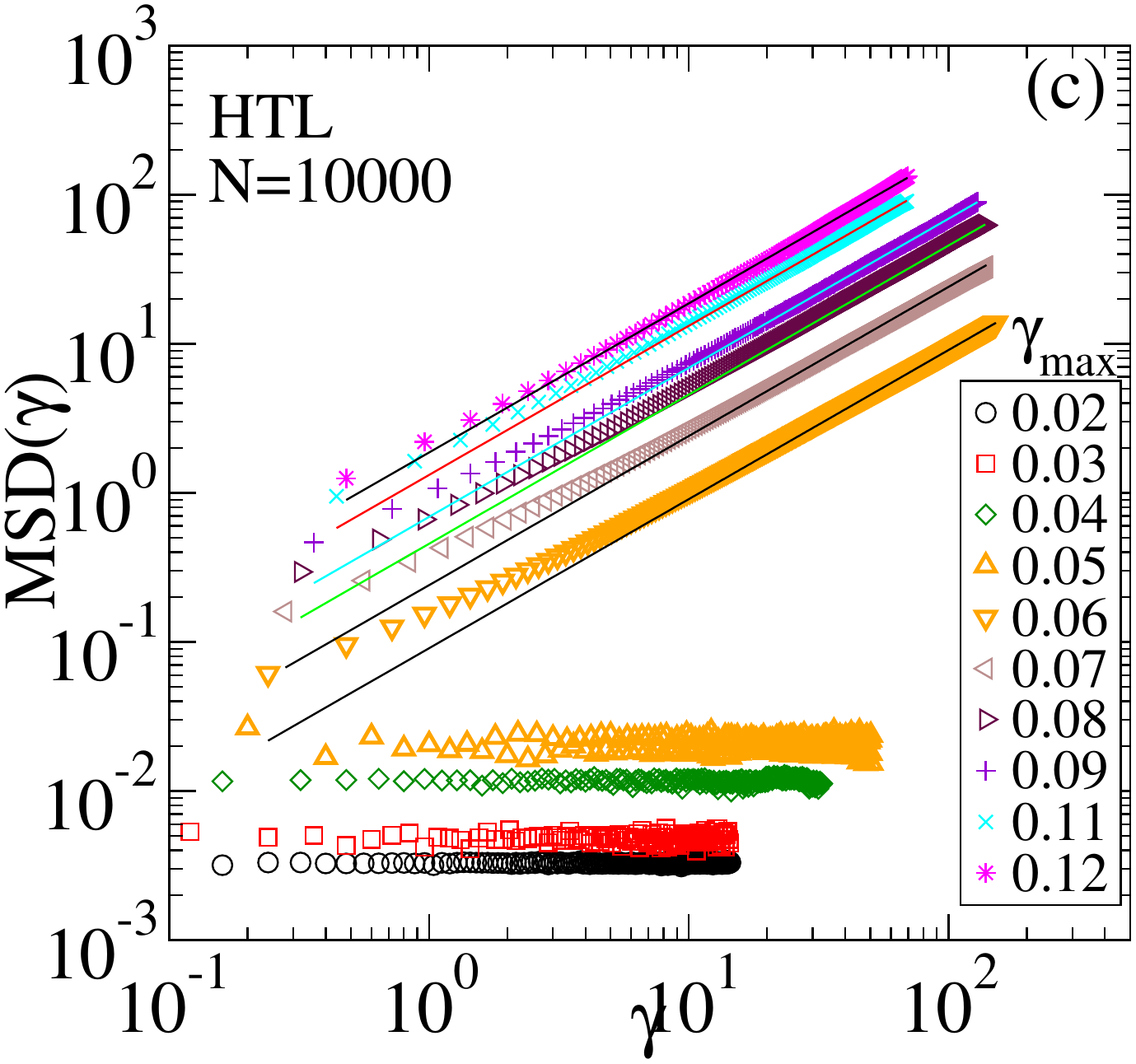}
\includegraphics[width = 0.30\textwidth]{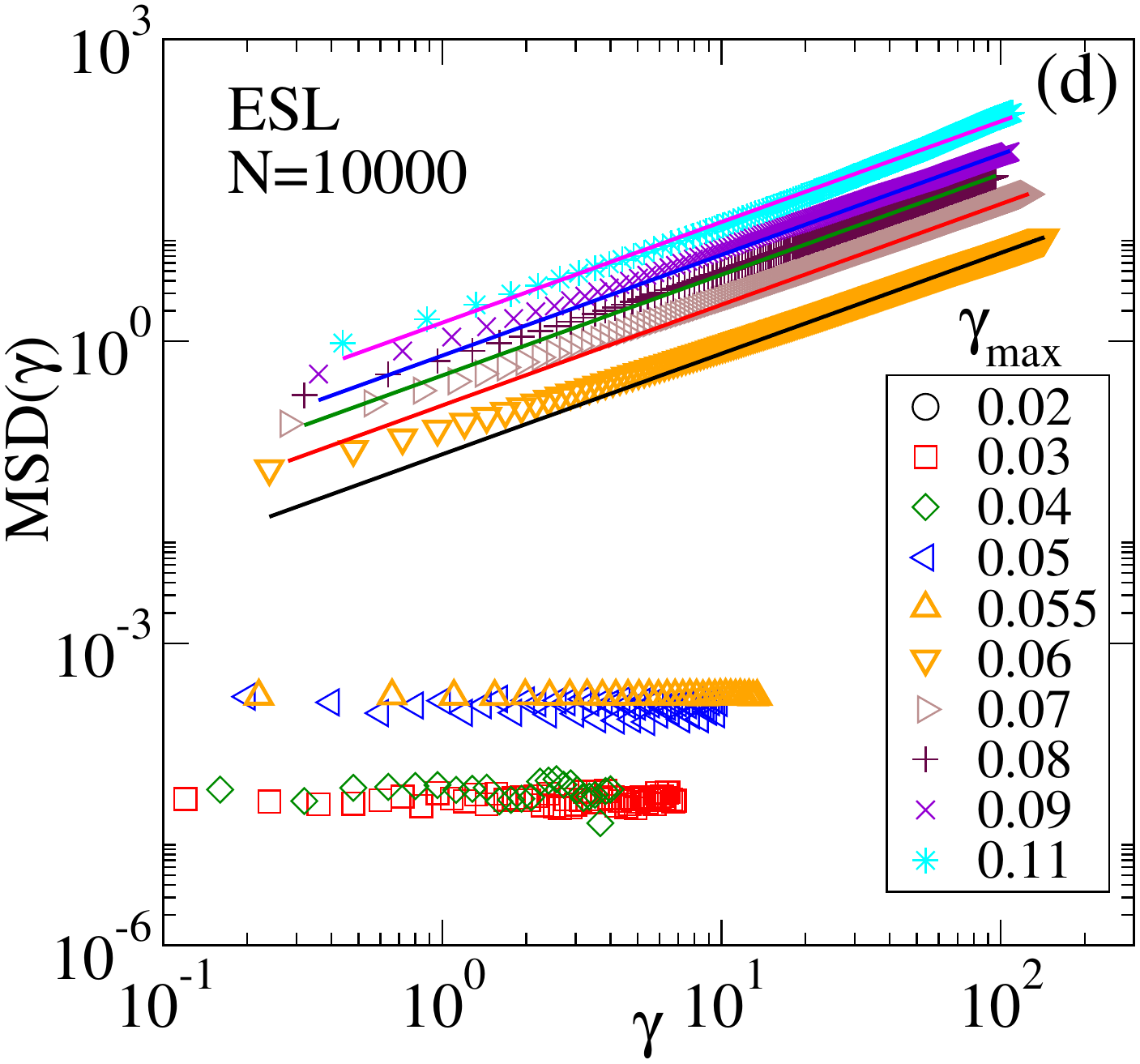}
\includegraphics[width = 0.30\textwidth]{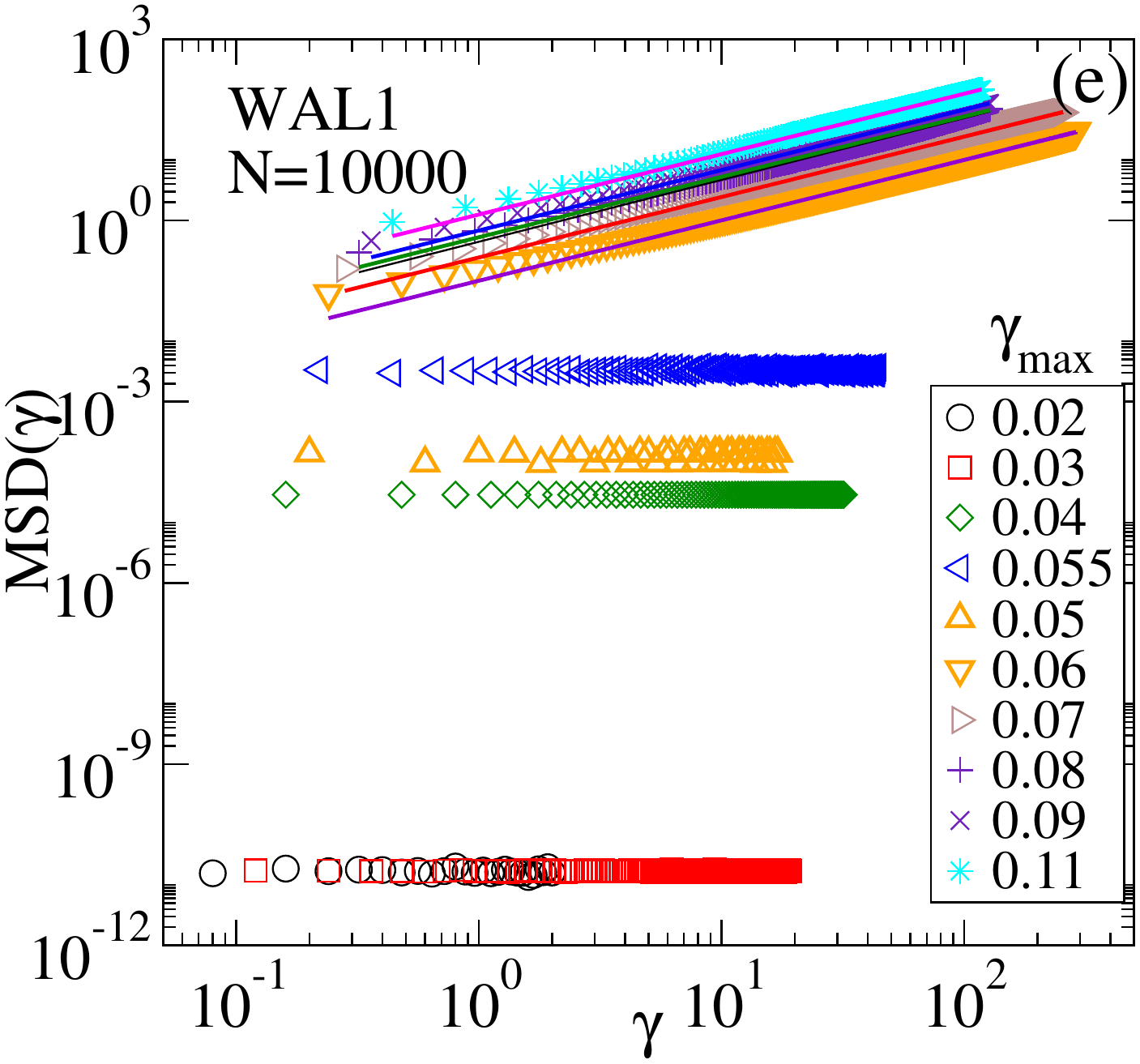}
\includegraphics[width = 0.30\textwidth]{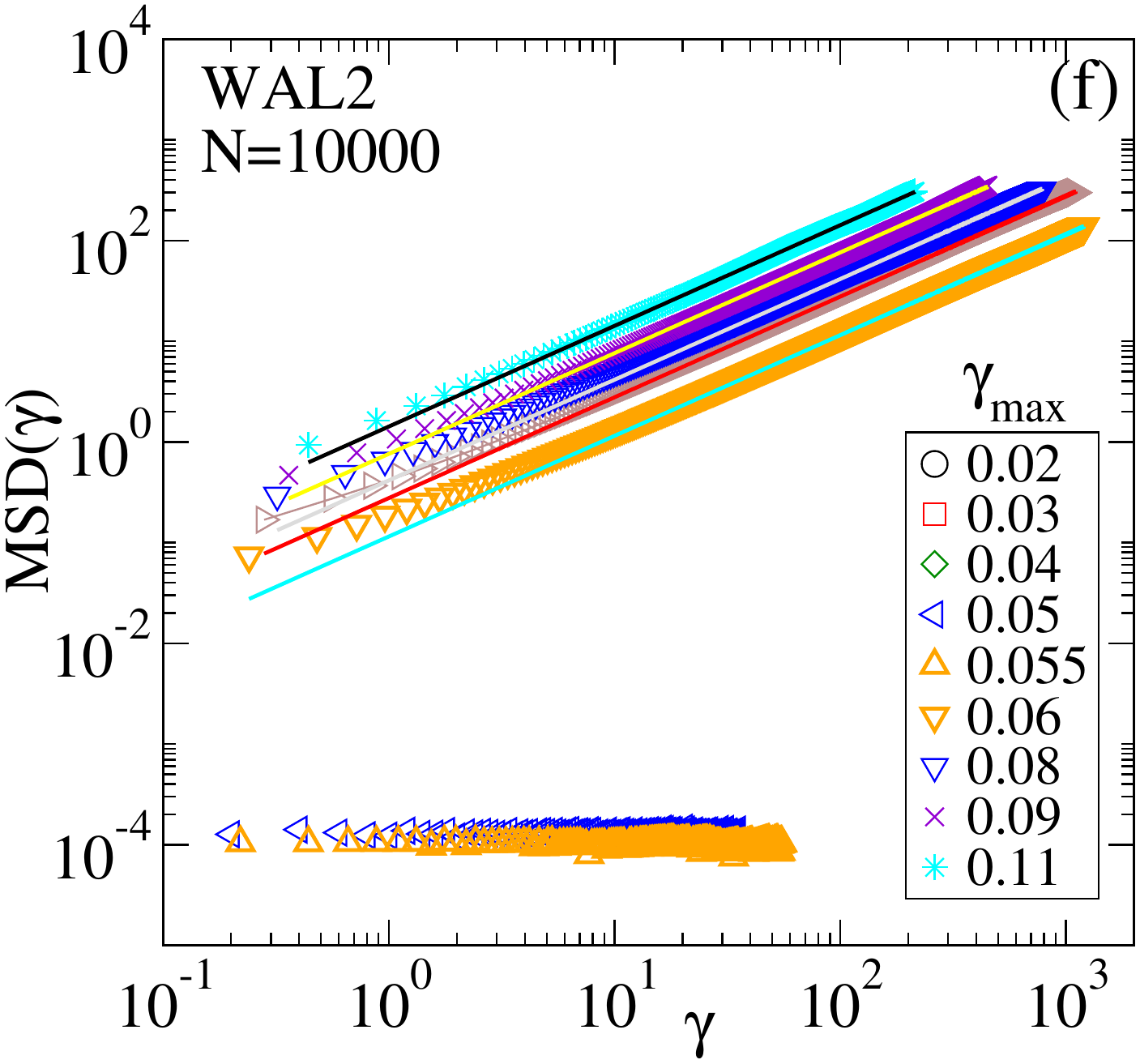}
}
\caption{\label{SI_msdavg} Mean squared displacements computed in the steady state against accumulated strain difference $\gamma$ for different strain amplitudes $\gamma_{max}$. Different panels are for different types of glasses and system sizes we study. To highlight the different behaviour across yield strain, we use the same colour (orange) for the highest $\gamma_{max}$ before yield and the lowest $\gamma_{max}$ above.
}
\end{figure*}

To characterize the motion of the particles, we calculate the mean square displacements (MSD) of the particles for stroboscopic configurations, as a function of the number of elapsed cycles.  We calculate this quantity in the steady state, using different initial configurations, averaging over them, for each accumulated strain difference $\gamma$:
\begin{equation}
\label{eq_avgmsd}
{\rm MSD}(\gamma)=\left \langle\frac{1}{N}\sum_i^{N} |\boldsymbol{\rm
  r}_i(\gamma_{acc}+\gamma)- \boldsymbol{\rm r}_i(\gamma_{acc})|^2 \right \rangle
\end{equation}
where the averaging is performed over several zero strain configurations as the initial configurations and over different samples. In Fig.~\ref{SI_msdavg}, we present the sample averaged data of MSD for different glasses we study. In Figs. \ref{SI_msdavg} (a)-(c) we show the MSDs for different systems sizes of the HTL system, for several strain amplitudes. In  Figs. \ref{SI_msdavg} (d)-(f) we present the data ESL, WAL1, and WAL2 glasses for system size $N = 10000$. Two  regimes are clearly observed. Above the yield strain amplitude, the MSD grows linearly with $\gamma$ whereas it remains substantially flat below it. This sharp change  corresponds to a jump  in diffusivity presented in Figs. \ref{fig_uss_sigma-max}(c) and \ref{fig_uss_sigma-max}(f).

\subsection{Single particle displacement field and cluster size distribution}
\begin{figure*}[t]
%\begin{center}
\centering{
 \includegraphics[width =
    0.3\textwidth]{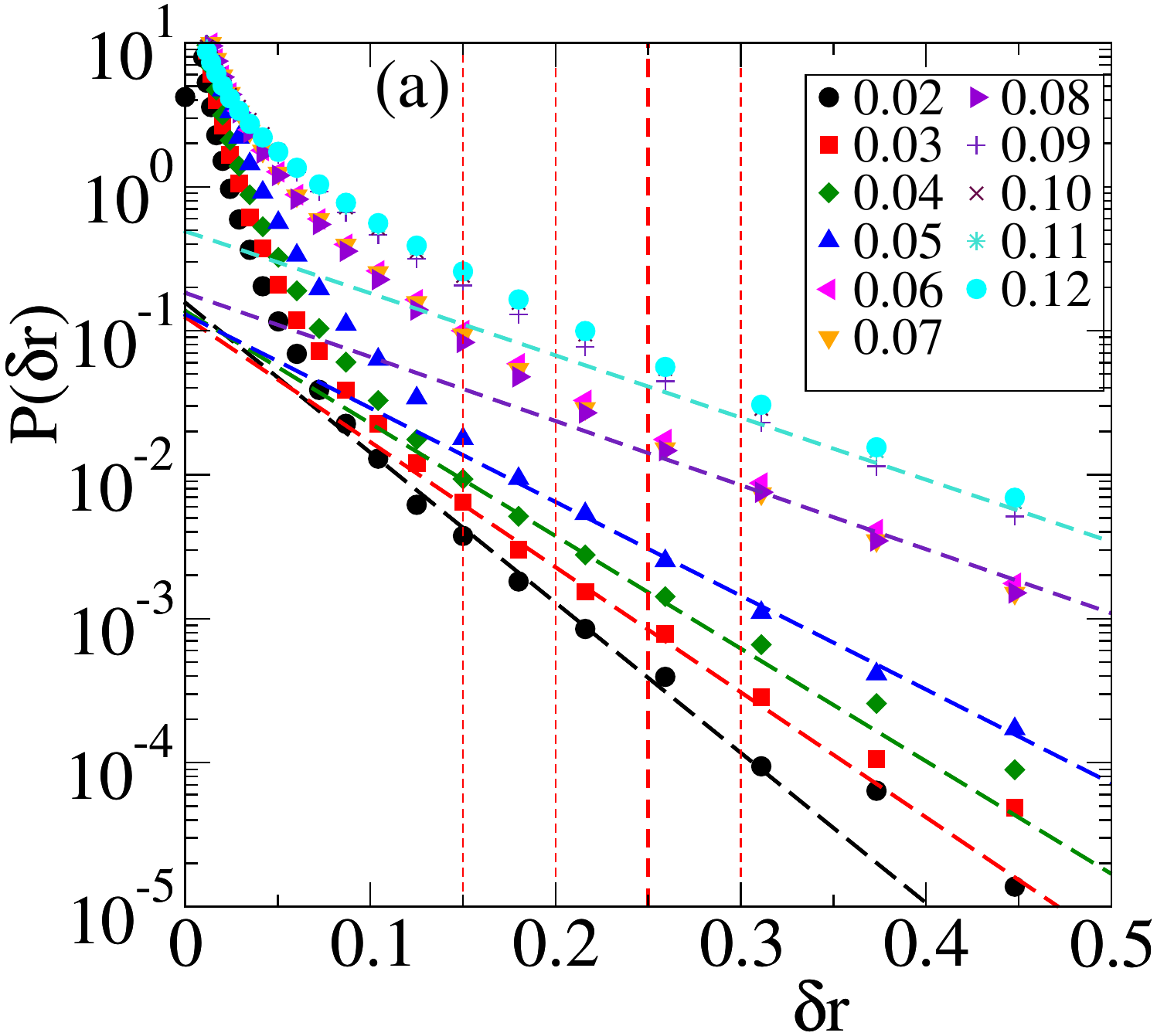}
  \includegraphics[width =
    0.3\textwidth]{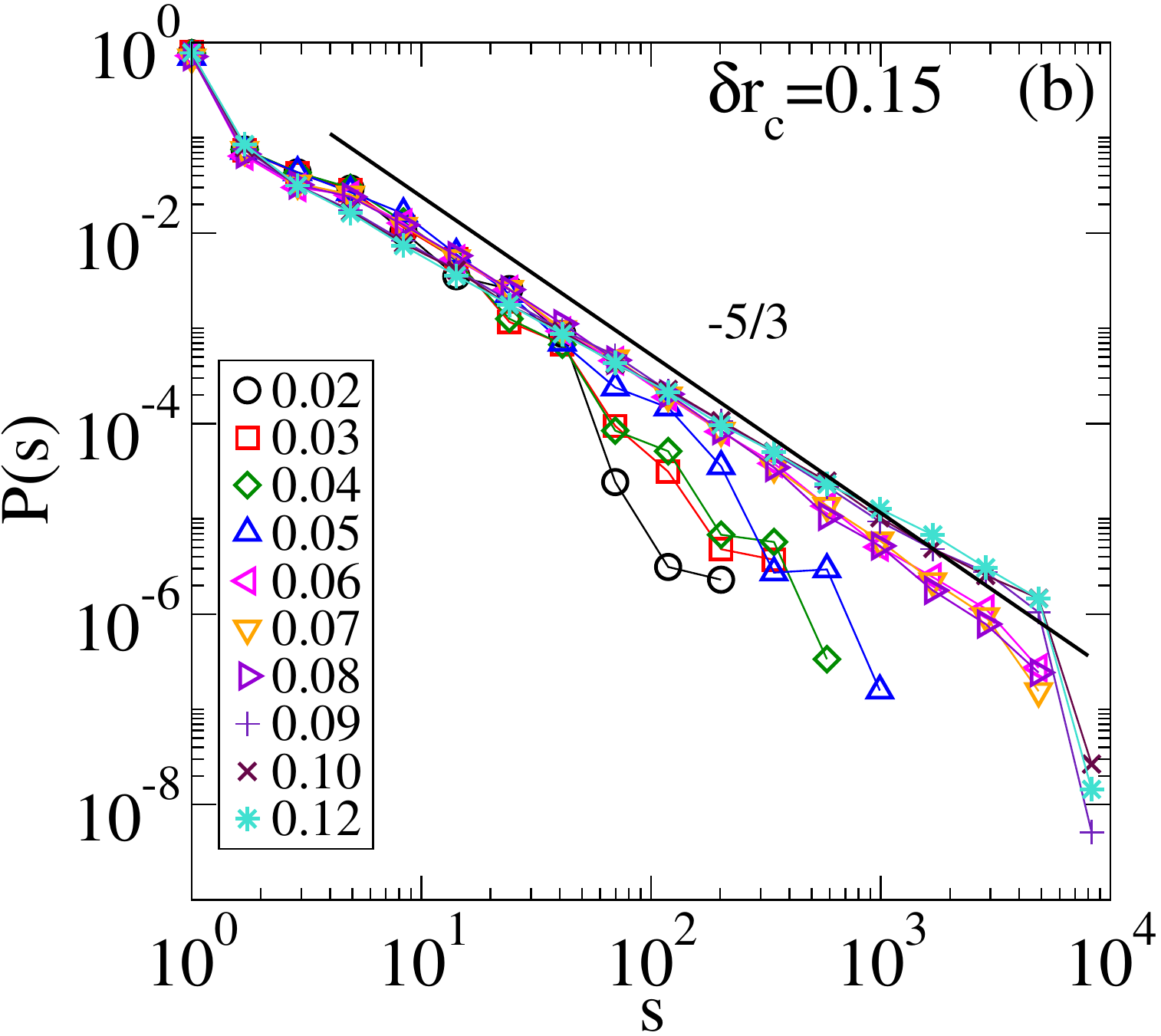}
    }
 %   \end{center}
     \vspace{-.3cm}
  \begin{center}
  \includegraphics[width =
    0.3\textwidth]{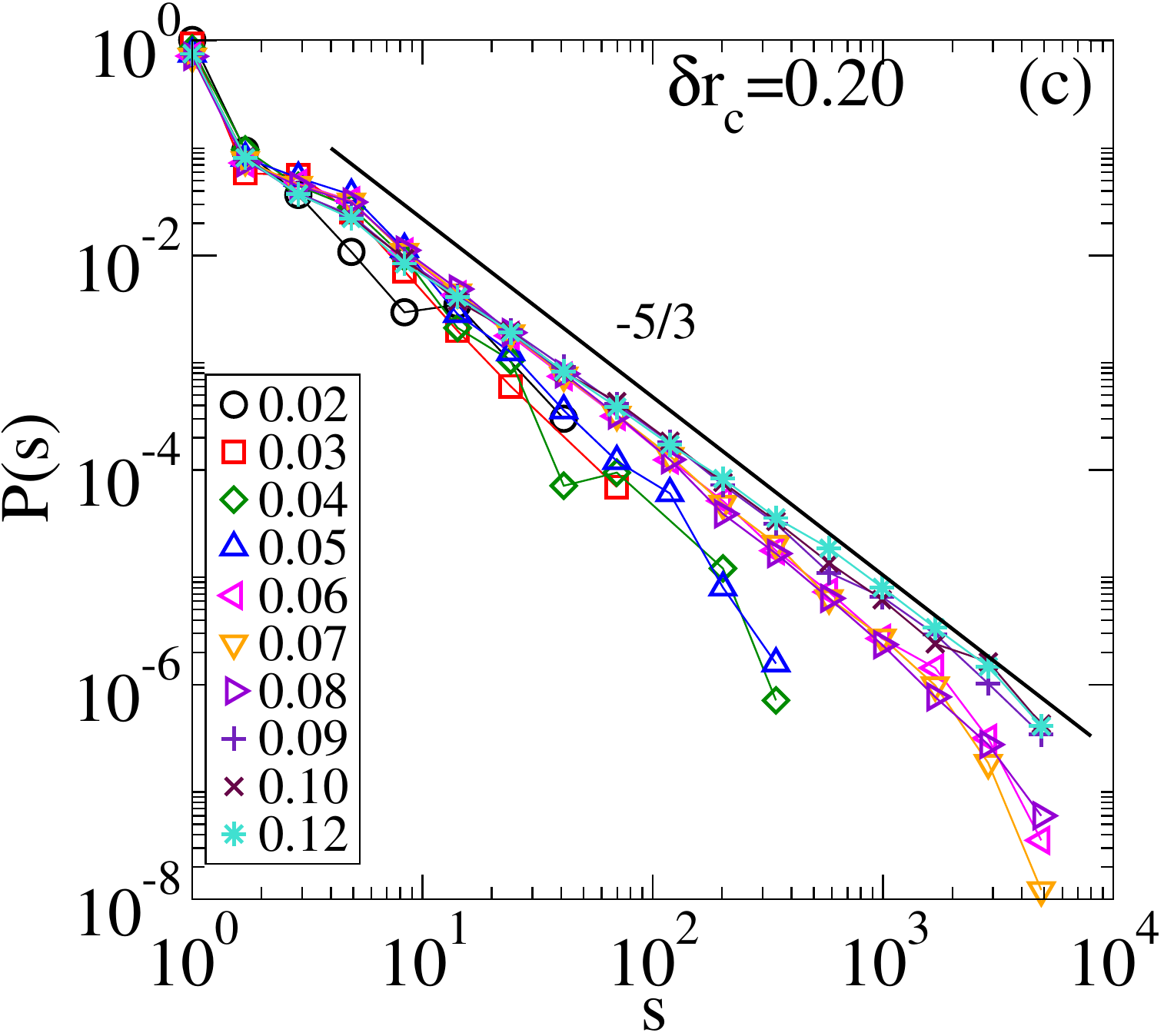}
     \includegraphics[width =
    0.3\textwidth]{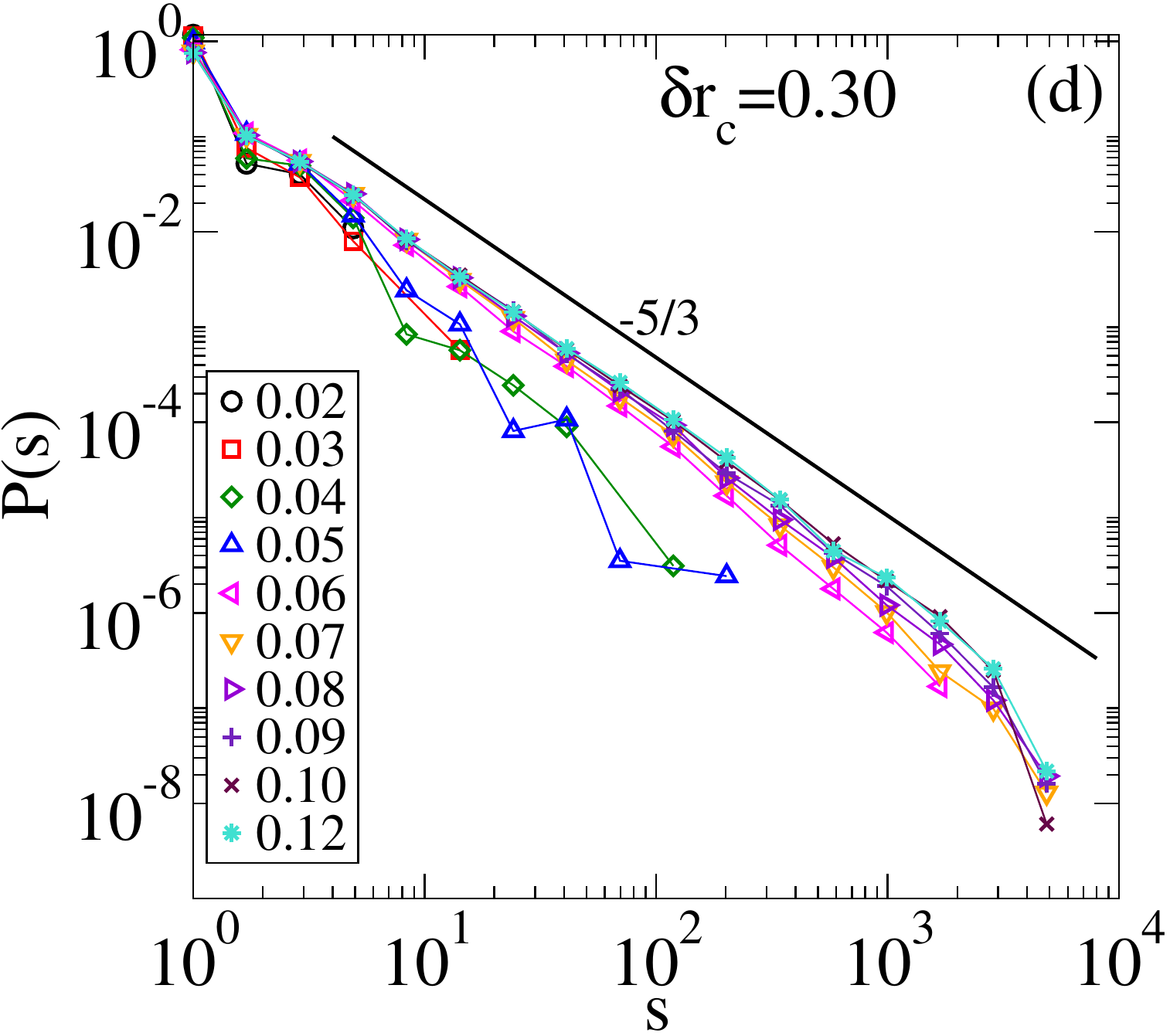}
    \end{center}
     \vspace{-.5cm}
\caption{\label{SI_distdelr_ps} (a) Distribution of single particle non-affine  displacements $\delta r$ during an avalanche for several strain amplitudes. The fitted dashed lines for $\delta r>\delta r_c $, represent the exponential tail. Vertical lines indicate the choice of various $\delta r_c$ for which we perform avalanche analysis. (b)-(d) The cluster size distribution of active particles ($\delta r> \delta r_c$) for several strain amplitudes. Different panels are for different choice of $\delta r_c=0.15,0.20,0.30$. Data for $\delta r_c=0.25$ is shown in main text.}
%\end{figure*}
%\begin{figure*}[!h]
   \vspace{.5cm}
\centering{ \includegraphics[width =
    0.3\textwidth]{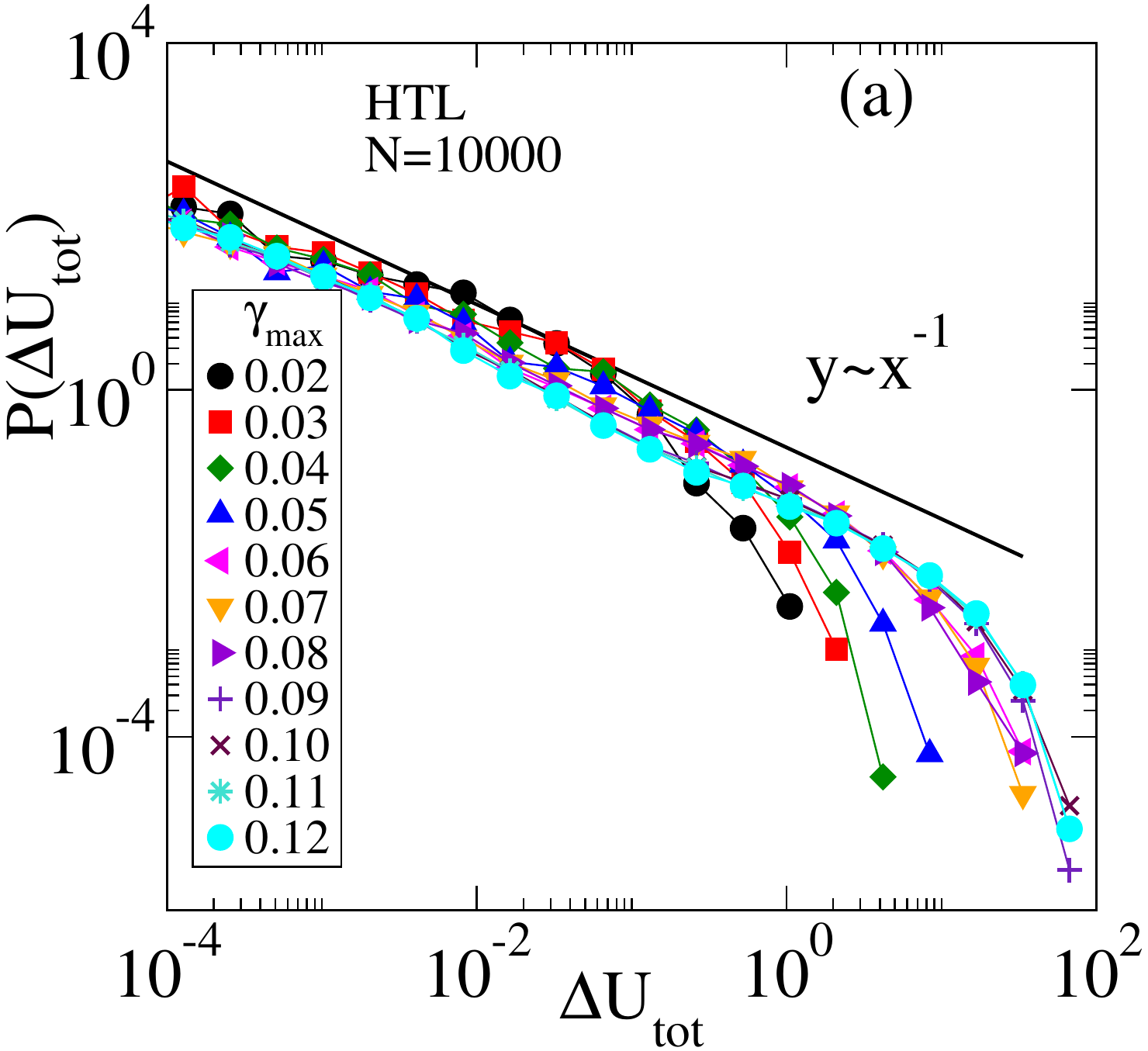}
  \includegraphics[width =
    0.3\textwidth]{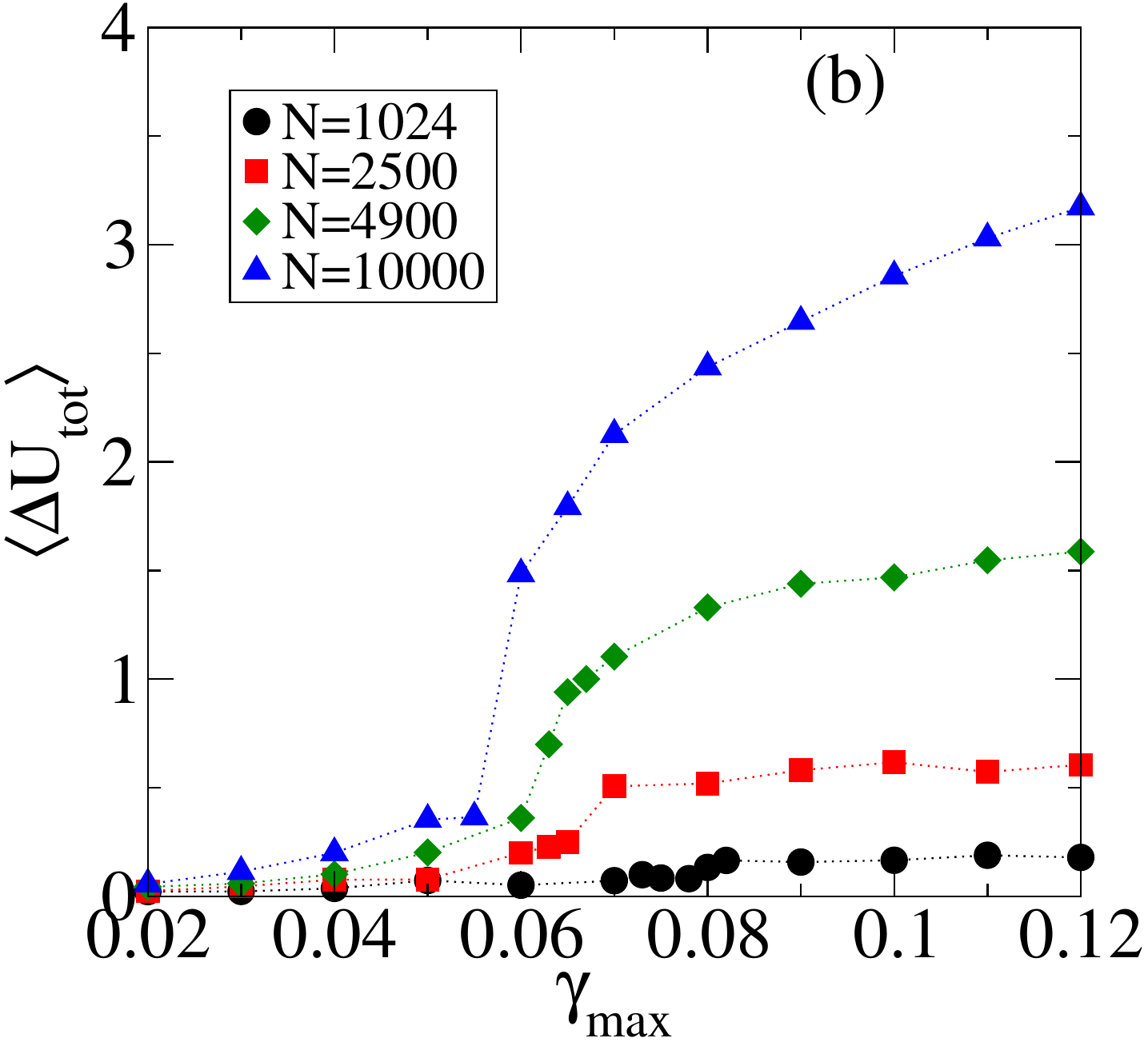}
    \includegraphics[width =
    0.3\textwidth]{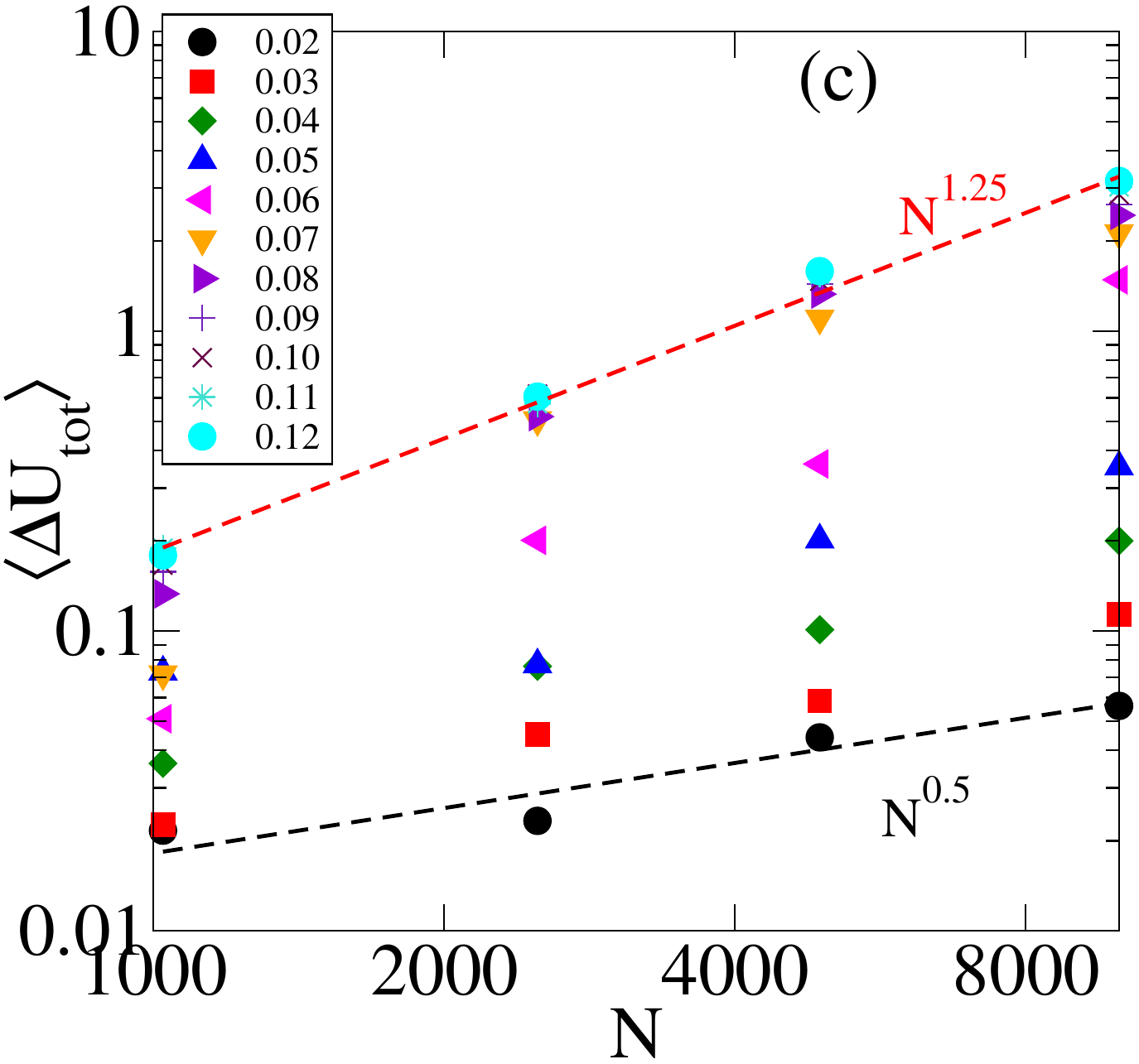}
    }
\centering{ \includegraphics[width =
    0.3\textwidth]{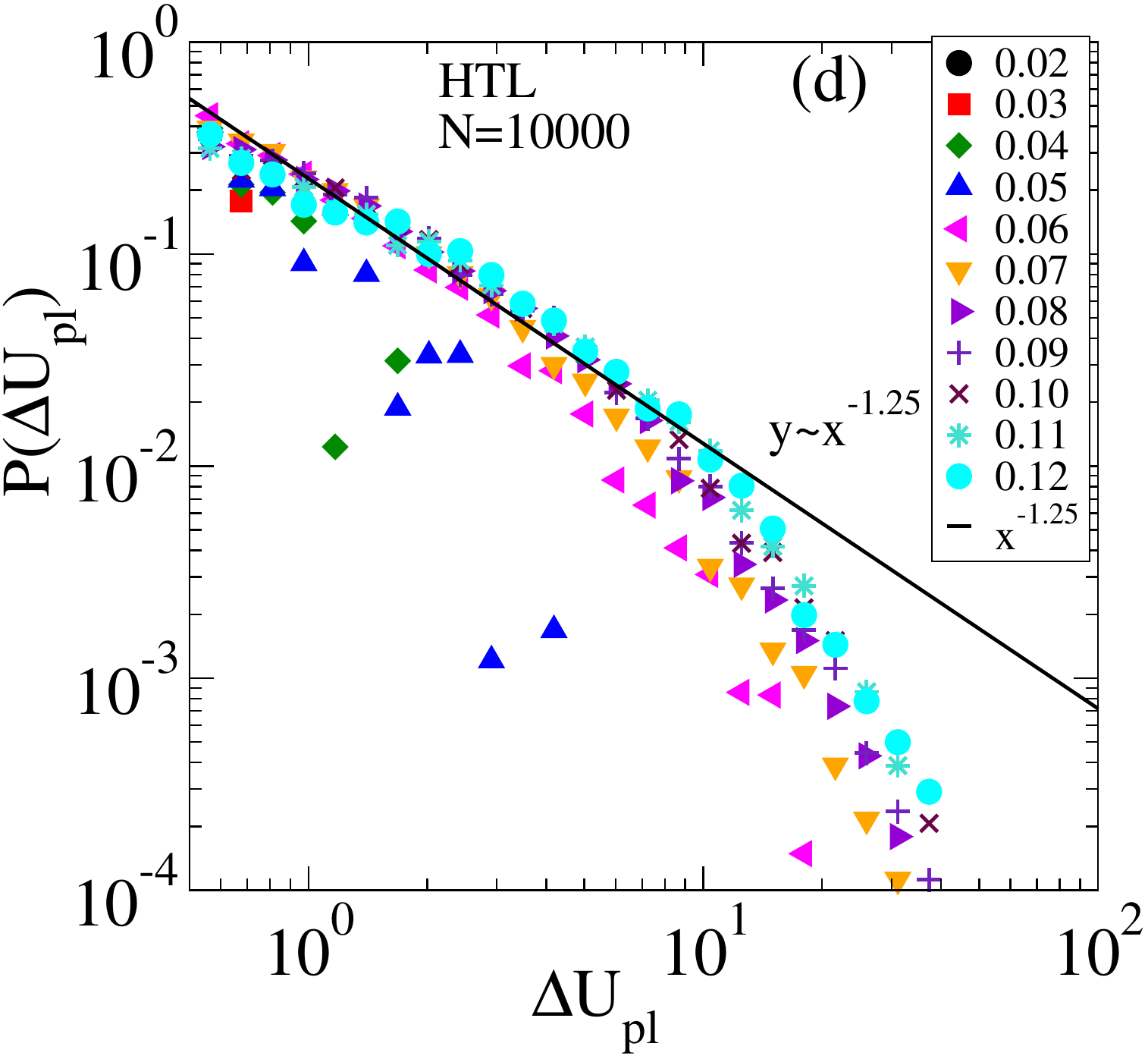}
  \includegraphics[width =
    0.3\textwidth]{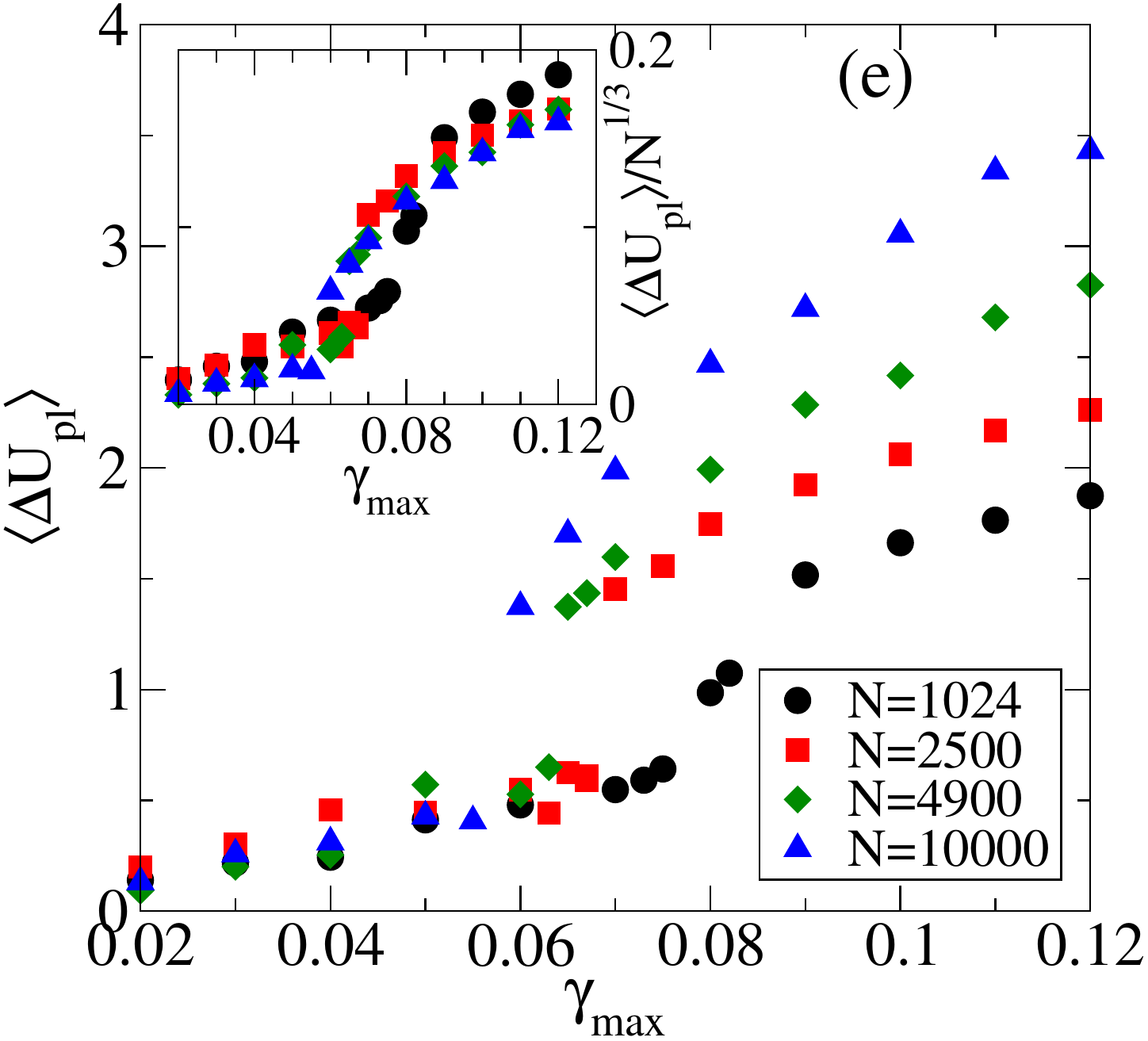}
    \includegraphics[width =
    0.3\textwidth]{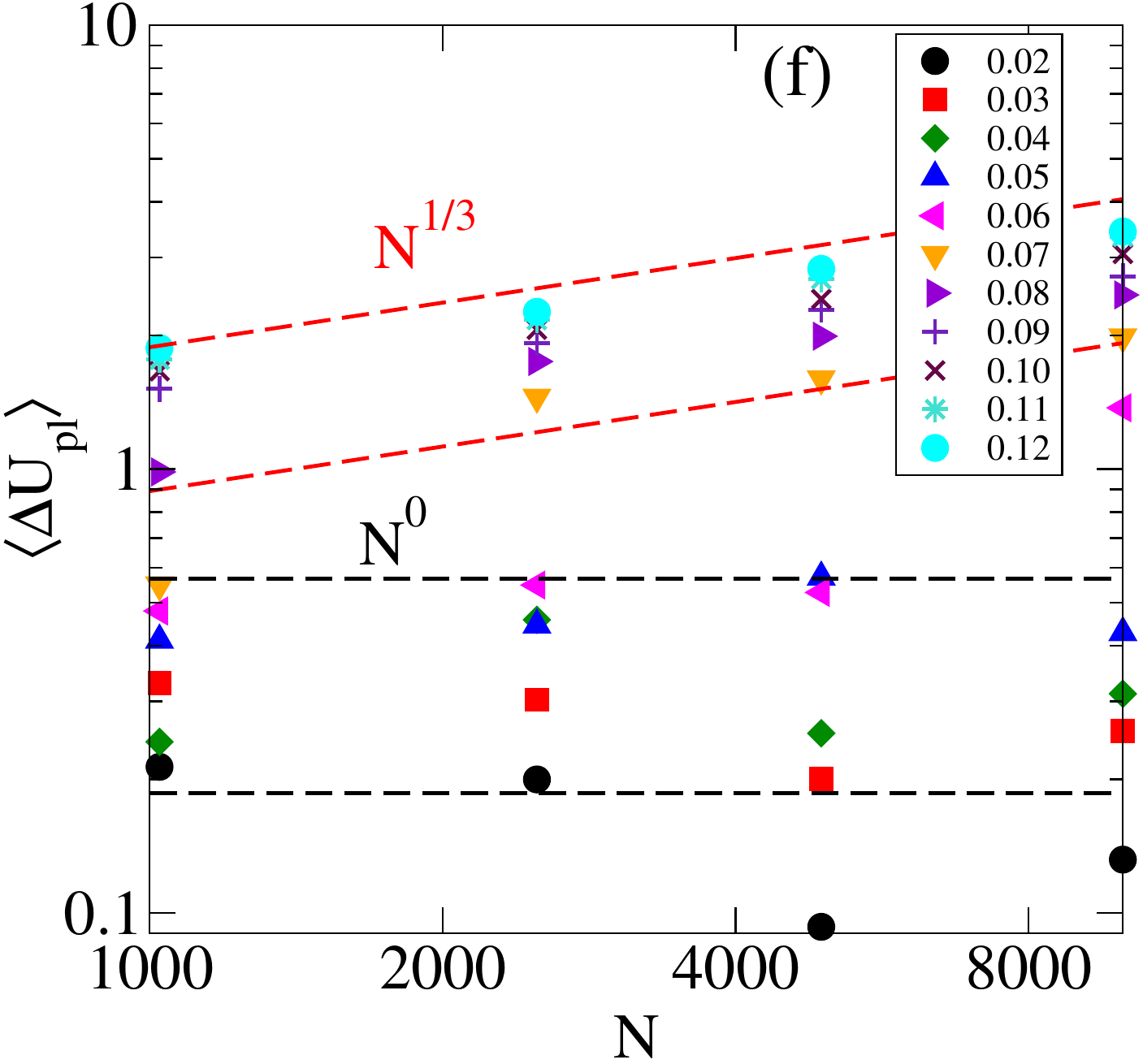}
    }
    \vspace{-.3cm}
\caption{\label{SI_delE} (a) Distribution of the total energy drop $\Delta U_{tot}$ during a plastic rearrangement, for different strain amplitudes, for $N=10000$. (b) Average $\langle \Delta U_{tot} \rangle$ against $\gamma_{max}$ for different system sizes. (c)  $\langle \Delta U_{tot} \rangle$ against system size for different strain amplitudes. (d) Distribution of the plastic component of the energy drop $\Delta U_{pl}$ for different $\gamma_{max}$. (e) Variation of $ \langle \Delta U_{pl} \rangle$ against $\gamma_{max}$ for different system sizes. Inset shows the scaling behaviour of $ \langle \Delta U_{pl}\rangle \sim N^{1/3}$ in the post-yield regime. (f) $\langle \Delta U_{pl} \rangle$ against system size for different strain amplitudes.}
\end{figure*}

In Fig. \ref{SI_distdelr_ps}(a) we show the distribution of single particle non affine displacements $P(\delta r)$ during a plastic event for various strain amplitudes. Such distributions are expected to have a power-law form with exponent $-3$ followed by an exponential tail at large $\delta r$, separated by a cutoff $\delta r_c$. The power-law part corresponds to elastic response whereas the exponential tail is associated with the plastic rearrangements. We fit the exponential part for the large $\delta r$ range for different $\gamma_{max}$ and observe that the cutoff value $\delta r_c$ varies approximately from $0.15$ to $0.35$ for the range of strain amplitudes we consider. Choosing various cut-off $\delta r_c=0.15,0.20,0.25$, and $0.30$ we identify the active particles undergoing plastic rearrangements that have $\delta r> \delta r_c$. For each choice of $\delta r_c$ we compute the cluster size distributions, as shown in Figs. \ref{SI_distdelr_ps} (b)-(d), which reveal that the results are almost independent of the choice of $\delta r_c$ in this range. We use $\delta r_c = 0.25$ in the analysis presented.


\subsection{Statistics of Energy Drops}
The plastic rearrangements lead to drops in energy that are often used as a quantifier of avalanche size. However, the total energy drop $\Delta U_{tot}$ for the
whole system during a plastic rearrangement contains both an elastic and a plastic component. The energy drop arising from the plastic core, $\Delta U_{pl}$, is computed considering only active particles that are identified by the procedure described before. In Figs. \ref{SI_delE} (a)-(c) we present the statistics of $\Delta U_{tot}$ and in \ref{SI_delE} (d)-(f) we show the results for $\Delta U_{pl}$. Both the distributions $P(\Delta U_{tot})$ and $P(\Delta U_{pl})$ display power-law regimes followed by a $\gamma_{max}$ dependent cutoff. However, the associated exponent describing the power-law regime is found to be different, $-1$ for $\Delta U_{tot}$ and $-1.25$ for $\Delta U_{pl}$. The mean energy drops, $\langle \Delta U_{tot}\rangle$ and $\langle \Delta U_{pl}\rangle $, show a similar dependence on strain amplitude $\gamma_{max}$, but with $\langle \Delta U_{tot}\rangle$ displaying a system size dependence below yielding, which is not present for $\langle \Delta U_{pl}\rangle $. 
The system size dependence of $\langle \Delta U_{tot}\rangle$ can be described by power laws ranging from $N^{1/2}$ to $N^{1.25}$, whereas $\langle \Delta U_{pl}\rangle $ does not display a system size dependence for $\gamma_{max} < \gamma_{y}$ and a $N^{1/3}$ dependence, similar to the mean cluster size, above  $\gamma_{y}$. 

\clearpage

\paragraph*{Acknowledgements.} 
We thank J. Horbach, S. Zapper and especially A. Rosso for useful discussions and comments on the manuscript. We acknowledge Indo-French Center for the Promotion of Advanced Research (IFCPAR/CEFIPRA) Project 5704-1 for support, the Thematic Unit of Excellence on Computational Materials Science, and the National Supercomputing Mission facility (Param Yukti) at the Jawaharlal Nehru Center for Advanced Scientific Research for computational resources. S.S. acknowledges support through the J. C. Bose Fellowship  (JBR/2020/000015)  SERB, DST (India).

\bibliographystyle{apsrev4-1} 
\bibliography{ref}
\clearpage

\end{document}